\pdfoutput=1
\documentclass{article}

\usepackage{iclr2027_conference,times}
\usepackage{float}
\usepackage[utf8]{inputenc} %
\usepackage[T1]{fontenc}    %
\usepackage[hidelinks]{hyperref}
\usepackage{url}            %
\usepackage{tabularx}

\usepackage{booktabs}       %
\usepackage{amsfonts}       %
\usepackage{nicefrac}       %
\usepackage{siunitx}
\usepackage{xcolor}
\usepackage{colortbl}
\definecolor{ours}{rgb}{0.94,0.96,1.0}
\arrayrulecolor{black!55}

\newcommand{\dmidp}[1]{{$(#1\%)$}}
\newcommand{\dsevp}[1]{{\textcolor{red!72!black}{$(#1\%)$}}}

\usepackage{microtype}      %
\usepackage{amsmath}
\usepackage{amssymb}
\usepackage{newtxmath} %
\usepackage{graphicx}
\usepackage{multirow}
\usepackage{makecell}
\usepackage[list=false]{subcaption}
\usepackage{adjustbox}
\usepackage{tikz}
\usepackage{needspace}
\usepackage{placeins}
\providecommand{\FloatBarrier}{\clearpage}
\usetikzlibrary{arrows.meta, positioning, shapes.geometric, fit, calc, backgrounds, patterns}
\usepackage{etoolbox}
\DeclareMathVersion{tabletext}
\DeclareSymbolFont{tabledigits}{OT1}{ptm}{m}{n}
\SetSymbolFont{tabledigits}{bold}{OT1}{ptm}{b}{n}
\SetMathAlphabet{\mathbf}{tabletext}{OT1}{ptm}{b}{n}
\newcount\tabledigitcode
\newcommand{\tablemathdigits}{%
  \mathversion{tabletext}%
  \tabledigitcode=48\relax
  \loop\mathcode\tabledigitcode=\numexpr"7000+\symtabledigits*256+\tabledigitcode\relax
    \advance\tabledigitcode by1
  \ifnum\tabledigitcode<58\repeat}
\AtBeginEnvironment{table}{\tablemathdigits}
\AtBeginEnvironment{tabular}{\tablemathdigits}
\AtBeginEnvironment{tabularx}{\tablemathdigits}
\graphicspath{{figures/}}
\newcommand{\colhead}[1]{{\bfseries\boldmath #1}}

\title{What File-Level Hits Hide in Bug\\\mbox{Localization:} Paths, Functions, and Repair}
\newcommand{\PathSwapAuthorBlock}{%
  {\normalsize\bfseries \mbox{Libin Chen\textsuperscript{1}}\quad \mbox{Quanyou Lao\textsuperscript{2}}\quad \mbox{Luyao Wang\textsuperscript{3}}\quad \mbox{Hongfu Liu\textsuperscript{1}}\par
   \mbox{Jing Chen\textsuperscript{1}}\quad \mbox{Chuan Shi\textsuperscript{2}}\quad \mbox{Muhan Zhang\textsuperscript{4,5}}\par}
  \smallskip
  {\footnotesize\begin{tabular}{@{}p{1em}@{}p{\dimexpr\textwidth-1em\relax}@{}}
    \textsuperscript{1} & College of Intelligence Science and Technology, National University of Defense Technology, Changsha, China \\
    \textsuperscript{2} & School of Computer Science, Beijing University of Posts and Telecommunications, Beijing, China \\
    \textsuperscript{3} & College of Systems Engineering, National University of Defense Technology, Changsha, China \\
    \textsuperscript{4} & Institute for Artificial Intelligence, Peking University, Beijing, China \\
    \textsuperscript{5} & Norizon AI
   \end{tabular}\par}}
\author{\PathSwapAuthorBlock}
\hypersetup{pdftitle={What File-Level Hits Hide in Bug Localization: Paths, Functions, and Repair},pdfauthor={Libin Chen, Quanyou Lao, Luyao Wang, Hongfu Liu, Jing Chen, Chuan Shi, Muhan Zhang}}

\iclrfinalcopy
\renewcommand{\headrulewidth}{0pt}
\makeatletter
\def\@maketitle{\vbox{\hsize\textwidth
  {\LARGE\sc \@title\par}
  \vskip .13in
  {\setlength{\parindent}{0pt}\@author\par}
  \vskip .19in minus .05in}}
\makeatother

\begin{document}
\raggedbottom

\maketitle

\begin{abstract}
When a bug localizer ranks a reference-changed file first, benchmarks record a success. For a developer or repair agent, however, finding that file leaves open what evidence drove the choice, which function needs to change, and whether it leads to a working fix. Changing only displayed paths---to opaque identifiers or plausible names from unrelated repositories---while holding issue reports, candidate pools, and short code previews fixed separates naming cues from code use. Rerankers trained with and without code previews score alike on original paths yet diverge sharply once names vanish, and plausible wrong paths redirect code-visible rerankers while code stays in view. Localizers that search and open complete files also diverge under opaque paths: a path-first pipeline collapses and some tool-using systems drop sharply, whereas another holds steady. Downstream, file hits routinely miss the changed function under name-and-summary scoring even when an eligible target sits in the candidate pool, with accuracy shifting markedly across scoring interfaces and code visibility. In separate repair settings, finding that function strongly tracks passing tests when a fixer must edit only a supplied function, yet remains uncertain when it receives the whole file and chooses where to edit. File-level recall therefore conflates matching filenames with reading source code, and leaves both the function to edit and a working fix unresolved.
\end{abstract}

\section{Introduction}\label{sec:intro}

Suppose a developer reports that login fails after a token refresh, and a bug localizer ranks \texttt{auth/login.py} first. Standard benchmarks award a file hit whenever that file is modified by the reference fix~\citep{jimenez2024swebench,xia2024agentless,zhang2024autocoderover,chen2025locagent,reddy2025swerank}. That signal is genuinely useful: it narrows a large repository to somewhere worth opening. Yet for a developer or repair agent, a file hit leaves two practical questions unanswered. First, why did the model choose that file? Given both a path and a short code preview, a reranker may have recognized the token-validation logic, matched the words \texttt{auth} and \texttt{login}, or blended the two---and top-one recall gives every route the same credit. Second, finding the file still leaves open which function needs to change and whether locating that function helps produce a working patch.

This ambiguity matters because file-level scores guide which models, prompts, and training recipes the field adopts. Informative paths are legitimate evidence: developers name files after their responsibilities, and many localization pipelines inspect directory trees before opening source files~\citep{xia2024agentless}. For a reranker given both names and code, however, high accuracy on original paths never reveals how much the model leans on naming conventions, or how it behaves when names are generic, outdated, or misleading. Ranking objectives reward any cue that lifts a changed file above distractors, so a path shortcut can improve benchmark scores while hiding sensitivity to that cue~\citep{geirhos2020shortcut}.

A controlled counterfactual makes this reliance testable. On a candidate pool retrieved once from the original repository, we hold the issue report, candidate files, short top-of-file code previews, and model weights fixed while changing only the displayed paths---replacing them with opaque identifiers or plausible names borrowed from unrelated repositories, alongside single-pair path exchanges and real Git renames. Input-ablation and perturbation studies have used similar controls to probe what benchmark accuracy actually reflects~\citep{poliak2018hypothesis,gururangan2018annotation,gupta2025sacl,logicloc2026,repomirage2026}. Fixing the candidate pool isolates the reranking decision itself, and matched training controls show why original-path recall alone can mislead: on SWE-bench Lite, Qwen2.5-14B rerankers trained with and without 50-line code previews score almost identically on original paths (57.0\% vs.\ 55.7\% Recall@1), yet diverge sharply once path names become opaque (39.7\% vs.\ 0.3\%).

Because a 50-line header preview withholds most of a file, we also test what happens when localizers can search a repository and open complete source files. Full-file tool access does not guarantee path invariance: on 193 SWE-bench Lite issues, opaque handles collapse a path-first pipeline (Agentless) and cause a steep drop in GPT-4.1-mini, yet leave a stronger reasoning configuration (\texttt{gpt-6.1-sol}) steady. We then trace file hits through two downstream stages. First, following GREPO file hits into name-and-summary function scoring separates candidate availability from scorer choices, showing that adapted scorers often miss the changed function even when an eligible target sits in their pool---and that stronger listwise scoring or complete-file reading changes that picture substantially in separate comparisons. Second, in separate SWE-bench Lite repair cohorts, we examine how localized function correctness relates to executed test outcomes when a fixer receives either a whole file or only a single supplied function.

\paragraph{Contributions.}
\begin{enumerate}
\setlength{\itemsep}{2pt}\setlength{\parskip}{0pt}
\item \textbf{Exposing what file hits conceal (\S\ref{sec:setup}--\ref{sec:training_main}, \S\ref{sec:revision-source-tests}).} Controlled displayed-path interventions and matched code-preview recipes show how nearly identical original-path Recall@1 can hide deep differences in code reliance, while full-file repository-navigation tests show that tool access yields widely different path robustness across models and pipelines.
\item \textbf{Separating file hits from function localization (\S\ref{sec:metric_validity}).} Following file hits into function selection shows that target availability is rarely the bottleneck under name-and-summary scoring, and demonstrates how strongly function accuracy responds to the scoring interface, explicit function names, and full-file context.
\item \textbf{Showing how editing interfaces shape the value of function localization (\S\ref{sec:consequence}).} Separate whole-file and single-function repair cohorts clarify why pinpointing the right function is decisive for function-scoped editors yet far harder to separate from editing headroom when a fixer receives the whole file.
\end{enumerate}

\section{Testing Paths on a Fixed Candidate Pool}\label{sec:setup}\label{sec:protocol}

\textbf{Scoring interface.} We use CodeGRIP as our experimental testbed for file reranking; its full pipeline proposes candidate files from an issue report, expands them using retrieval and repository relations, and reranks the pool (Appendix~\ref{app:impl}). To isolate the ranking step in our SWE-bench path experiments, we build a BM25 top-100 file pool once from the original repository and hold it fixed across conditions. A LoRA-adapted Qwen2.5 reranker scores each report--file pair independently from the bug report, the displayed file path, and---for code-visible variants---a preview of up to 50 opening source lines (capped at 1{,}500 characters inside a 768-token prompt), outputting the logit margin $s_\theta(q,f)=z_{\mathrm{Yes}}-z_{\mathrm{No}}$. Because a 50-line prefix often covers imports, docstrings, and headers rather than a bug deeper in the file, this interface specifically measures how short-preview rerankers weigh displayed paths against opening source lines; full-file inspection is evaluated separately with tool-using localizers in \S\ref{sec:revision-source-tests}.

\textbf{Training the reranker.} Although inference scores each candidate file independently, training couples a positive file $f_0$ with $m$ hard negatives $f_1,\ldots,f_m$ through a group-softmax objective:
\begin{equation}
\ell(q;f_0,\ldots,f_m)=-\log\frac{\exp s_\theta(q,f_0)}{\sum_{j=0}^{m}\exp s_\theta(q,f_j)}.
\label{eq:group_softmax_main}
\end{equation}
Our primary 14B recipes train fresh LoRA adapters from Qwen2.5-14B-Instruct on 5{,}916 training groups (1{,}870 from SWE-bench and 4{,}046 from GREPO), each paired with eight BM25-hard negatives. Code-visible training includes up to 50 opening lines. To test whether reducing path reliance during training changes behavior, a matched path-augmented recipe (\textbf{Codeaware}) shuffles displayed paths on half of its training examples while keeping code previews and target labels intact. Separate 7B factorial and adaptation recipes follow the same scoring interface and are detailed in Table~\ref{tab:hyperparams}.

\Needspace{6\baselineskip}
\textbf{Four displayed-path interventions.} Each intervention alters only displayed file paths while leaving issue text, candidate files, and source previews untouched (construction details in Appendix~\ref{app:pathalias}):
\begin{itemize}
\setlength{\itemsep}{2pt}\setlength{\parskip}{0pt}
\item \textbf{PathAlias} replaces repository-specific path components with readable names drawn from unrelated repositories, matching extension, directory depth, separators, casing, and character- and tokenizer-length bins while retaining common structural names such as \texttt{tests}. Each mapping is one-to-one within a repository and applied to every candidate; evaluation averages across ten independent mappings.
\item \textbf{PathSwap} replaces every path component with a deterministic opaque identifier while preserving depth, extension, and shared folders (turning \path{cirq/value/linear_dict.py} into \path{d_f4e8/d_a406/m_008e.py}). This fixed-pool control strips readable naming cues; the interactive navigation study in \S\ref{sec:revision-source-tests} uses a separate alphanumeric handle map built for tool use.
\item \textbf{Path exchange} swaps the displayed paths of just two files---a correctly ranked gold file and one distractor (the highest-ranked distractor, a same-directory sibling, or a random file)---to test which cue wins when path and preview directly conflict.
\item \textbf{Historical renames} display each gold file under its earliest path recorded in Git history across the 862 GREPO issues whose gold file was actually renamed.
\end{itemize}

\begin{figure*}[!htbp]
\centering
\includegraphics[width=\textwidth]{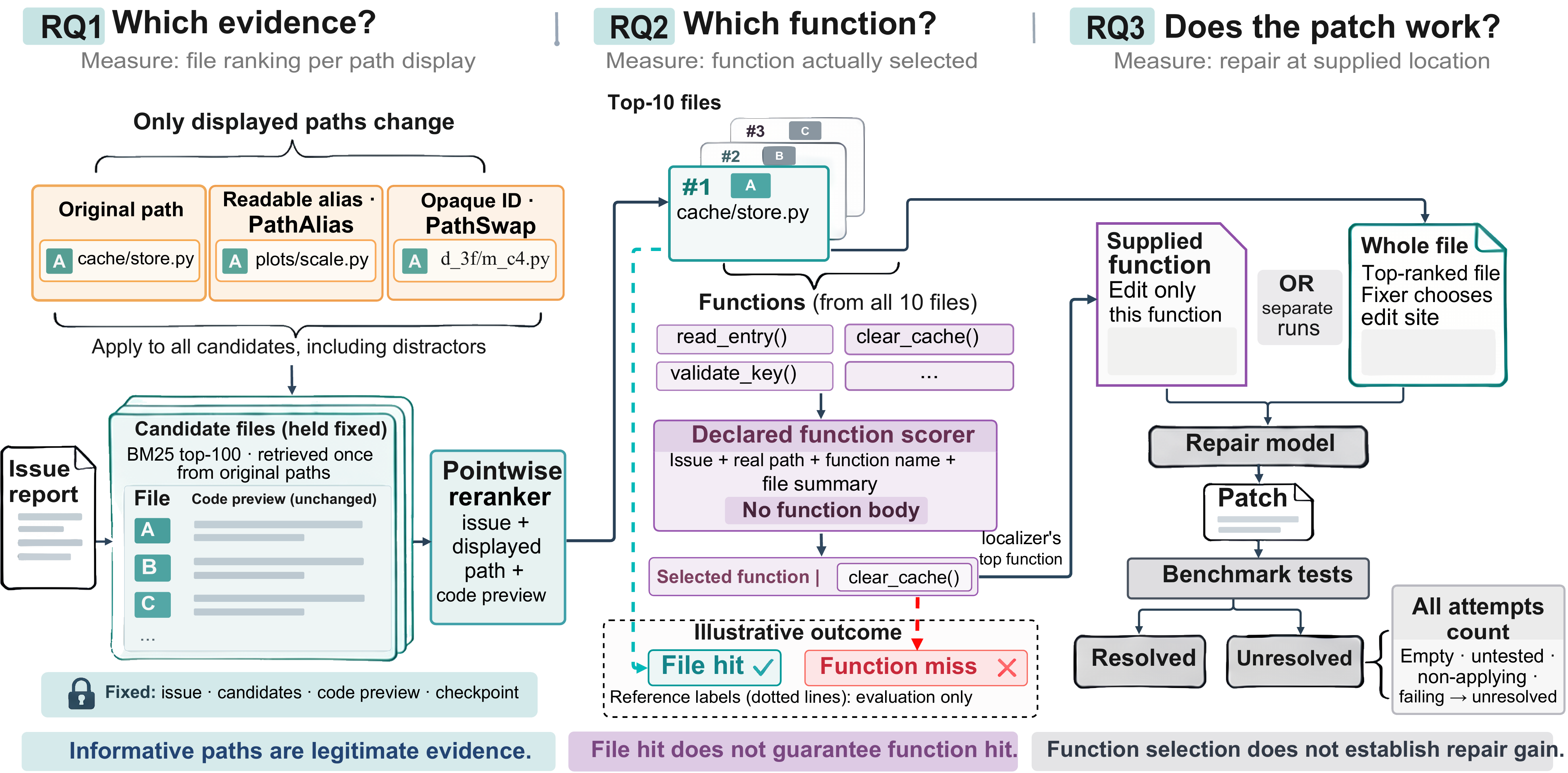}
\caption{A file hit is only the first checkpoint. The three columns illustrate our three main measurements on separate populations (paths, functions, and code shown are illustrative). \textbf{RQ1 (Path sensitivity, SWE-bench Lite primary; Verified sensitivity):} On a fixed BM25 top-100 pool with unchanged issue text and code previews, only displayed paths change---to readable unrelated aliases (PathAlias) or opaque structure-preserving IDs (PathSwap)---before pointwise reranking. \textbf{RQ2 (Function transfer, GREPO):} Given the top-10 predicted files, a declared function scorer ranks candidate functions using the issue, real path, function name, and a short file summary (without function bodies), so a file hit can still coexist with a function miss. \textbf{RQ3 (Executed repair, SWE-bench Lite):} Separate runs supply a fixer with either the entire top-ranked file (where the fixer chooses where to edit) or only the localizer's top function; all attempts, including empty, untested, and non-applying patches, enter test-resolution rates. Dotted lines mark evaluation against reference labels rather than model inputs; arrows trace workflow stages rather than causal effects.}
\label{fig:overview}
\end{figure*}

Figure~\ref{fig:overview} maps these three stages and their distinct populations.

\textbf{Populations, checkout protocols, and metrics.} We compare the primary unaugmented and path-augmented (Codeaware) 14B rerankers alongside a matched no-code 14B control and a three-seed 7B factorial (\S\ref{sec:training_main}). Checking normalized repository and issue IDs against the 5{,}916-group training set shows zero overlap with SWE-bench Lite (0/300) and heavy overlap with SWE-bench Verified (384/500, all from the 1{,}870-group SWE-bench training slice; 93 Lite issues also appear in Verified). Lite is therefore issue-ID held out within its 12 repositories---without claiming unseen repositories, unseen issue text, or exclusion from base-model pretraining---and serves as our primary SWE-bench evaluation; Verified acts strictly as an overlap-limited sensitivity check. Function-transfer studies evaluate GREPO's 1{,}704 Python test issues.

Our headline Lite path evaluations rebuild both BM25 pools and 50-line previews from each issue's exact base commit (reported on all 300 issues and on a 259-issue date-filtered subset). Several auxiliary controls---including path exchange, structural masking, matched training controls, and Verified panels---come from an earlier single-snapshot repository checkout and are labeled by checkout protocol and inference precision (bf16 vs.\ 4-bit), with every comparison kept paired within its own evaluation run. File accuracy is fractional Recall@1 (crediting $1/|G|$ when the top file belongs to gold set $G$); analyses conditioned on a file hit use the binary event that the top-ranked file was changed. Uncertainty intervals are paired 95\% bootstrap CIs (protocol summary in Appendix~\ref{app:moved_maintext}).

\section{Readable Wrong Paths Redirect Code-Visible Rerankers}\label{sec:shortcut}\label{sec:analysis}

\subsection{Readable aliases versus opaque identifiers}

\begin{figure}[!htbp]
\centering
\includegraphics[width=\linewidth]{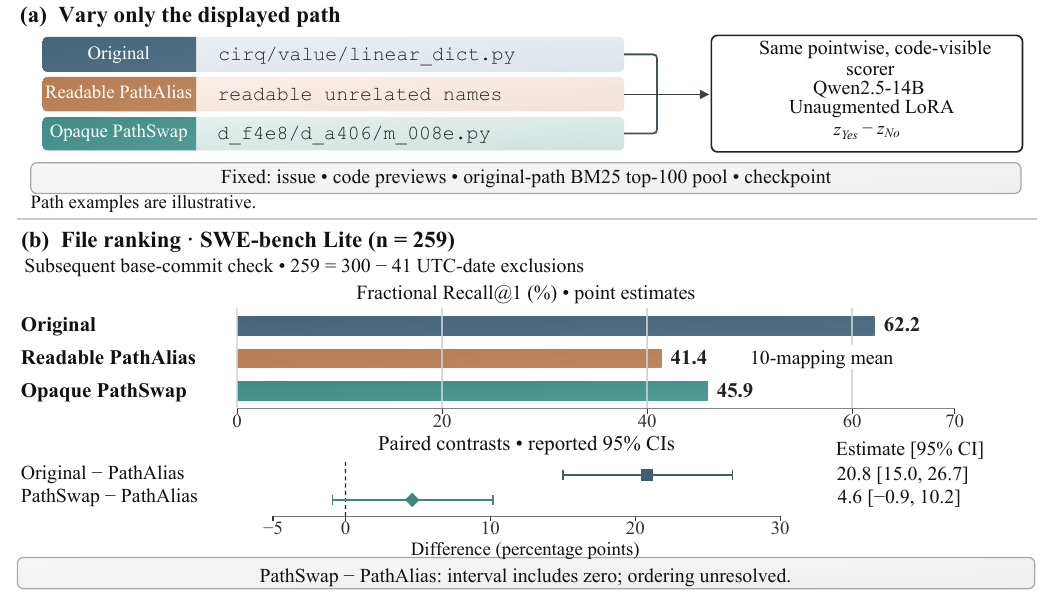}
\caption{Displayed paths and file ranking with code visible. (a) Each candidate receives its original path, a readable unrelated PathAlias, or an opaque PathSwap ID while the issue, candidate pool, 50-line previews, and reranker stay fixed (paths and code are illustrative). (b) Fractional Recall@1 for unaugmented Qwen2.5-14B on 259 date-filtered SWE-bench Lite issues rebuilt from each issue's base commit (ten-mapping mean for PathAlias; 95\% crossed issue-by-mapping CIs). Both interventions cause large drops from original paths, while the PathSwap-minus-PathAlias interval includes zero.}
\label{fig:main_pathalias}
\end{figure}

Both PathAlias and PathSwap erase a file's original name while holding the issue report, candidate pool, and 50-line code previews fixed, yet they test distinct failures: PathSwap strips lexical meaning altogether, whereas PathAlias supplies plausible names borrowed from unrelated repositories. Comparing both against original paths measures how strongly a code-visible reranker relies on displayed path text, and asks whether misleading readable names disrupt rank-one recall more than opaque identifiers (Figure~\ref{fig:main_pathalias}).

On SWE-bench Lite rebuilt from each issue's exact base commit, replacing original paths cuts top-one accuracy sharply even though all opening source lines remain visible. On the 259 date-filtered issues in Figure~\ref{fig:main_pathalias}b---a conservative timestamp screen excluding 41 issues whose legacy snapshot postdated PR creation, without assuming post-fix code in all of them---unaugmented Qwen2.5-14B loses roughly 21 percentage points of Recall@1 under PathAlias and 16 under PathSwap. All 300 base-commit issues show nearly identical losses (Appendix~\ref{app:pathalias}). Both interventions clearly hurt relative to original paths, whereas whether readable wrong aliases disrupt rank-one recall more than opaque identifiers remains unresolved: under base-commit checkouts, both alias-versus-opaque confidence intervals span zero.

\paragraph{Single-snapshot baselines and Verified sensitivity.} Because several auxiliary controls---path exchange, structural masking, matched training controls, and Verified panels---come from an earlier single-snapshot checkout, each is compared against its matching single-snapshot baseline (Figure~\ref{fig:pathalias_fullmaps} and Table~\ref{tab:pathalias_cells}). Large losses persist across both checkouts and on the overlap-limited Verified sensitivity panel. Although PathAlias lowers Recall@1 slightly more than PathSwap on single-snapshot Lite, moving to per-issue base commits rebuilt candidate pools and previews together and lifted all arms by 5--6 points without isolating either factor. A complementary test-time structural-masking control (MaskPath) produces a matching drop (Table~\ref{tab:maskpath_v8}).

\subsection{Exchanging two files' paths moves the top prediction}\label{sec:path_exchange}

PathAlias alters every candidate in the pool at once. To put path name and short code preview into direct head-to-head conflict, we keep the rest of the pool untouched and swap only the displayed paths of a correctly ranked gold file and one distractor under the single-snapshot bf16 protocol (Table~\ref{tab:ps1_transplant}).

\begin{table}[!htbp]\centering\small
\caption{Single-pair path exchange (single-snapshot bf16 fixed-pool protocol). The gold file and one distractor swap displayed paths while all code previews and other candidates remain fixed. Capture is the share of a recipe's original correct-file predictions ($G$) that move to the distractor, with 95\% bootstrap CIs. Verified uses the 116 adaptation-disjoint issues. Full tier counts and path-margin shares appear in Table~\ref{tab:ps1_transplant_details}.}
\label{tab:ps1_transplant}
\setlength{\tabcolsep}{3.4pt}
\begin{tabular}{@{}llcc@{}}\toprule
{\colhead{Distractor}} & {\colhead{Recipe}} & {\colhead{Lite capture [CI]; $G$}} & {\colhead{Verified capture [CI]; $G$}}\\\midrule
Top-ranked & No augmentation & $77.9\ [71.5,83.7];172$ & $84.1\ [72.7,93.2];44$\\
Top-ranked & Codeaware & $18.6\ [13.1,24.6];183$ & $4.9\ [0.0,12.2];41$\\
Same directory & No augmentation & $80.2\ [74.1,86.4];162$ & $79.5\ [66.7,92.3];39$\\
Same directory & Codeaware & $11.7\ [7.0,17.0];171$ & $11.1\ [2.8,22.2];36$\\
Random & No augmentation & $63.4\ [55.8,70.3];172$ & $77.3\ [63.6,88.6];44$\\
Random & Codeaware & $1.1\ [0.0,2.7];183$ & $0.0\ [0.0,0.0];41$\\\bottomrule
\end{tabular}\end{table}

When the unaugmented 14B reranker ranks the gold file first on original Lite paths, handing that file's path to a single distractor pulls the top prediction onto the distractor in roughly two-thirds to four-fifths of cases---including 63.4\% for a random distractor that ranked first on original paths only 0.4\% of the time (Table~\ref{tab:ps1_transplant}; full breakdown in Table~\ref{tab:ps1_transplant_details}). Adaptation-disjoint Verified issues show comparable capture. Under the path-augmented Codeaware recipe, capture on its own original wins stays far lower. Because each recipe conditions on its own original wins ($G$ in Table~\ref{tab:ps1_transplant}) and the top-ranked distractor can vary by recipe, the cross-recipe contrast is descriptive rather than paired on shared wins; within each recipe, however, the swap directly reveals which cue governs the top rank when name and short preview disagree.

\subsection{Real renames and other rerankers}\label{sec:path_scope}

Synthetic swaps could overstate real-world impact, so we also evaluate historical file renames recorded in Git. Across 862 GREPO issues whose gold file was renamed, showing it under its earliest recorded path---with source code unchanged---lowers Recall@1 by 3.5 percentage points for unaugmented 14B and 2.2 points for Codeaware (both confidence intervals exclude zero), while Recall@5 and Recall@10 rise (Table~\ref{tab:organic_drift} in Appendix~\ref{app:organic_drift}). Outdated names alone therefore depress top-one accuracy, though far more modestly than wholesale renaming.

Strong path reliance also generalizes beyond one backbone or architecture. Path-only cross-encoders across Qwen2.5-7B, Llama-3.1-8B, and Qwen3-8B lose roughly three-quarters of Recall@1 under filename shuffling and collapse under PathSwap (Table~\ref{tab:path_controls_main}), and filename shuffling similarly hurts a bi-encoder, a listwise generator, frontier models prompted with 100-line previews, and SweRank's dense encoder when evaluated on path strings (Appendix~\ref{app:scope_studies}). By contrast, SweRank's native function-body retrieval-and-reranking pipeline reproduces 216/274 (78.8\%) strict file Acc@1 on unperturbed Lite issues and was not perturbed under that reproduced operating point (Appendix~\ref{app:swerank_repro}). Localizers that actively search and open full source files are examined next in \S\ref{sec:revision-source-tests}.

\section{Original-Path Recall@1 Can Conceal Code Use}\label{sec:training_main}\label{sec:mitigation}\label{sec:recipe_sensitivity}

If displayed paths pull code-visible rerankers so strongly, do those rerankers use their 50-line code previews at all, or mainly fall back on opening source lines when readable names disappear? Matched training and inference controls separate the two behaviors under the single-snapshot fixed-pool protocol (Table~\ref{tab:code_controls_main}).

\begin{table}[!htbp]
\centering\small
\caption{Code-preview controls on single-snapshot fixed BM25 top-100 pools (fractional Recall@1 in \%). Panel A compares matched Qwen2.5-14B LoRAs trained with and without 50-line previews on Lite ($n=300$). Panel B redacts 80\% of body lines from previews given to one fixed 7B Codeaware checkpoint on Lite and Verified ($n=500$, adaptation-overlap sensitivity). Full counts and paired tests appear in Tables~\ref{tab:matched_14b_codepreview} and~\ref{tab:body_deletion}.}
\label{tab:code_controls_main}
\begin{tabularx}{\linewidth}{@{}Xrrr@{}}
\toprule
\multicolumn{4}{l}{\colhead{\textbf{A. Matched training inputs: code present minus code absent}}}\\
{\colhead{Condition}} & {\colhead{With code}} & {\colhead{Without code}} & {\colhead{$\Delta$ pp}}\\\midrule
Normal & 57.00 & 55.67 & $+1.33$\\
PathSwap & 39.67 & 0.33 & $+39.33$\\\midrule
\multicolumn{4}{l}{\textbf{B. Fixed checkpoint: redacted minus original preview}}\\
{\colhead{Split / condition}} & {\colhead{Original}} & {\colhead{Redacted}} & {\colhead{$\Delta$ pp}}\\\midrule
Lite / Normal & 49.33 & 48.33 & $-1.00$\\
Verified / Normal & 43.32 & 41.80 & $-1.52$\\
Lite / PathSwap & 40.67 & 26.67 & $-14.00$\\
Verified / PathSwap & 32.23 & 24.53 & $-7.70$\\\bottomrule
\end{tabularx}
\end{table}

At a matched 768-token prompt budget on single-snapshot Lite, two Qwen2.5-14B LoRAs trained identically except for whether they receive 50-line code previews look virtually indistinguishable on original paths, yet separate by nearly 40 points once paths turn opaque (Table~\ref{tab:code_controls_main}A; paired tests in Appendix~\ref{app:codeprompt_14b_matched}). Because Panel A compares two complete training recipes, Panel B isolates inference-time code use inside a single fixed 7B Codeaware checkpoint by redacting most body lines while leaving imports, headers, and the first non-blank line intact: original-path accuracy barely moves, whereas opaque-path recall drops sharply on both Lite and Verified (Table~\ref{tab:code_controls_main}B). Together, the two controls show that close original-path scores hide substantial fallback use of opening source lines once readable paths vanish. Every contrast stays paired within its own run; Appendix~\ref{app:codeprompt_14b_matched} reconciles baseline counts across all four Lite runs of the unaugmented 14B checkpoint.

\begin{table}[!htbp]
\centering\small
\caption{Code-preview by path-augmentation factorial on single-snapshot Lite ($n=300$). Three-seed Qwen2.5-7B models (SWE-only training; fractional Recall@1 mean $\pm$ SD) are evaluated under the fixed 50-line preview interface; training setup, full effects, and paired seed-42 tests appear in Tables~\ref{tab:cross_cut}, \ref{tab:mcnemar}, and~\ref{tab:hyperparams}.}
\label{tab:lever_2x2}
\setlength{\tabcolsep}{5pt}
\begin{tabular}{@{}lrrrr@{}}
\toprule
\multirow{2}{*}{\colhead{Path augmentation}} & \multicolumn{2}{c}{\colhead{Normal Recall@1 (\%)}} & \multicolumn{2}{c}{\colhead{PathSwap Recall@1 (\%)}}\\
\cmidrule(lr){2-3}\cmidrule(l){4-5}
& {\colhead{No code}} & {\colhead{Code preview}} & {\colhead{No code}} & {\colhead{Code preview}}\\\midrule
0\% & $53.00\pm0.88$ & $54.33\pm1.20$ & $1.44\pm0.51$ & $30.22\pm3.24$\\
50\% & $50.00\pm1.00$ & $48.56\pm2.59$ & $3.11\pm1.58$ & $36.78\pm4.86$\\\bottomrule
\end{tabular}
\end{table}

Crossing code previews with 50\% path-shuffling augmentation across three 7B seeds confirms this blind spot and highlights an accuracy--robustness trade-off (Table~\ref{tab:lever_2x2}). Code previews drive nearly all opaque-path resilience while leaving original-path Recall@1 flat, whereas path augmentation buys additional opaque-path recall at the cost of several points on original paths---so standard benchmarks and opaque-path stress tests favor different recipes. The same divergence appears on 14B, where Codeaware cuts the single-snapshot Lite PathAlias drop from 21.8 to 4.9 points while keeping original-path accuracy comparable or slightly higher (Table~\ref{tab:pathalias_cells}; paired 4-bit contrasts in Appendix~\ref{sec:codeaware:mcnemar}), and on GREPO, where path-only and code-preview rerankers achieve nearly identical original-path recall (Tables~\ref{tab:code_centric} and~\ref{tab:shared_scorer_mde}A).

\paragraph{Training with code under path-only inference.}
Does supervising with retrieved functions during training still help when the model sees only paths at test time? On 116 weak-path-cue GREPO issues (Code-Crucial, Appendix~\ref{sec:where_code}), training with up to three TF--IDF-retrieved functions per file---while feeding only the issue report and candidate path at evaluation---moves binary Hit@1 in opposite directions across two seeds (Table~\ref{tab:selective-code-main}), showing no consistent carryover once code is withheld at inference.

\begin{table}[!htbp]\centering\small
\caption{Training with retrieved functions on GREPO under path-only file scoring. Full-test reports fractional Recall@1 ($n=1{,}704$; three-seed mean $\pm$ SD for path-only); Code-Crucial reports binary Hit@1 on 116 weak-path-cue issues. All evaluations use original paths without code at inference.}
\label{tab:selective-code-main}
\begin{tabularx}{\linewidth}{@{}Xrr@{}}\toprule
{\colhead{Training recipe}} & {\colhead{Full-test Recall@1 (\%)}} & {\colhead{Code-Crucial Hit@1 (\%)}}\\\midrule
Path-only reference & $28.92\pm0.56$ & 41/116 (35.34)\\
Selective retrieval, seed 42 & 28.87 & 47/116 (40.52)\\
Selective retrieval, seed 123 & 28.00 & 37/116 (31.90)\\\bottomrule
\end{tabularx}\end{table}

\section{File Hits Often Leave the Function Unlocalized}\label{sec:metric_validity}\label{sec:metric_main}

\subsection{How often a file hit reaches the function}

\begin{figure}[t]
\centering
\includegraphics[width=\linewidth]{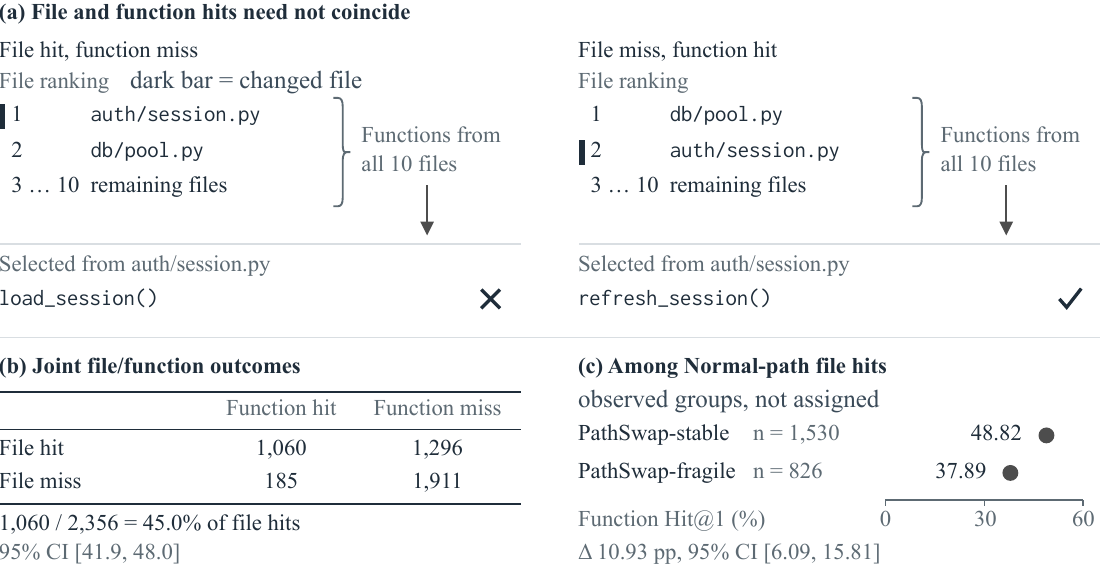}
\caption{File hits and selected functions on GREPO. (a) Illustrative example where the fix edits \texttt{refresh\_session()} in \texttt{auth/session.py}: a file hit (dark bar) can still end in a function miss. (b) Joint file and function outcomes across four Qwen2.5 rerankers using their own scorers on 1{,}113 common GREPO issues (4{,}452 records, issue-clustered 95\% CI). (c) Function Hit@1 among Normal-path file hits split by whether the file hit survives PathSwap (792 issue clusters).}
\label{fig:main_function_followthrough}
\end{figure}

Ranking the right file first solves only the coarse half of localization: a real source file often holds dozens of functions, and the developer or repair agent still needs to know which one to edit. We trace each reranker's ten highest-ranked GREPO files into a function-scoring stage that sees the bug report, real file path, a 200-character AST file summary, and each candidate function name, without candidate function bodies or parameter lists. Because candidates come from all ten files, a function hit can also come from a file ranked below first (Figure~\ref{fig:main_function_followthrough}a).

\suppressfloats[t]

Across four Qwen2.5 rerankers using their own adapted function scorers, fewer than half of original-path file hits---45.0\% (Figure~\ref{fig:main_function_followthrough}b)---yield a top-one changed function. Missing targets are rarely the bottleneck: on more than nine out of ten misses, an eligible changed function already sits inside the top-ten file pool. A separate SWE-bench Lite replication without file summaries shows a comparable rate (47.5\%, remaining stable after a manual label check; Appendix~\ref{app:metric_validity_confusion}). Path stability also tracks downstream transfer: file hits that survive PathSwap convert into function hits roughly 11 percentage points more often than fragile hits that flip to misses (Figure~\ref{fig:main_function_followthrough}c).

\subsection{Separating file ranking from function scoring}

\begin{table}[t]
\centering\small
\caption{File and function localization on GREPO. Panel A compares complete file-to-function pipelines on 1{,}113 common issues. Panel B holds each 14B reranker's file hits fixed and varies only the function scorer. Panel C holds a shared scorer fixed and compares each 14B reranker against the matched bf16/top-100 path-only snapshot on their common file wins ($\Delta$ is 14B minus path-only, paired 95\% bootstrap CI).}
\label{tab:shared_scorer_mde}
\begin{tabularx}{\linewidth}{@{}Xrr@{}}
\toprule
\multicolumn{3}{l}{\colhead{\textbf{A. Complete procedures on the same 1,113 GREPO issues}}}\\
{\colhead{System}} & {\colhead{File Recall@1 (\%)}} & {\colhead{Function Hit@1 (\%)}} \\
\midrule
Path-only & 27.62 & 33.96 \\
14B Codeaware & 29.65 & 33.24 \\
14B no-augmentation & 29.15 & 33.33 \\
\bottomrule
\end{tabularx}
\par\smallskip
\begin{tabularx}{\linewidth}{@{}Xrrrr@{}}
\toprule
\multicolumn{5}{l}{\colhead{\textbf{B. Within-system scorer changes on fixed file-win records}}}\\
{\colhead{14B reranker}} & {\colhead{$n$}} & {\colhead{Own (\%)}} & {\colhead{Fixed neutral (\%)}} & {\colhead{Name-blind (\%)}}\\\midrule
Codeaware & 649 & 47.77 & 27.73 & 8.63\\
No augmentation & 652 & 48.93 & 26.84 & 8.44\\\bottomrule
\end{tabularx}
\par\smallskip
\begin{tabularx}{\linewidth}{@{}Xrrl@{}}
\toprule
\multicolumn{4}{l}{\colhead{\textbf{C. Paired reranker contrasts on common file wins (matched bf16/top-100)}}}\\
{\colhead{Shared scorer / 14B reranker}} & {\colhead{$n$}} & {\colhead{$\Delta$ (pp)}} & {\colhead{95\% CI (pp)}}\\\midrule
Fixed neutral / Codeaware & 472 & $-0.21$ & $[-2.33,1.91]$\\
Fixed neutral / no augmentation & 481 & $-0.21$ & $[-2.29,1.87]$\\
Name-blind / Codeaware & 472 & $-0.85$ & $[-2.54,0.85]$\\
Name-blind / no augmentation & 481 & $0.00$ & $[-1.66,1.66]$\\\bottomrule
\end{tabularx}
\end{table}

Does a stronger file reranker automatically localize better functions, or does the function scorer dominate? Table~\ref{tab:shared_scorer_mde} separates the two stages. As complete pipelines on the 1{,}113 GREPO issues (Panel A), the path-only, 14B Codeaware, and 14B no-augmentation systems all converge around 33--34\% function Hit@1 (here ``path-only'' refers to the file stage; its function stage still sees real paths, AST summaries, and candidate function names). Holding each 14B reranker's file hits fixed and swapping only the function scorer (Panel B) reveals how much of the 48--49\% conversion rate rests on scorer adaptation and explicit function names: an unadapted \texttt{Qwen2.5-7B-Instruct} scorer cuts transfer nearly in half, and masking candidate names with opaque tokens such as \texttt{fn\_0000} drops it into single digits. Holding those shared scorers fixed on common file wins against the matched bf16/top-100 path-only snapshot (Panel C) shows nearly identical gold-function coverage ($\sim$96\%) and sub-point differences whose intervals all span zero (legacy 4-bit/top-200 sensitivity in Appendix~\ref{app:metric_validity_confusion}). While inconclusive rather than establishing equivalence, these comparisons detect no clear function-level advantage from the 50-line file preview under a shared name-and-summary scorer.

Importantly, low transfer in Panel B reflects the tested scoring system rather than a ceiling on the candidate pools. Simply appending candidate bodies to lexical or unadapted pointwise scorers hurts accuracy because of incomplete body recovery and prompt sensitivity---to the point that masking the explicit candidate name alongside bodies paradoxically helps the zero-shot pointwise scorer (Appendix~\ref{app:body_aware_2048}). Switching from that unadapted pointwise scorer to a hosted listwise scorer (\texttt{gpt-6.1-sol-2026-09-29}) on the same candidate shortlists and nominal no-body fields changes the picture markedly, lifting function Hit@1 by roughly 36 percentage points in both 14B cohorts (Table~\ref{tab:body_aware_2048}). Because this switch changes the underlying model, listwise-versus-pointwise format, context packing, and compute budget together---and because the two 14B file-hit cohorts overlap heavily in both issues and top-ranked files---it is a system-level contrast within each cohort rather than an isolated ablation, independent replication, or cross-reranker test (full counts and overlap details in Appendix~\ref{app:body_aware_2048}). Including brief body snippets inside the listwise prompt adds no further gain (Table~\ref{tab:body_aware_2048}), leaving unrestricted full-file code reading a separate question.

\subsection{Code access beyond fixed previews}\label{sec:revision-source-tests}

Because a 50-line preview hides most of each file, a natural next question is whether letting a localizer search a repository and open complete source files removes path sensitivity. Across 193 SWE-bench Lite issues, we evaluate four code-inspecting systems under original paths and under opaque handles built for interactive navigation (\texttt{PathSwap-agent-v1}; handle construction, cohort exclusions, and call budgets in Appendix~\ref{app:revision-navigation}). Pre-fix source code, embedded identifiers, issue reports, and the read-only search and file-reading interface stay identical across arms, and any run that fails to return a file counts as a miss.

\begin{table}[!htbp]
\begingroup
\centering\small
\caption{Read-only repository navigation with GPT-4.1-mini on 193 SWE-bench Lite issues, comparing opaque handles (\texttt{PathSwap-agent-v1}) against original paths on first-file Hit@1 (paired 95\% issue-bootstrap CI). Trajectories without a returned file---whether from hitting the request limit or from unresolved infrastructure errors---count as misses in the fixed 193-issue denominator (missing-run counts and bounds in Appendix~\ref{app:revision-source-studies}).}\label{tab:revision-source-main}
\begin{tabularx}{\linewidth}{@{}Xrll@{}}\toprule
{\colhead{Condition}} & {\colhead{Successes}} & {\colhead{Change (pp)}} & {\colhead{95\% CI (pp)}}\\\midrule
Original handles & 142/193 & reference & ---\\
Opaque handles & 92/193 & $-25.9$ & $[-32.1,-19.2]$\\\bottomrule
\end{tabularx}
\endgroup
\end{table}

Giving a localizer read-only tools to search and open complete files does not automatically make it path-invariant. Under opaque handles, GPT-4.1-mini loses 25.9 percentage points on first-file Hit@1 (Table~\ref{tab:revision-source-main}). Because some runs terminated without a final file after hitting the request limit or encountering unresolved infrastructure errors, we also bound the effect by leaving every unresolved infrastructure outcome unrestricted while retaining request-limit misses: even at the most favorable extreme, opaque handles still cut accuracy by at least 13 percentage points (identification bounds $-35.2$ to $-13.0$ points), and finalized runs show a similar gap (Appendix~\ref{app:revision-source-studies}). Missing trajectories therefore cannot explain away the drop.

\begin{table}[!tbp]
\begingroup
\centering\small
\caption{Further code-inspecting localizers on the same 193 Lite issues, comparing opaque handles (\texttt{PathSwap-agent-v1}) against original paths on first-file Hit@1 with paired 95\% issue-bootstrap CIs. Shuffled-filename arms and component diagnostics appear in Appendix~\ref{app:revision-code-inspecting}.}\label{tab:code_inspecting_localizers}
\begin{tabularx}{\linewidth}{@{}Xlrrll@{}}\toprule
{\colhead{System}} & {\colhead{Model}} & {\colhead{Original}} & {\colhead{Opaque}} & {\colhead{Change (pp)}} & {\colhead{95\% CI (pp)}}\\\midrule
Tool-using localizer (interface of Table~\ref{tab:revision-source-main}) & \texttt{gpt-6.1-sol} & 188/193 & 188/193 & $0.0$ & $[-1.6,1.6]$\\
Agentless & GPT-4o & 135/193 & 0/193 & $-69.9$ & $[-76.2,-63.2]$\\
Moatless & GPT-4o & 143/193 & 131/193 & $-6.2$ & $[-11.4,-1.0]$\\\bottomrule
\end{tabularx}
\endgroup
\end{table}

Broadening to a stronger reasoning model and two established localization pipelines shows why full-file access produces no single verdict (Table~\ref{tab:code_inspecting_localizers}). In the identical read-only interface and budget, the high-reasoning configuration \texttt{gpt-6.1-sol} finalizes every trajectory and matches its original-path hit rate under opaque handles---showing that a stronger model can rely on code search and file inspection when path names disappear. Established GPT-4o pipelines behave very differently depending on how heavily their workflow filters by path structure before reading code. Agentless v1.5.0~\citep{xia2024agentless}, which narrows the repository tree first, collapses to zero first-file hits under opaque handles (and loses most hits under shuffled filenames), even though every pipeline run completed and a handful of opaque top-three lists still place a gold file below rank one. By contrast, Moatless v0.0.2~\citep{orwall2024moatless} leans more directly on iterative code search and loses only 6.2 points under opaque handles, with a similar drop under shuffled filenames (shuffled-arm scores and component diagnostics in Appendix~\ref{app:revision-code-inspecting}). Because we did not run a within-model tools-on versus tools-off ablation, these contrasts reflect differences across tested models and architectures rather than isolating a causal effect of tool access alone; clearly, opening full source files does not by itself eliminate path reliance.

\paragraph{Complete-file function selection after an upstream miss.}
In a separate fixed-candidate comparison on SWE-bench Lite file hits (where upstream top functions come from rankings produced before \S\ref{sec:metric_validity}'s left-padding index fix; Appendix~\ref{app:repair_outcomes}), letting Qwen3-14B and Qwen3-8B read the complete source file recovered 65 and 40 of the 146 upstream function misses, confirming that full-file visibility can help substantially once the file is found.

\section{Function Correctness and Executed Repair}\label{sec:consequence}\label{sec:outcome}

Even when a localizer pinpoints the right function, does that actually translate into more passing patches? Intuitively, the answer hinges on the editing interface. A repair model handed an entire file can still read around an upstream function miss and locate the bug itself, whereas a model restricted to rewriting a single supplied function lives or dies by the localizer's choice.

We examine both interfaces on SWE-bench Lite using archived supplied-location repair runs scored under the official test harness, counting empty outputs and incomplete test runs as unresolved (Table~\ref{tab:realres}; exclusions and execution coverage in Appendix~\ref{app:repair_outcomes}). Two boundaries govern interpretation. First, the supplied functions and hints in our in-house cohorts come from archived rankings produced before \S\ref{sec:metric_validity}'s left-padding index fix (with repair-input labels checked directly against the actual inputs supplied to the fixer), so these runs measure observational associations under fixed historical inputs rather than estimating a causal effect of the selector fix. Second, whole-file and single-function panels evaluate separate observational cohorts in which an issue can contribute up to two localizer attempts, so all intervals resample issue clusters.

\begin{table}[!tbp]\centering\small
\caption{Executed repair on SWE-bench Lite by supplied editing interface and localized function correctness. Panel A reports source-checked whole-file attempts across 105 issues alongside the full 210-attempt sensitivity; Panel B reports separate single-function cohorts across three fixers. Counts show resolved over eligible attempts, treating empty or untested outputs as unresolved. Differences are function-correct minus function-incorrect (pp) with issue-clustered 95\% bootstrap CIs; issue multiplicity, repository coverage, label checks, and regrouping appear in Appendix~\ref{app:repair_outcomes}.}\label{tab:realres}
\begin{tabularx}{\linewidth}{@{}Xrrrr@{}}\toprule
{\colhead{Input / fixer}} & {\colhead{Correct}} & {\colhead{Incorrect}} & {\colhead{Difference}} & {\colhead{95\% CI}}\\\midrule
\multicolumn{5}{@{}l}{\textbf{A. Whole-file input}}\\
All 184 attempts (primary) & 18/67 & 15/117 & $+14.0$ & $[-1.2,29.7]$\\
\quad Non-empty patches & 18/51 & 15/75 & $+15.3$ & $[-4.9,35.4]$\\
All 210 (sensitivity) & 18/68 & 16/142 & $+15.2$ & $[0.4,30.6]$\\\midrule
\multicolumn{5}{@{}l}{\textbf{B. Function input, all attempts}}\\
Qwen3-14B & 23/90 & 1/176 & $+25.0$ & $[13.6,37.3]$\\
Qwen3-8B & 20/85 & 0/172 & $+23.5$ & $[11.9,36.0]$\\
Llama-3.1-8B & 17/68 & 1/136 & $+24.3$ & $[11.4,38.2]$\\\bottomrule
\end{tabularx}\end{table}

In the whole-file cohort (Table~\ref{tab:realres}A), localized function correctness shows a positive point estimate, yet its issue-clustered confidence interval crosses zero and the link proves fragile under closer scrutiny. Retrospectively regrouping those same archived attempts and unchanged test outcomes under \S\ref{sec:metric_validity}'s corrected selector labels shrinks the contrast to $+3.0$ percentage points with a wide interval spanning both directions---an exploratory sensitivity on fixed supplied files that does not simulate supplying corrected functions to a fixer---and the small observational subset of issues where two localizers received opposite labels yields zero discordant resolutions (Appendix~\ref{app:repair_outcomes}). Under single-function input (Table~\ref{tab:realres}B), where certification almost coincides with the supplied function being reference-changed, the pattern is stark: across all three fixers, function-incorrect attempts almost never pass tests, yielding strong positive associations of roughly 24 percentage points. Because Panels A and B come from separate observational cohorts rather than matched assignment on a common issue set, they describe associations within each supplied interface rather than proving a causal interface interaction; practically, however, they show why function accuracy is critical for function-scoped editors yet far harder to isolate when an editor can search the whole file.

\paragraph{Hints, matched diffs, and fixer headroom.}
Two follow-ups show how easily weak patch generators mask localization differences. In the matched file-hit cohort of \S\ref{sec:revision-source-tests}, adding three pre-correction function hints alongside the whole file improves changed-function naming by 6--7 percentage points; yet when Qwen3-14B generates unified diffs across the corresponding repair records, every view stalls at a floor of just 4--6 passing patches (with over a third failing to apply), leaving too little headroom to evaluate hints or establish equivalence across views (Appendix~\ref{app:repair_outcomes}). Likewise, without localizer input, a weak free-location agent (mini-swe-agent~\citep{minisweagent2025} with Qwen3-14B) resolves only 3 of the 105 primary-cohort issues, whereas a stronger \texttt{gpt-6.1-sol} free-location configuration resolves 80/105 on the same issues (Appendix~\ref{app:free_location})---underscoring that downstream repair evaluations require fixers with enough headroom to reflect localization quality, without attributing a localizer or selector effect to that cross-configuration contrast.

\section{Discussion, Related Work, and Conclusion}\label{sec:related_main}\label{sec:discussion}

\paragraph{Related work.}
Our experimental design draws on input-ablation and counterfactual-edit methods developed to test which cues actually drive model predictions~\citep{poliak2018hypothesis,gururangan2018annotation,jia2017adversarial,yefet2020adversarial,rabin2021understanding}. Within software localization, SACL probes textual bias~\citep{gupta2025sacl}, LogicLoc evaluates localization under keyword-agnostic queries~\citep{logicloc2026}, and RepoMirage perturbs repository context inside multi-step agents~\citep{repomirage2026}. Our fixed-pool interventions isolate displayed paths at a single scoring stage---separating readable unrelated aliases, opaque identifiers, single-pair swaps, and historical Git renames---while our repository-navigation evaluations test how full-file tool use behaves under opaque handles across models and pipelines. Downstream, prior work on hierarchical localization and repair shows that finding a file is only an intermediate milestone~\citep{chang2025bugcerberus,loc2repair2026,flcontext2026}; we connect file hits to scorer-declared function localization and executed repair under the exact edit interface supplied to the fixer. Appendix~\ref{app:relatedwork} reviews the wider retrieval, repository-graph, and agent literature.

\paragraph{Implications for evaluation and remaining boundaries.}
Three practical lessons follow for how bug localization should be evaluated. First, original-path file Recall@1 should be paired with a displayed-path stress test whenever models see both paths and code: two rerankers can look interchangeable on original paths while relying on entirely different cues, and even full-file tool access spans complete collapse to empirical invariance depending on the model and pipeline. Second, function-level localization cannot be inferred from file hits alone and should always be reported alongside its scoring model, ranking interface, and input scope; on the same candidate pools, switching from an unadapted pointwise scorer to a stronger listwise scorer shifts accuracy dramatically. Third, downstream repair results must be read strictly within their editing interface, issue clustering, and fixer headroom.

Several empirical boundaries remain open. Under per-issue base-commit checkouts, whether readable wrong aliases hurt rank-one accuracy more than opaque identifiers is unresolved; SweRank's reproduced native function-body pipeline was not perturbed under PathSwap; our active-navigation experiments compare distinct models and pipelines rather than isolating a within-model tools-on/off effect; and our supplied-location repair cohorts remain observational evaluations on archived pre-correction inputs rather than causal end-to-end demonstrations of improved repair.

\paragraph{Conclusion.}\label{sec:conclusion}
A file-level hit shows that a localizer narrowed a repository to a relevant file, not that it understood the code or found the edit site. Evaluating localization by original-path file recall alone treats path shortcuts and genuine code use alike, overlooks wide differences in how tool-using systems cope when naming cues fail, and leaves function selection unmeasured. Developers and benchmark designers should therefore test path reliance directly, separate file ranking from function scoring, and judge downstream repair strictly within the editing interface and fixer headroom actually deployed.

\section*{Reproducibility Statement}
The main text and appendices specify the fixed-pool estimands, path interventions, model configurations, checkout protocols, and uncertainty procedures. Supplementary analysis scripts, result summaries, and a claim-to-evidence index document the reported tables and figures; large raw prediction logs, transformed candidate pools, and model checkpoints are designated for public release upon publication. Appendix~\ref{app:metric_validity_confusion} documents the own and shared function-scorer populations, and Appendix~\ref{app:pathalias} records SWE-bench Verified adaptation overlap.

\section*{Ethics Statement}
All data come from public software-engineering benchmarks---SWE-bench Lite and Verified, GREPO, BeetleBox, Bench4BL, and BEIR-SciFact---and all evaluated models are public checkpoints or hosted API services. The study involves no human participants or personal data. PathAlias, PathSwap, \texttt{PathSwap-agent-v1}, and MaskPath are controlled evaluation perturbations, and PathMask is applied during training only; none modifies live repositories.

\section*{AI Use Statement}
Generative AI tools assisted manuscript organization, language editing, translation, figure preparation, supporting analysis code, and verification of supplementary controls reported in this work. All outputs were checked against study protocols and experimental records, and the authors retain full responsibility for the manuscript and its scientific claims.

\bibliography{references_iclr}
\bibliographystyle{iclr2027_conference}
\clearpage
\appendix
\raggedbottom

\section{Evaluation Protocols and Supplementary Guide}
\label{app:moved_maintext}

\subsection{Questions and reading routes}
The appendices follow the order of the main questions: displayed-path interventions in Appendix~\ref{app:pathalias}, file-to-function transfer in Appendix~\ref{app:metric_validity_confusion}, and executed repair inputs, exclusions, and outcomes in Appendix~\ref{app:repair_outcomes}. Training-recipe controls, metadata and code-use diagnostics, and cross-setting scope studies follow in Appendices~\ref{app:cross_cut}, \ref{app:path_diagnostics}, and~\ref{app:scope_studies}. Exploratory training and inference interventions appear in Appendix~\ref{app:exploratory}, the native SweRank reproduction in Appendix~\ref{app:swerank_repro}, CodeGRIP implementation and reference analyses in Appendix~\ref{app:impl}, extended related work in Appendix~\ref{app:relatedwork}, and the read-only repository-navigation and code-inspecting pipeline studies in Appendix~\ref{app:revision-source-studies}. Each appendix states the population, checkout protocol, precision, and candidate pool of its comparisons. 

\begin{table}[!htbp]
\centering
\caption{Evidence and primary boundary for each check.}
\label{tab:audit_map}
\small
\setlength{\tabcolsep}{4pt}
\renewcommand{\arraystretch}{1.08}
\begin{tabular}{@{}>{\raggedright\arraybackslash}p{0.14\linewidth}>{\raggedright\arraybackslash}p{0.34\linewidth}>{\raggedright\arraybackslash}p{0.43\linewidth}@{}}
\toprule
{\colhead{\textbf{Check}}} & {\colhead{\textbf{Evidence}}} & {\colhead{\textbf{Boundary}}} \\
\midrule
Sensitivity & On per-issue base-commit Lite ($n=259$), PathAlias drops Recall@1 by $20.8$pp ($[15.0,26.7]$); on single-snapshot Lite ($n=300$), the drop is $21.8$pp ($4.9$pp for Codeaware; Table~\ref{tab:pathalias_cells}). Active tool navigation spans $-69.9$pp (Agentless) and $-25.9$pp (GPT-4.1-mini) to $0.0$pp (\texttt{gpt-6.1-sol}). &
Alias-versus-opaque ordering at rank one is unresolved under per-issue base-commit checkouts (both 300- and 259-issue CIs include zero). Verified has 384/500 adaptation overlap and is strictly an overlap-limited sensitivity. \\
Transfer & Under matched bf16/top-100 shared scorers (Table~\ref{tab:shared_scorer_mde}C), paired 14B-minus-path-only contrasts span $-0.85$ to $0.00$pp (all CIs include zero). &
The 45.0\% own-scorer level depends strongly on the scorer and function names; zero-shot body prompts do not generalize to trained body scorers. Matched cross-reranker nulls are inconclusive, not equivalence. \\
Consequence & The primary source-checked whole-file gap is $+14.0$pp ($[-1.2,29.7]$) on 184 attempts (105 issues); the 210-record sensitivity gives $+15.2$pp ($[0.4,30.6]$). &
Repair cohorts use archived pre-index-correction inputs and do not estimate the causal effect of \S\ref{sec:metric_validity}'s index fix. Function-given and whole-file cohorts differ in input and inclusion criteria. \\
Reporting & Pair file Recall@1 with a displayed-path intervention, active-inspection bounds where applicable, and scorer-declared function transfer. &
Both-scorer certification measures agreement across shared name-visible and name-blind scorers, not downstream resolution. \\
\bottomrule
\end{tabular}
\end{table}

\begin{table}[!htbp]
\centering
\caption{Comparison objects, units and outcomes for the three checks.}
\label{tab:protocol_registry}
\small
\begin{tabular}{@{}>{\raggedright\arraybackslash}p{.12\linewidth}>{\raggedright\arraybackslash}p{.28\linewidth}>{\raggedright\arraybackslash}p{.19\linewidth}>{\raggedright\arraybackslash}p{.31\linewidth}@{}}
\toprule
{\colhead{Check}} & {\colhead{Fixed or supplied inputs}} & {\colhead{Analysis unit}} & {\colhead{Outcome and boundary}} \\ \midrule
Path sensitivity & Fixed report, code, candidates and pool; changed displayed path & Paired evaluation record & Fractional file Recall@1 change under the specified path intervention \\
Function transfer & Functions from top-10 predicted files; named scorer & File-win record in a declared label/overlap set & Scorer-referenced conditional function Hit@1; common-win contrast \\
Repair outcome & Localized function or file; source-exclusion rule & Repair attempt or non-empty patch & Certified--uncertified resolution difference; association \\
\bottomrule
\end{tabular}
\end{table}

\subsection{Populations and scoring instruments}

GREPO~\citep{wang2026grepo} contains 47{,}166 Python issue reports across 86 repositories. Its original training split has 7{,}883 examples; our primary 14B recipes combine a cleaned 4{,}046-group GREPO subset with 1{,}870 SWE-bench groups (5{,}916 groups in total). GREPO evaluation uses the 1{,}704 Python-changing examples from its 2{,}175-example test split. Matching normalized repository--issue pairs against the 5{,}916-group training set yields 0/300 overlap with SWE-bench Lite and 384/500 overlap with SWE-bench Verified (all from the SWE-bench training slice). Lite is therefore issue-held-out within its 12 repositories and serves as our primary SWE-bench evaluation, while Verified serves as an overlap-limited sensitivity panel. On GREPO, candidate pools merge BM25 top-500 retrieval with graph-expanded neighbors.

The primary SWE-bench path evaluations compare an unaugmented 14B reranker against \textbf{Codeaware}, a matched recipe trained with 50\% path-shuffled examples; both rerank fixed BM25 top-100 pools unless noted otherwise (retrieval details in Appendix~\ref{app:impl}). Because GREPO and SWE-bench candidate pools define different ranking tasks, recall percentages are compared within each pool rather than across pools. Unless noted, we report fractional Recall@1 with paired percentile bootstrap 95\% CIs ($B{=}10{,}000$), using exact two-sided McNemar tests for binary top-1 hit or flip comparisons. Evaluated rerankers span Qwen2.5-7B/14B-Instruct LoRAs~\citep{yang2024qwen25}, with Llama-3.1-8B~\citep{dubey2024llama3}, Qwen3-8B~\citep{yang2025qwen3}, and Mistral-7B~\citep{jiang2023mistral} breadth checks (hyperparameters and data separation in Appendices~\ref{app:hyperparams}--\ref{app:leakage}).

To calibrate synthetic path changes against real repository evolution, \textbf{organic rename drift} displays each gold file under its earliest recorded \texttt{git} path across 862 genuinely renamed GREPO issues while holding code and candidate pools fixed (Appendix~\ref{app:organic_drift}). Throughout the appendix tables, ``path-only baseline'' denotes the three-seed Qwen2.5-7B GREPO mean ($28.92\pm0.56\%$), whereas cross-model perturbation tables use the single-checkpoint baseline ($27.01\%$) so relative drops stay paired within their own baseline. Analysis code and aggregate summaries accompany the supplementary material; transformed paths, candidate pools, and model checkpoints are designated for public release upon publication.

\begin{table}[!htbp]\centering\small
\caption{Summary of analysis populations and evaluation settings. Counts refer to each analysis's own unit; separate analyses draw on distinct cohorts rather than one shared sample.}\label{tab:analysis_registry}
\begin{tabular}{@{}>{\raggedright\arraybackslash}p{.24\linewidth}>{\raggedright\arraybackslash}p{.33\linewidth}>{\raggedright\arraybackslash}p{.35\linewidth}@{}}\toprule
{\colhead{Analysis}} & {\colhead{Population / unit}} & {\colhead{Instrument / boundary}} \\ \midrule
PathAlias Lite (base commit) & 259 date-filtered / 300 full Lite; 0 adaptation IDs & 14B no-aug; bf16; per-issue top-100 \\
PathAlias Lite (single snapshot) & 300 issues; 0 adaptation IDs & 14B no-aug / Codeaware; bf16; fixed top-100 \\
PathAlias Verified & 500 issues; 384 adaptation IDs & overlap-limited sensitivity; ten alias mappings \\
Active navigation & 193 Lite issues; 11 repositories & \texttt{PathSwap-agent-v1}; GPT-4.1-mini, \texttt{gpt-6.1-sol}, Agentless, Moatless \\
Own-scorer transfer & 1{,}113 issues; 4{,}452 system--instance pairs & 2{,}356 wins; 1{,}060 joint hits; 792 issue clusters \\
Shared-scorer transfer & 472 / 481 matched bf16/top-100 common wins & fixed-neutral / name-blind; legacy 468/476 in App.~\ref{app:metric_validity_confusion} \\
Source-checked whole-file repair & 184 attempts; 105 issues; 126 non-empty & excludes 24 mismatches and 2 untested; 79 paired issues \\
All-attempt whole-file repair & 210 attempts; 115 issues & 24 input mismatches and 2 untested records included \\
Function-given Qwen3-14B & 266 eligible file wins; 162 non-empty & pre-index-fix supplied function; untested unresolved \\
\bottomrule\end{tabular}\end{table}

\subsection{Evaluation roles and material availability}
SWE-bench Lite (0/300 adaptation overlap, issue-held-out within its 12 repositories) serves as the primary PathAlias evaluation, whereas Verified (384/500 overlap) serves strictly as an overlap-limited sensitivity check (Appendix~\ref{app:pathalias}). Repair exclusions and provenance boundaries appear in Appendix~\ref{app:repair_outcomes}. Supplementary code and summary tables document all reported figures and intervals; full per-instance reconstruction additionally uses raw prediction logs, transformed pools, and checkpoints slated for release upon publication.

\section{Displayed-Path Interventions and Calibration}
\label{app:pathalias}
Under the per-issue base-commit rebuild of SWE-bench Lite (\S\ref{sec:shortcut}, bf16, BM25 top-100), unaugmented Qwen2.5-14B scores 62.33\% (187/300) on original paths, 41.97\% under PathAlias (ten-mapping mean), and 46.67\% (140/300) under PathSwap across all 300 issues---drops of 20.4 and 15.7 percentage points, with PathSwap minus PathAlias at $+4.70$ points ($[-0.50,9.97]$). On the 259 date-filtered issues (Figure~\ref{fig:main_pathalias}b), the corresponding scores are 62.2\%, 41.4\%, and 45.9\%, with Normal minus PathAlias at 20.8 points (crossed 95\% CI $[15.0,26.7]$), Normal minus PathSwap at 16.3 points, and PathSwap minus PathAlias at $+4.6$ points ($[-0.9,10.2]$). Unless marked otherwise, the remaining tables and diagnostics in this appendix report the single-snapshot 300-issue Lite and 500-issue Verified checkout protocols.

\subsection{PathAlias construction, populations and uncertainty}

PathAlias replaces every nonstructural path component with a readable, form- and tokenizer-matched alias
drawn from unrelated repositories.

Components appearing in ${\ge}10\%$ of
repositories across the full benchmark corpus (e.g., \texttt{src}, \texttt{tests},
\texttt{\_\_init\_\_}) keep their original form, except that a component equal to any target
repository's canonical name is never treated as structural.

Most aliases are whole components from held-out donor repositories; for 14/2{,}721 Lite and 140/5{,}739 Verified components per seed, a fallback instead composes the alias from held-out donor subtokens. Donor components are matched to the original component on extension, path depth, separator style, case morphology (underscore, hyphen, trailing-digit, and uppercase, checked independently), character-length bin, and BPE-token-count bin under the evaluated model's tokenizer. In this way, the construction preserves natural path form while breaking the correspondence between a path and the target repository's content.

Each mapping is one-to-one within a repository. The same original component receives the same alias across all instances of that repository, and for each seed the mapping is shared across benchmarks. The mapping is applied to \emph{all} candidate paths, including distractors.

The donor filters reject components that occur in the target repository's own path vocabulary, including lexical/BPE subtokens of its canonical name. Accidental overlap between aliases and issue reports is not removed by hand; instead, it is retained and measured at three levels.

Exact matches of a whole alias in the issue text are rare: measured over component--instance pairs, the ten-seed means are $0.43\%$ for Verified and $0.59\%$ for Lite. Lexical subtokens of length ${\ge}3$ occur in the issue text for $4.6$--$5.4\%$ of such pairs on Lite and $8.7$--$9.8\%$ on Verified, mainly through generic tokens such as \texttt{port} or \texttt{remove}. The broader diagnostic, which counts overlap through either lexical subtokens or BPE pieces, gives $15$--$22\%$.

With $\sim$300 aliased components per instance, nearly every instance contains \emph{some} generic-token coincidence. An instance-level sensitivity comparison that excludes every instance with a coincidence would therefore be uninformative. Residual coincidences can shift the measured drop in either direction. Because aliases are applied to every candidate, overlap with a gold alias can reduce the drop, whereas overlap with a distractor alias can increase it.

Across ten independent alias draws, every aggregate no-augmentation drop is positive, and the largest range across seeds within a cell is $5.1$pp. The drop therefore appears under every recorded mapping.

\FloatBarrier
\begin{figure}[!htbp]
\centering
\includegraphics[width=\linewidth]{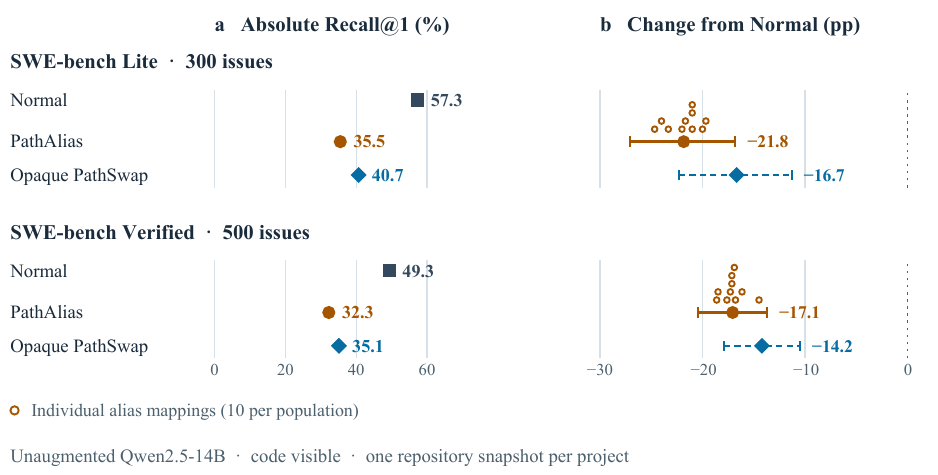}
\caption{Single-snapshot sensitivity. File ranking under the original single-snapshot protocol. The unaugmented, code-visible Qwen2.5-14B reranker scores fixed BM25 top-100 file candidates on all 300 Lite issues and 500 Verified issues. Panel a gives fractional Recall@1. Panel b shows changes from Normal. Open circles mark ten alias mappings (seeds 42--51), offset vertically for visibility. The solid interval around their mean resamples issues and mappings. The dashed interval for the deterministic opaque mapping resamples issues. Both are 95\% intervals. The primary Lite analysis in main Figure~\ref{fig:main_pathalias} uses 259 issues and each issue's base commit. Verified includes 384 issues also used in adaptation.}
\label{fig:pathalias_fullmaps}
\end{figure}

For the additional
PathAlias intervals, a crossed bootstrap resamples both instances and the ten alias seeds ($B{=}10{,}000$,
seed 42), yielding $[-27.1,-16.8]$pp on Lite and $[-20.4,-13.7]$pp on Verified. Every per-benchmark estimand uses the full population,
matching the PathSwap anchors.
\begin{table}[!htbp]
\centering
\caption{Single-snapshot PathAlias per-cell results (paired drop vs.\ the same-batch Normal run; outcomes averaged
over ten preselected seeds). The two 14B no-augmentation rows use the crossed instance-by-alias-seed intervals described above. Lite is the issue-held-out primary population; Verified rows are
sensitivities limited by adaptation overlap (384/500 evaluation IDs occur in the adaptation set).
The single-snapshot bf16 PathAlias/path-exchange harness Normal baselines (57.3 for Lite no-augmentation, batch 16, and 61.0 for Lite Codeaware)
use 50-line, 1{,}500-character, 768-token previews and differ by at most 1pp from the 4-bit seed-42 rows (Table~\ref{tab:codeaware_main}) and the matched bf16 control (Table~\ref{tab:matched_14b_codepreview}).}
\label{tab:pathalias_cells}
\footnotesize
\setlength{\tabcolsep}{3pt}
\begin{tabular}{@{}llrrrr@{}}
\toprule
{\colhead{\textbf{Cell}}} & {\colhead{\textbf{Benchmark}}} & {\colhead{\textbf{Normal}}} & {\colhead{\textbf{PathAlias}}} & {\colhead{\makecell{\textbf{Alias$-$Normal}\\\textbf{(pp; 95\% CI)}}}} & {\colhead{\makecell{\textbf{Normal$-$Alias}\\\textbf{(pp; per-mapping range)}}}} \\
\midrule
\makecell[l]{14B no-aug\\(single-snapshot ref.)} & Lite & $57.3$ & $35.5$ & $-21.8$ $[-27.1,-16.8]$ & $19.7$--$24.7$ \\
\makecell[l]{14B no-aug\\(adaptation-overlap\\sensitivity)} & Verified & $49.3$ & $32.3$ & $-17.1$ $[-20.4,-13.7]$ & $14.5$--$18.6$ \\
\makecell[l]{7B SWE-only\\unaugmented recipe\\(replication)} & Lite & $56.0$ & $25.0$ & $-31.0$ $[-36.0,-26.0]$ & $30.0$--$32.3$ \\
\makecell[l]{7B SWE-only\\unaugmented recipe\\(adaptation-overlap\\sensitivity)} & Verified & $46.9$ & $27.5$ & $-19.4$ $[-22.7,-16.2]$ & $17.2$--$22.3$ \\
\makecell[l]{14B Codeaware\\(recipe sensitivity)} & Lite & $61.0$ & $56.1$ & $-4.9$ $[-8.3,-1.7]$ & $3.7$--$6.3$ \\
\makecell[l]{14B Codeaware\\(recipe/adaptation-\\overlap sensitivity)} & Verified & $49.2$ & $48.7$ & $-0.5$ $[-2.2,+1.1]$ & $-0.1$--$1.1$ \\
\bottomrule
\end{tabular}
\end{table}

Of the 300 Lite issues, 93 also occur in Verified; on the remaining 207, the same seed-mean paired
estimator gives a $-21.5$pp drop, CI $[-27.5,-15.6]$.

\begin{center}
\small
\begin{tabular}{@{}lrrr@{}}
\toprule
{\colhead{\textbf{Verified stratum}}} & {\colhead{\textbf{$n$}}} & {\colhead{\textbf{Drop (pp)}}} & {\colhead{\textbf{95\% CI}}} \\
\midrule
Adaptation-disjoint & 116 & $-14.2$ & $[-20.2,-8.8]$ \\
In-adaptation & 384 & $-17.9$ & $[-21.7,-14.3]$ \\
\bottomrule
\end{tabular}
\end{center}
The adaptation-membership stratification uses the same seed-mean paired estimator as the
primary analysis; the in-adaptation-minus-disjoint stratum-difference CI is $[-10.4,+3.3]$ and crosses
zero. The Verified drop is therefore not attributable to adaptation overlap alone.

\paragraph{Repository-composition sensitivity.}
Repository-stratified standardization supports this reading. For both PathAlias
and the PathSwap reference, every adjusted and unadjusted in-adaptation-minus-disjoint 95\% CI includes zero,
while the adjusted point-estimate sign changes across the disjoint, common-support in-adaptation-mix,
and pooled-common-support targets. Exact standardization of the disjoint stratum to the full
in-adaptation repository mix is not estimable: $3/384$ target cases occur in repositories absent from the disjoint stratum. Sign instability reflects repository mix.

Across all ten alias mappings, every unaugmented drop is positive, spanning 19.7--24.7 percentage points on Lite and 14.5--18.6 points on Verified (a maximum spread of 5.1 points across mappings within a cell; Figure~\ref{fig:pathalias_fullmaps} and Table~\ref{tab:pathalias_cells}). Restricting to the gold-reachable-in-top-100 subset (excluding deterministic misses in both arms) gives larger drops as expected: $-25.2$ points ($[-30.9,-19.7]$) on Lite ($n{=}260$) and $-25.6$ points ($[-30.2,-21.0]$) on Verified ($n{=}333$).

\phantomsection
\label{par:pathalias_recipe_interaction}
\paragraph{Recipe interaction in the Alias--PathSwap loss difference.}
For recipe $m$, define $a_m=(R_m^N-\overline{R_m^A})-(R_m^N-R_m^W)=R_m^W-\overline{R_m^A}$, where $R$ is fractional Recall@1, $N$ denotes original paths, $A$ denotes PathAlias averaged over the ten mappings, and $W$ denotes the deterministic opaque PathSwap mapping. A positive $a_m$ indicates that readable aliases cost more than opaque identifiers, and $a_{\rm noaug}-a_{\rm Codeaware}$ measures how differently the two recipes respond.

Averaging alias mappings first and resampling matched instances for both recipes gives no-augmentation asymmetries of $+5.17$ points ($[+0.67,+9.70]$) on Lite and $+2.86$ ($[-0.03,+5.72]$) on Verified, versus Codeaware asymmetries of $-2.10$ ($[-4.57,+0.37]$) and $-0.68$ ($[-1.99,+0.54]$). The resulting recipe interactions are $+7.27$ points ($[+1.87,+12.73]$) on Lite and $+3.54$ ($[+0.47,+6.58]$) on Verified. A crossed instance-by-mapping bootstrap yields within-recipe no-augmentation intervals of $[+0.4,+10.0]$ on Lite and $[-0.2,+5.9]$ on Verified (against single-snapshot Normal-minus-PathSwap drops of $16.7$ points $[11.3,22.3]$ on Lite and $14.2$ $[10.5,17.9]$ on Verified). Strictly larger Alias than PathSwap loss at rank one thus appears only in the single-snapshot unaugmented Lite run; both per-issue base-commit Lite intervals (\S\ref{sec:shortcut}) and all Codeaware intervals span zero.

\paragraph{Zero-shot replication with Llama.}
Evaluating zero-shot Llama-3.1-8B-Instruct (code visible) on Verified ($n{=}500$) starts at $15.40\%$ Normal Recall@1---below a 20\% baseline floor---and drops to $8.78\%$ across the ten PathAlias mappings ($-6.62$ points, paired 95\% CI $[4.06,9.26]$, positive on every mapping).

\paragraph{Leave-one-repository-out robustness.}
Omitting any single repository leaves both aggregate drops intact (Table~\ref{tab:pathalias_repository_drops}). For instance, excluding Django ($n{=}269$ remaining Verified issues) raises the PathAlias and PathSwap drops from $+17.06$ and $+14.20$ points to $+19.46$ and $+18.40$ points.

\begin{table}[!htbp]
\centering
\caption{Leave-one-repository-out robustness. Positive values are Normal-minus-arm Recall@1 drops (pp). Repository rows give 95\% paired-bootstrap CIs; the Lite/Verified counts are shown in the Lite panel. LORO (leave-one-repository-out) rows give instance-weighted point estimates after one repository is omitted; LORO CIs are not reported. Repository CIs are not adjusted for multiplicity and can be unstable for small $n$.}
\label{tab:pathalias_repository_drops}
\footnotesize
\setlength{\tabcolsep}{3pt}
\begin{tabular}{@{}lrrr@{}}
\toprule
\multicolumn{4}{c}{\colhead{\textbf{Lite}}} \\
\cmidrule(lr){1-4}
{\colhead{\textbf{Repository}}} & {\colhead{\textbf{Lite/Verified $n$}}} & {\colhead{\textbf{PathAlias}}} & {\colhead{\textbf{PathSwap}}} \\
\midrule
astropy      & $6/22$    & $+45.00\ [+11.67,+78.33]$ & $+50.00\ [+16.67,+83.33]$ \\
django       & $114/231$ & $+20.44\ [+11.84,+29.04]$ & $+12.28\ [+2.63,+21.93]$ \\
matplotlib   & $23/34$   & $+23.48\ [+7.39,+39.13]$  & $+8.70\ [-8.70,+26.09]$ \\
seaborn      & $4/2$     & $-17.50\ [-45.00,+5.00]$  & $+25.00\ [-50.00,+100.00]$ \\
flask        & $3/1$     & $+53.33\ [+50.00,+60.00]$ & $+33.33\ [+0.00,+100.00]$ \\
requests     & $6/8$     & $+36.67\ [+10.00,+66.67]$ & $+33.33\ [+0.00,+66.67]$ \\
xarray       & $5/22$    & $-6.00\ [-38.00,+16.00]$  & $+0.00\ [+0.00,+0.00]$ \\
pylint       & $6/10$    & $+51.67\ [+23.33,+80.00]$ & $+66.67\ [+33.33,+100.00]$ \\
pytest       & $17/19$   & $+21.76\ [+2.94,+42.35]$  & $+29.41\ [+0.00,+52.94]$ \\
scikit-learn & $23/32$   & $+9.57\ [-4.35,+24.35]$   & $+8.70\ [+0.00,+21.74]$ \\
sphinx       & $16/44$   & $+45.00\ [+23.12,+66.25]$ & $+25.00\ [+0.00,+50.00]$ \\
sympy        & $77/75$   & $+19.61\ [+9.61,+29.87]$  & $+15.58\ [+5.19,+25.97]$ \\
\midrule
\textbf{LORO range} & not applicable & $+20.53$--$+22.85$ & $+15.65$--$+19.35$ \\
\textbf{Max shift} & not applicable & sphinx $-1.31 \to +20.53$ & django $+2.69 \to +19.35$ \\
\bottomrule
\end{tabular}

\medskip
\begin{tabular}{@{}lrr@{}}
\toprule
\multicolumn{3}{c}{\colhead{\textbf{Verified}}} \\
\cmidrule(lr){1-3}
{\colhead{\textbf{Repository}}} & {\colhead{\textbf{PathAlias}}} & {\colhead{\textbf{PathSwap}}} \\
\midrule
astropy      & $+6.21\ [-6.52,+19.24]$   & $+13.64\ [-4.55,+31.82]$ \\
django       & $+14.26\ [+9.86,+18.76]$  & $+9.31\ [+4.44,+14.18]$ \\
matplotlib   & $+20.34\ [+6.52,+33.73]$  & $+20.59\ [+5.88,+35.29]$ \\
seaborn      & $+5.00\ [+0.00,+10.00]$   & $+25.00\ [+0.00,+50.00]$ \\
flask        & $+60.00\ [+60.00,+60.00]$ & $+100.00\ [+100.00,+100.00]$ \\
requests     & $+36.25\ [+10.00,+66.25]$ & $+25.00\ [+0.00,+62.50]$ \\
xarray       & $+31.36\ [+13.64,+50.00]$ & $+25.00\ [+0.00,+50.00]$ \\
pylint       & $+11.25\ [-0.50,+28.50]$  & $+15.00\ [-10.00,+45.00]$ \\
pytest       & $+42.37\ [+22.11,+63.16]$ & $+47.37\ [+26.32,+68.42]$ \\
scikit-learn & $+9.69\ [+3.75,+16.56]$   & $+4.69\ [+0.00,+12.50]$ \\
sphinx       & $+21.44\ [+10.68,+33.33]$ & $+8.71\ [-1.89,+20.08]$ \\
sympy        & $+15.81\ [+8.40,+23.68]$  & $+19.56\ [+10.44,+29.12]$ \\
\midrule
\textbf{LORO range} & $+16.06$--$+19.46$ & $+12.89$--$+18.40$ \\
\textbf{Max shift} & django $+2.40 \to +19.46$ & django $+4.20 \to +18.40$ \\
\bottomrule
\end{tabular}
\end{table}

\paragraph{Concentration and lexical-coincidence checks.}
On Verified, django accounts for
$0.386$ of the summed absolute repository contributions, exceeding the $0.35$ concentration cap; on Lite, the small seaborn
($n{=}4$) and xarray ($n{=}5$) repositories have mean drops of $-17.5$pp and $-6.0$pp. We also attempted
a stricter lexical-exclusion construction that excluded all evaluator-visible issue vocabulary. It failed its
feasibility criterion because Sphinx required $23.66\%$ composed aliases, above the $20\%$
cap; no scoring model was loaded and no outcome was produced.

\paragraph{Single-pair path exchange details.}
For each eligible issue, single-pair path exchange swaps the displayed paths of one gold--distractor pair while holding their source code and the rest of the fixed candidate pool unchanged (main summary in Table~\ref{tab:ps1_transplant}; full counts in Table~\ref{tab:ps1_transplant_details}). On Lite, the three distractor tiers (top-ranked distractor, same-directory sibling, and random distractor) score $n{=}260/248/260$ issues; on the 116 adaptation-disjoint Verified issues, they score $n{=}54/49/54$. Letting $G_i=1$ indicate that the recipe ranked the gold file first on original paths, capture is $P(\text{swapped top-1}=d\mid G_i=1)$ over the conditional denominator $G=\sum_i G_i$, whereas the Normal reference reports how often $d$ already ranked first across the full tier (a distinct denominator). Path share measures the displayed path's share of the winning score margin on original wins.

\begin{table}[!htbp]
\centering
\caption{Detailed single-pair path exchange results (all entries in \% except $n$ and $G$). Here $n$ is the tier size and $G=\sum_i G_i$ is the number of original gold wins conditioning capture, $P(\text{swapped top-1}=d\mid G_i=1)$ $[95\%\ \mathrm{CI}]$. Ref./path reports the full-tier Normal top-1 rate of $d$ followed by the path share of the winning margin on $G_i=1$ wins. Lite is primary; adaptation-disjoint Verified is secondary.}
\label{tab:ps1_transplant_details}
\small
\setlength{\tabcolsep}{5pt}
\begin{tabular}{@{}lrlrr@{}}
\toprule
{\colhead{\textbf{Tier}}} & {\colhead{\textbf{$n$}}} & {\colhead{\textbf{Arm}}} & {\colhead{\textbf{Capture $[\mathrm{CI}]$; $G$}}} & {\colhead{\textbf{Ref./path}}} \\
\midrule
\multicolumn{5}{@{}l}{\textbf{Lite (single-snapshot primary)}} \\
\multirow{2}{*}{Top-ranked distractor} & \multirow{2}{*}{260} & No-aug & $77.9\ [71.5,83.7];\ 172$ & $33.8/80.8$ \\
 & & Codeaware & $18.6\ [13.1,24.6];\ 183$ & $29.6/15.3$ \\
\multirow{2}{*}{Same-directory sibling} & \multirow{2}{*}{248} & No-aug & $80.2\ [74.1,86.4];\ 162$ & $18.5/79.3$ \\
 & & Codeaware & $11.7\ [7.0,17.0];\ 171$ & $15.3/16.4$ \\
\multirow{2}{*}{Random distractor} & \multirow{2}{*}{260} & No-aug & $63.4\ [55.8,70.3];\ 172$ & $0.4/72.1$ \\
 & & Codeaware & $1.1\ [0.0,2.7];\ 183$ & $0.4/5.8$ \\
\midrule
\multicolumn{5}{@{}l}{\textbf{Verified (adaptation-disjoint secondary)}} \\
\multirow{2}{*}{Top-ranked distractor} & \multirow{2}{*}{54} & No-aug & $84.1\ [72.7,93.2];\ 44$ & $18.5/82.1$ \\
 & & Codeaware & $4.9\ [0.0,12.2];\ 41$ & $24.1/21.8$ \\
\multirow{2}{*}{Same-directory sibling} & \multirow{2}{*}{49} & No-aug & $79.5\ [66.7,92.3];\ 39$ & $16.3/83.1$ \\
 & & Codeaware & $11.1\ [2.8,22.2];\ 36$ & $20.4/16.5$ \\
\multirow{2}{*}{Random distractor} & \multirow{2}{*}{54} & No-aug & $77.3\ [63.6,88.6];\ 44$ & $0.0/75.8$ \\
 & & Codeaware & $0.0\ [0.0,0.0];\ 41$ & $1.9/8.0$ \\
\bottomrule
\end{tabular}
\end{table}

On Lite, no-augmentation capture reaches $77.9/80.2/63.4\%$ across the three tiers (versus full-tier Normal rates of $33.8/18.5/0.4\%$), with displayed paths accounting for $72$--$81\%$ of the winning margin. By contrast, Codeaware's descriptive capture is $18.6/11.7/1.1\%$ with path-margin shares of $6$--$16\%$, and full-tier paired McNemar tests for Codeaware do not reach Holm-adjusted significance. As a score-distribution check on the swapped prompts, path--code incongruent pairs exhibit larger Kolmogorov--Smirnov shifts from the Normal score distribution than congruent pairs in 12/12 cells for the swapped gold-code item and in 11/12 cells when both swapped items are pooled.

\subsection{Structural-Masking Contrast}

\begin{table}[!htbp]
\centering
\caption{\footnotesize Test-time MaskPath contrasts on single-snapshot Lite ($n=300$; distinct from training-time PathMask in Table~\ref{tab:pathmask_invariance}). $N$ denotes original paths, $M$ structural masking of displayed path text, and $X$ cross-repository PathAlias (ten mappings). $\delta_{\textstyle NM}{=}N-M$ uses an issue bootstrap; $\delta_{\textstyle MA}$ compares $M$ against the $X$ mean with a crossed issue-by-mapping bootstrap ($B{=}10{,}000$). Two planned within-repository-donor contrasts lacked token-length equivalence and were not estimable; they are omitted from the rows below and held conservatively at $p=1.0$ in the four-test Holm adjustment. Codeaware rows are descriptive sensitivities.}
\label{tab:maskpath_v8}
\footnotesize
\setlength{\tabcolsep}{3pt}
\renewcommand{\arraystretch}{1.1}
\begin{tabular}{@{}llrrrrl@{}}
\toprule
{\colhead{\textbf{Cell}}} & {\colhead{\textbf{Contrast}}} & {\colhead{\makecell{\textbf{Estimate}\\\textbf{(pp)}}}} & {\colhead{\textbf{95\% CI}}} & {\colhead{\textbf{Raw $p$}}} & {\colhead{\textbf{Holm $p$}}} & {\colhead{\makecell{\textbf{Reading}}}} \\
\midrule
\multicolumn{7}{@{}l}{\emph{Primary unaugmented contrasts}} \\
14B no-aug (primary) & $\delta_{\textstyle NM}$ & $+21.7$ & $[16.3,27.0]$ & $0.0002$ & $0.0008$ & \makecell[l]{supported (positive)} \\
14B no-aug (primary) & $\delta_{\textstyle MA}$ & $+0.2$ & $[-3.6,3.9]$ & $0.92$ & $1.00$ & \makecell[l]{not detected} \\
\midrule
\multicolumn{7}{@{}l}{\emph{Secondary descriptive contrasts}} \\
14B Codeaware (secondary) & $\delta_{\textstyle NM}$ & $+5.7$ & $[2.3,9.0]$ & \makecell[r]{not\\tested} & \makecell[r]{not\\tested} & \makecell[l]{not\\tested} \\
14B Codeaware (secondary) & $\delta_{\textstyle MA}$ & $-0.8$ & $[-2.6,1.0]$ & \makecell[r]{not\\tested} & \makecell[r]{not\\tested} & \makecell[l]{not\\tested} \\
\bottomrule
\end{tabular}
\end{table}

On single-snapshot Lite ($n=300$), test-time MaskPath lowers unaugmented 14B Recall@1 by $21.7$ points ($[16.3,27.0]$, Table~\ref{tab:maskpath_v8}), closely matching PathAlias ($-21.8$ points in Figure~\ref{fig:pathalias_fullmaps}). The $0.2$-point MaskPath-minus-PathAlias gap ($[-3.6,3.9]$) is not statistically distinguishable from zero, which neither separates the two interventions nor establishes their equivalence.

\subsection{Opaque-reference construction}
\label{app:path_anon}

To test sensitivity to readable path content, we replace path components with deterministic opaque identifiers.
Within each repository, every path component is mapped to the same identifier wherever it occurs. This preserves stable component
identity, tree depth, separators and the \texttt{.py} extension while removing readable lexical content, as in the first PathSwap definition.

\begin{table}[!htbp]
\centering
\caption{Opaque-identifier path control (GREPO, 1704 examples). ``Opaque identifiers'' replaces each path component with a deterministic opaque identifier; ``Shuffled'' randomly permutes file names across the repository.}
\label{tab:path_anon}
\small
\setlength{\tabcolsep}{6pt}
\renewcommand{\arraystretch}{1.15}
\begin{tabular}{l S[table-format=2.2] S[table-format=2.2] S[table-format=2.2] S[table-format=2.2]}
\toprule
{\colhead{\textbf{Condition}}} & {\colhead{{\textbf{Recall@1}}}} & {\colhead{{\textbf{R@5}}}} & {\colhead{{\textbf{R@10}}}} & {\colhead{{\textbf{Cond.\ Acc@1 (\%)}}}} \\
\midrule
Original paths & 24.51 & 44.18 & 52.12 & 46.51 \\
Opaque identifiers &  1.19 &  3.94 &  7.07 &  3.82 \\
Shuffled paths &  1.10 &  3.43 &  5.77 &  3.56 \\
\bottomrule
\end{tabular}
\end{table}

For this path-only scorer, opaque identifiers reduce Recall@1 from 24.5\% to $\sim$1.2\%. PathSwap preserves component identity and hierarchy, so this result measures sensitivity to removing readable lexical content while structural information remains available. Opaque identifiers and shuffled paths both score near $1\%$ in these cells. The residual is of the same order as the inverse average pool size ($\sim$163 candidates); no calibrated random baseline was run.

\subsection{Historical-rename calibration}
\label{app:organic_drift}

\textbf{Construction.} From GREPO we identify each repository's rename/move history with Git's rename tracking and chain-walk it to map each current gold path to its earliest historical path. The organic-drift subset is the $n{=}862$ test instances whose gold file was genuinely renamed and remains in the BM25 top-100 candidate pool. For the drift arm we replace the gold's displayed path with its historical path in both the candidate list and the ground-truth key; a lookup map redirects code reads back to the file's real on-disk content, so \emph{only} the displayed path token changes. Both arms use the 14B checkpoints of Table~\ref{tab:codeaware_main} (path-augmented ``Codeaware'', aug$=$0.5, and the matched no-augmentation LoRA), bf16 inference, code-in-prompt, BM25 top-100, seed 42.

\textbf{Historical-path effects depend on the cutoff.} Recall@1 falls under historical-path drift, but top-$k$ recall for $k{>}1$ rises for both models. Recall@5 rises from $44.22$ to $54.91$ for Codeaware and from $43.71$ to $50.54$ for the no-augmentation reranker. Recall@10 rises from $50.22$ to $63.14$ and from $49.90$ to $59.81$, respectively. At both cutoffs, Codeaware gains more than the no-augmentation reranker; this paired difference-in-differences is significant: $+3.85$pp (CI $[+2.02,+5.78]$) at Recall@5 and $+3.01$pp (CI $[+1.59,+4.59]$) at Recall@10. The robustness conclusion in \S\ref{sec:analysis} is based on Recall@1, while the higher-cutoff gains describe complementary outcomes.

\textbf{Power.} The subset size is limited to the number of gold files that were genuinely renamed ($n{=}862$); enlarging it would require changing the perturbation. At this $n$, the Recall@1 difference-in-differences ($+1.25$pp) has a 95\% CI of $[-0.37,+2.87]$ and is of the same magnitude as floating-point nondeterminism induced by batch size.

\begin{table}[!htbp]
\centering
\caption{Organic rename drift (GREPO, $n{=}862$ genuinely renamed gold files) calibrates the path intervention against repository rename history. The displayed current path of each gold file is replaced by its historical path in the repository; content, candidate pool, and all other paths are held fixed. Recall@1 (\%) with paired bootstrap 95\% CIs ($B{=}10{,}000$). Historical-path drift significantly lowers Recall@1 for both the 14B path-augmented (Codeaware) and no-augmentation LoRAs, providing a check based on actual repository renames alongside the opaque-identifier condition. The Codeaware$-$baseline difference-in-differences is positive but \emph{not} significant. Recall@$5$/Recall@$10$ increase under drift and are reported as complementary cutoff-dependent outcomes in Appendix~\ref{app:organic_drift}.}
\label{tab:organic_drift}
\small
\setlength{\tabcolsep}{7pt}
\renewcommand{\arraystretch}{1.2}
\begin{tabular}{@{}lrrr@{}}
\toprule
{\colhead{\textbf{Model}}} & {\colhead{\textbf{Current path}}} & {\colhead{\textbf{Historical path}}} & {\colhead{\textbf{Paired drop (95\% CI)}}} \\
\midrule
No-augmentation      & 22.13 & 18.66 & $-3.47$ \; $[-5.12,-1.82]$ \\
Codeaware (path-aug) & 23.16 & 20.94 & $-2.22$ \; $[-3.64,-0.84]$ \\
\midrule
\multicolumn{3}{@{}l}{\shortstack[l]{\emph{Difference-in-differences}\\(magnitude: $|$no-aug drop$|-|$Codeaware drop$|$)}} & $+1.25$ \; $[-0.37,+2.87]$ \;\textit{n.s.} \\
\bottomrule
\end{tabular}
\end{table}

\section{File-to-Function Transfer}
\label{app:metric_validity_confusion}
Let scorer $g$ rank function candidates drawn from the top-10 files predicted by system $s$, and let $F_{is}$ and $H_{is}^{(g)}$ be the file-win and function-hit indicators. A function hit is a prediction that matches both the gold file path and the bare function name. We call a file win that also yields a function hit \emph{certified}. On a declared label-defined set $\mathcal E$, the certification rate is
\begin{equation}
\widehat c_s(g;\mathcal E)=\frac{\sum_{i\in\mathcal E}F_{is}H_{is}^{(g)}}{\sum_{i\in\mathcal E}F_{is}}.\label{eq:certification}
\end{equation}

We specify each function scorer and record set before comparing certification rates. The sensitivity analyses of historical configurations, the missing-label rules, the replication and the PathSwap strata are each reported on their own populations rather than pooled.
\subsection{Transfer Populations and Scoring Instruments}

\begin{table}[!htbp]
\centering\small
\caption{Function outcomes within file-level success. Panel A contains 4,452 system and instance records from 1,113 common label-defined GREPO issues and four Qwen rerankers. Here $1{,}060/2{,}356=45.0\%$ $[41.9,48.0]$ of file wins yield an own-scorer function hit. Panel B compares Normal-path file wins that remain hits under PathSwap with those that become misses. The difference is $10.93$pp $[6.09,15.81]$, clustered over 792 issues. Function candidates come from top-10 files.}
\label{tab:file_func_confusion}
\label{tab:fragility_main}
\begin{tabularx}{.60\linewidth}{@{}Xrr@{}}
\toprule
\multicolumn{3}{l}{\colhead{\textbf{A. Joint file/function outcomes}}}\\
& {\colhead{Function correct}} & {\colhead{Function wrong}}\\\midrule
File correct & 1,060 & 1,296\\
File wrong & 185 & 1,911\\\bottomrule
\end{tabularx}
\par\smallskip
\begin{tabularx}{.60\linewidth}{@{}Xrr@{}}
\toprule
\multicolumn{3}{l}{\colhead{\textbf{B. Own-scorer transfer among Normal file wins}}}\\
{\colhead{Group}} & {\colhead{File-win records}} & {\colhead{Function Hit@1 (\%)}}\\\midrule
PathSwap-stable & 1,530 & 48.82\\
PathSwap-fragile & 826 & 37.89\\\bottomrule
\end{tabularx}
\end{table}

\paragraph{Evaluation records and denominators.}
For Table~\ref{tab:file_func_confusion}, each reranker has 1,704 file-localization records and 1,574 paired function-localization records. Pairing the file and function localization records across the four rerankers gives 6,296 system and issue pairs. A common set of 1,113 issue keys has defined function judgments under all four rerankers, giving 4,452 pairs for the confusion table. Among those pairs, 2,356 are file wins and 1,060 also achieve function Hit@1. The file-win records span 792 issue clusters.

\paragraph{Left-padding index correction in function scoring.}
An earlier scoring pass read the Yes-minus-No logit at the unpadded prompt length minus one inside left-padded batches, which fell before the final prompt token on shorter prompts. Correcting the extraction index to the final token column leaves all 2{,}356 file-hit flags, file rankings, gold sets, and candidate counts unchanged while shifting 597 function misses to hits and 122 hits to misses---raising pooled own-scorer transfer from 585/2{,}356 (24.8\%) to 1{,}060/2{,}356 (45.0\%). The same correction applies to the GREPO shared scorers in Table~\ref{tab:shared_scorer_mde}B, the SWE-bench Lite replication (123/509 before vs.\ 242/509 after), and the PathSwap-stable versus PathSwap-fragile transfer rates (28.30\% and 18.40\% before vs.\ 48.82\% and 37.89\% after).

\paragraph{Complete file-to-function procedures.}
On the 1,113 common issues in Table~\ref{tab:shared_scorer_mde}A, file Recall@1 is 27.62\% for path-only, 29.65\% for 14B Codeaware, and 29.15\% for 14B no-augmentation. The paired path-only differences are $-2.02$pp (95\% CI $[-3.63,-0.41]$) and $-1.52$pp ($[-3.07,+0.05]$). Codeaware therefore has higher file recall on this subset, whereas the no-augmentation interval includes zero. The full 1,704-issue comparison in \S\ref{sec:training_main} uses a different population.

Function Hit@1 on all 1,113 examples is 378/1113 (33.96\%) for the path-only procedure, 370/1113 (33.24\%) for 14B Codeaware, and 371/1113 (33.33\%) for 14B no-augmentation. The paired path-only differences are $+0.72$pp (95\% CI $[-1.71,3.14]$) and $+0.63$pp ($[-1.71,2.96]$), and both intervals include zero. These results compare complete procedures using their own function scorers. The common-file-win comparisons below instead hold the function scorer fixed.

The scores of the three complete procedures appear in Table~\ref{tab:shared_scorer_mde}A of the main text. This section gives the paired comparisons and population definitions.

The Lite replication evaluates the four rerankers, each with its own function scorer, on 173 issues whose changed files enter the initial top-100 candidate pool. It keeps the corresponding adapted weights and top-10 file construction but supplies no file summaries. There are 509 file-level wins spanning 153 issue clusters: 14B Codeaware certifies 63/133 (47.37\%), 14B no-augmentation 68/133 (51.13\%), 7B Codeaware 50/115 (43.48\%), and 7B no-augmentation 61/128 (47.66\%). The pooled rate is 242/509 (47.54\%, issue-clustered 95\% CI $[40.28,54.79]$). All 509 hit files contain a gold changed function, yet in 267 records the top-ranked function is not a gold changed function. The full 272-issue evaluation yields the same conditional rate because all file hits belong to the 173-issue set.

The manual label check on 100 sampled labels flags 3 (3.0\%). Excluding all eight flagged issues from the 279-label check population leaves 230/491 file-hit records certified (46.84\%, issue-clustered 95\% CI $[39.43,54.31]$). Those eight issues affect 18 file-hit records; assigning all 18 as misses or hits bounds the pooled rate to 45.19--48.72\%.

\paragraph{Shared-scorer instantiation.}
Both shared scorers use the base \texttt{Qwen2.5-7B-Instruct} checkpoint without an adapter, bf16 inference, a 768-token context, file summaries, and the top 10 predicted files. For each candidate function, the prompt contains the report, real file path, file summary and candidate-name field, but no candidate function body. The fixed-neutral scorer retains real function names. These inputs are documented for the shared scorers. The name-blind probe uses the same weights and prompt, but replaces candidate names within each instance with deterministic identifiers such as \texttt{fn\_0000} and \texttt{fn\_0001}. In the candidate-name field, real names are used only for evaluation.

\begin{sloppypar}
The verbatim prompt is ``\texttt{Given the bug report, is this function likely to need modification?}
\texttt{Bug Report: \{issue\_text\}} \texttt{File: \{file\_path\}}
\texttt{File Content: \{file\_summary\}} \texttt{Function: \{function\_name\}} \texttt{Answer:}'',
with blank lines between the question, report, and file block. The score is the final-token \texttt{Yes} logit minus the \texttt{No} logit.
\end{sloppypar}

The file summary is computed once per file from its source with a Python syntax-tree parse and reused for every candidate function in that file. It lists the names of up to 5 top-level classes, up to 8 top-level functions and class methods (methods written as \texttt{Class.method}), up to 6 imported top-level modules, and the first line of the module docstring (at most 80 characters). These parts are joined and the string is cut to 200 characters. Parameter lists, decorators and function bodies are not included. When a file yields no summary, the \texttt{File Content} line is omitted from the prompt. The summary can therefore contain real function names of the file, and the name-blind probe replaces only the candidate-name field, so names in the summary stay visible.

For Table~\ref{tab:shared_scorer_mde}C, both the fixed-neutral and name-blind cross-reranker comparisons use the matched bf16/top-100 path-only file-ranking snapshot (598 label-available wins), yielding 472 common wins with 14B Codeaware and 481 with 14B no augmentation. The two 14B systems retain 649 and 652 label-available wins. The separate legacy 4-bit/top-200 path-only snapshot (603 wins; 468 and 476 common wins) is reported below as a protocol-sensitivity check. Within a given system and snapshot, masking replaces only the explicit candidate-name field.

Under the top-10 bf16 function-scoring protocol, the transfer rates of the four Qwen rerankers with their own scorers range from 38.8\% to 48.9\%, and the pooled rate is 45.0\% (issue-clustered 95\% CI $[41.9,48.0]$).

Across 61 repositories, the per-system Spearman correlations between marginal fractional file and function Recall@1 range from $0.34$ to $0.44$. The pooled correlation is $0.44$. Restricting each system to the 32 repositories with at least ten defined evaluation instances gives correlations of $0.44$, $0.37$, $0.28$ and $0.38$, respectively.

In contrast, the correlations between file Recall@1 and function transfer conditioned on file hits range from $-0.21$ to $+0.29$, with all four $p$-values above $0.10$. Marginal score correlation and conditional transfer answer different questions. Among repositories with at least three instances, the 14B and 7B no-augmentation systems each reach 100\% binary file Hit@1 but 0\% function Hit@1 on PyPSA.

\paragraph{Filename-shuffling sensitivity analysis.}
All seven system variants retain 23--47\% of Recall@1 under filename shuffling (mean ratio $0.29 \pm 0.08$). Code-aware variants are not consistently less sensitive than path-only variants. The perturbation therefore measures within-system filename sensitivity across both system types. For the path-only baseline, the repository-level bootstrap intervals for Normal ([21.2, 33.5]) and shuffled ([5.3, 9.9]) paths do not overlap.

\paragraph{Candidate availability and function selection on GREPO}
We reconstructed the functions in each reranker's ten highest-ranked files from its saved file order and the Python syntax-tree index. A pair is eligible when its file belongs to the issue's changed Python files and the function's bare name occurs both in that file's index and in the issue's changed-function labels. This procedure measures coverage of the benchmark labels. Repeated bare names across changed files and unannotated edits limit what the labels establish about the true edit site.

For the four own scorers in order (14B Codeaware, 14B without augmentation, 7B Codeaware, 7B without augmentation), eligible file hits were 649, 652, 520, and 535. The top-ten pools contained an eligible pair on 626, 620, 494, and 506 records, and top-one function hits were 310, 319, 202, and 229. Thus 2{,}246 of the 2{,}356 file hits had a gold candidate; among 1{,}296 misses, 1{,}186 had an eligible pair and 110 did not. Candidate availability alone does not reveal which misses a trained body-aware scorer could recover.

For the matched bf16/top-100 intersections in Table~\ref{tab:shared_scorer_mde}C, code-visible versus path-only top-ten pool coverage is 455/472 (96.4\%) versus 452/472 (95.8\%) for Codeaware (both cover 450) and 460/481 (95.6\%) versus 459/481 (95.4\%) for no augmentation (both cover 456). In the legacy 4-bit/top-200 sensitivity intersections, coverage is 451/468 versus 448/468 (both 445) and 456/476 versus 454/476 (both 450). A common file win fixes the issue and the shared function scorer, while each file reranker supplies candidates from its own top-ten files.

Table~\ref{tab:instrument_decomp} shows how complete function-selection configurations change measured transfer for a trained path-only file reranker. On the 603 label-available wins from the legacy 4-bit/top-200 snapshot, a neutral top-10/bf16 function scorer reaches 28.9\% (174/603), while the trained top-50/4-bit configuration reaches 43.6\%. On the matched bf16/top-100 snapshot (598 wins), the trained top-10/bf16 procedure reaches 49.0\% (293/598), the neutral scorer reaches 27.1\% (162/598), and the name-blind probe reaches 8.0\% (48/598). Each operating point is defined jointly by its function scorer, pool depth, precision, and file-ranking snapshot; precision and candidate pool change jointly between the 603- and 598-win snapshots. The shared-scorer comparisons that follow fix the function scorer to compare file rerankers, and the name-field intervention fixes the candidate pool as well.
\begin{table}[!htbp]
\centering
\caption{Function-selection configurations for a trained path-only file reranker. $n$ is the number of label-available GREPO file wins. Cert.\ is conditional function Hit@1. Rows 1--2 share the legacy 4-bit/top-200 file ranking but differ in function scorer, top-$N$, and precision; Row 3 uses the matched bf16/top-100 file-ranking snapshot.}
\label{tab:instrument_decomp}
\small
\begin{tabular}{@{}lllr@{}}
\toprule
{\colhead{\makecell[l]{\textbf{Function-scoring}\\\textbf{configuration}}}} & {\colhead{\makecell[l]{\textbf{File-ranking}\\\textbf{snapshot}}}} & {\colhead{\textbf{$n$}}} & {\colhead{\textbf{Cert.}}} \\
\midrule
Neutral base-model scorer, top-10, bf16 & 4-bit, top-200 & 603 & $28.9\%$ \\
Trained LoRA scorer, top-50, 4-bit & same snapshot & 603 & $43.6\%$ \\
Trained LoRA scorer, top-10, bf16 & bf16, top-100 & 598 & $49.0\%$ \\
\bottomrule
\end{tabular}
\end{table}

\subsection{Shared-Scorer Comparisons}
We use the shared scorers for two comparisons. The name-field intervention changes displayed candidate function names while keeping the function pool and scorer weights fixed. The cross-system comparison holds the shared scorer fixed and applies it to each file reranker's top-ten candidate pool on common file wins. Table~\ref{tab:inversion} gives each system's certification rates within a single consistent file-ranking snapshot per row.

\begin{table}[!htbp]
\centering
\caption{Conditional function Hit@1 under three scorers within consistent bf16/top-100 file-ranking snapshots. Each record is a repository--issue file win with a qualified function label: 598 records for path-only (293, 162, and 48 hits), 649 for 14B Codeaware (310, 180, and 56), and 652 for 14B no augmentation (319, 175, and 55). The legacy 4-bit/top-200 path-only sensitivity gives 174/603 (28.86\%) under the fixed-neutral scorer (Table~\ref{tab:instrument_decomp}).}
\label{tab:inversion}
\small
\setlength{\tabcolsep}{6pt}
\begin{tabular}{@{}lrrr@{}}
\toprule
{\colhead{\textbf{File reranker}}} & {\colhead{\textbf{Own scorer (\%)}}} & {\colhead{\textbf{Fixed neutral (\%)}}} & {\colhead{\textbf{Name-blind (\%)}}} \\
\midrule
Path-only (598 wins)  & 49.00 & 27.09 & 8.03 \\
14B Codeaware (649)   & 47.77 & 27.73 & 8.63 \\
14B no-augmentation (652) & 48.93 & 26.84 & 8.44 \\
\bottomrule
\end{tabular}
\end{table}

\paragraph{Function-name evidence under a fixed scorer.}
Joining per-issue outcomes on identical eligible file wins gives 310, 180, and 56 gold-pair hits out of 649 for 14B Codeaware under own, fixed-neutral, and name-blind scoring, respectively, and 319, 175, and 55 of 652 for 14B without path augmentation. On the matched 598-win path-only snapshot, own, fixed-neutral, and name-blind scoring yield 293, 162, and 48 hits. Within each row, the two shared settings differ only in whether the explicit candidate-name field shows the real function name or an opaque identifier.

Replacing each candidate function's name with an opaque identifier reduces certification by 18.4--19.1 points across the three file-ranking configurations (the within-system name-field comparison below). The report, real file path, file summary, candidate pool and scorer weights remain fixed, and evaluation uses the original function identities. Function names therefore provide substantial usable evidence to the shared scorer, which continues to depend on this input even after the file has been found.

Within a fixed file ranking and function pool, masking the candidate name lowers shared-scorer success by 19.11 points for 14B Codeaware (95\% paired CI $[-22.65,-15.56]$, masked minus real), 18.40 points for 14B no-augmentation ($[-21.93,-14.88]$), and 19.06 points for the matched 598-win path-only snapshot ($[-22.74,-15.38]$).

\paragraph{Lexical function-body check.}\label{app:lexical_body_check}
We ran a deterministic lexical comparison on all 649 GREPO 14B Codeaware file wins. The baseline uses issue terms to score candidate names and paths. The second arm adds a BM25 score for function-body tokens. Both arms retain the same top-ten file lists, indexed candidates and labels. Bodies come only from the verified pre-fix Git commit; an unavailable body contributes no body evidence and the candidate stays in both arms.

Only three issues in three repositories met the primary recovery threshold of at least 80\% of candidate bodies; none had complete coverage. Name and path scoring selected a changed function on one of these three, and adding the body field selected one on none. In the sensitivity analysis including every issue with any recovered candidate body (604 issues, 46 repositories), the counts were 145/604 and 135/604, respectively. Adding body tokens gained 81 hits and lost 91 ($-1.66$ percentage points, paired 95\% CI $[-5.79,2.65]$). The primary population is too small to assess a high-coverage effect. The sensitivity measures partial body availability with lexical matching; a model that interprets code or actively inspects a repository could behave differently.

A follow-up restricted both lexical scorers to functions with recovered bodies. On 594 issues with at least two such candidates, this retained 129,141 of 306,026 candidate occurrences (42.2\%; median retention per issue 39.0\%). Names and paths alone gave 108/594 hits; adding body tokens gave 136/594, with 84 gains and 56 losses ($+4.71$ points, paired 95\% CI $[0.84,8.59]$). Restricting the pool removed every eligible gold pair on 105 previously covered issues and left 121 of the 594 without one. On the 473 restricted pools retaining a gold pair, the corresponding counts are 108/473 and 136/473 ($+5.92$ points, $[1.06,10.78]$). These gains describe a smaller candidate universe on which both scorers have complete body coverage.

\paragraph{Body-aware and stronger listwise shared scoring.}\label{app:body_aware_2048}
To test whether low shared-scorer transfer reflects the candidate pool or the scoring system, we evaluate two changes on the exact same 649 Codeaware and 652 no-augmentation file-hit cohorts (where an indexed gold function is present in 626/649 and 620/652 top-ten pools, Table~\ref{tab:shared_scorer_mde}B): (i) extending the unadapted pointwise \texttt{Qwen2.5-7B-Instruct} prompt to 2{,}048 tokens with candidate function bodies up to 200 lines, and (ii) switching to a hosted listwise scorer (\texttt{gpt-6.1-sol-2026-09-29}, Chat Completions, low reasoning effort, 3{,}000 max completion tokens). Because the two cohorts share 569 of 732 distinct repository-qualified issues (505 with the same top-ranked file), they provide overlapping within-cohort contrasts rather than independent replications or a cross-reranker comparison. Within each cohort, the listwise scorer ranks all indexed candidates from the same top-ten files (316{,}543 and 314{,}250 candidates across the two cohorts), packed on whole-file boundaries into chunks of at most 130{,}000 characters (each chunk returning top-2, or top-3 if single-chunk, with multi-chunk pools resolved by one final call over chunk winners). Its body arms include truncated 10-line / 400-character snippets---a different cap and packing scheme from the 200-line pointwise prompt. In both families, name masking replaces only the explicit candidate-name field, leaving file summaries and any signatures inside body text visible. Paired intervals use 10{,}000 issue resamples within each cohort (Table~\ref{tab:body_aware_2048}).

\begin{table}[!htbp]
\centering\small
\caption{Shared-scorer body and ranking-interface checks within the 649 Codeaware and 652 no-augmentation file-hit cohorts of Table~\ref{tab:shared_scorer_mde}B (569 shared issues of 732 in their union; 505 with the same first file). Entries are binary file-qualified function Hit@1 counts. Panel A shows the unadapted zero-shot pointwise \texttt{Qwen2.5-7B-Instruct} scorer (768-token rows come from a batch-8 reference run alongside the batch-16 base counts of 180/56 and 175/55; the 2{,}048-token no-body row is the primary pointwise comparator). Panel B shows the hosted listwise \texttt{gpt-6.1-sol-2026-09-29} scorer on the same candidate pools (0 incomplete cases). Only the no-body rows share the same candidate pool and nominal fields (issue, path, file summary, visible function name), while context packing (2{,}048 tokens itemwise vs.\ whole-file chunks up to 130{,}000 characters), model, and compute cost all differ; body arms use different snippet caps and are compared within their own system.}\label{tab:body_aware_2048}
\begin{tabularx}{\linewidth}{@{}Xrr@{}}\toprule
{\colhead{Shared-scorer configuration}} & {\colhead{Codeaware (649)}} & {\colhead{No augmentation (652)}}\\\midrule
\multicolumn{3}{@{}l}{\textbf{A. Unadapted pointwise \texttt{Qwen2.5-7B-Instruct} (Yes/No logit)}}\\
768 tokens, name visible, no body (batch-8 ref.) & 187 & 176\\
768 tokens, name masked, no body (batch-8 ref.) & 68 & 61\\
2048 tokens, name visible, no body & 188 & 176\\
2048 tokens, name visible, with body ($\le 200$ lines) & 137 & 139\\
2048 tokens, name masked, with body ($\le 200$ lines) & 151 & 161\\\midrule
\multicolumn{3}{@{}l}{\textbf{B. Listwise \texttt{gpt-6.1-sol-2026-09-29} (same top-10 file pools)}}\\
Name visible, no body & 419 & 410\\
Name visible, with capped body ($\le 10$ lines / 400 chars) & 407 & 405\\
Name masked, with capped body ($\le 10$ lines / 400 chars) & 395 & 396\\\bottomrule
\end{tabularx}
\end{table}

Comparing the two 768-token 7B runs on the exact same records shows noticeable item-level churn despite identical weights and prompts (for Codeaware, visible hits shift from 180 to 187 across 37 flips and masked hits from 56 to 68 across 44 flips; for no-augmentation, 175 to 176 and 55 to 61); batch size (16 vs.\ 8) and runtime differed across runs, without an isolated ablation. Within the batch-8 driver, lengthening context from 768 to 2{,}048 tokens without bodies changes Hit@1 by at most one case (187 to 188; 176 to 176). Relative to that 2{,}048-token no-body baseline, appending 200-line bodies with visible names lowers zero-shot 7B Hit@1 by $-7.86$ points ($[-12.02,-3.70]$) on Codeaware and $-5.67$ points ($[-9.97,-1.53]$) on no-augmentation, whereas masking the explicit name field alongside bodies paradoxically scores higher (151 vs.\ 137, $-2.16$ $[-4.93,0.46]$; 161 vs.\ 139, $-3.37$ $[-6.13,-0.61]$)---pointing to prompt sensitivity in an unadapted pointwise Yes/No logit scorer.

Switching from the 2{,}048-token pointwise 7B scorer (Panel A) to hosted listwise \texttt{gpt-6.1-sol-2026-09-29} (Panel B) over the same candidate pools and nominal visible-name, no-body fields yields a large system-level paired gain: Hit@1 rises from 188/649 to 419/649 ($+35.59$pp, $[+31.12,+40.06]$; 265 listwise-only vs.\ 34 pointwise-only) on Codeaware and from 176/652 to 410/652 ($+35.89$pp, $[+31.44,+40.18]$; 266 vs.\ 32) on no-augmentation (distinct from the 188/193 navigation count in \S\ref{sec:revision-source-tests}). Because model capability, joint-versus-itemwise format, context packing, and API compute cost all change together---and recorded Azure/OpenAI endpoint mixes differed across the two cohorts, with aggregate summaries verified rather than raw per-request payloads---this contrast does not isolate a model-only or format-only effect, nor does it compare rerankers across the overlapping cohorts. Within Panel B, adding short 10-line/400-character snippets shifts Hit@1 by $-1.85$pp ($[-4.31,+0.46]$, 407 vs.\ 419) on Codeaware and $-0.77$pp ($[-2.91,+1.38]$, 405 vs.\ 410) on no-augmentation, while masking explicit names alongside those snippets gives 395/649 ($-3.70$pp $[-6.32,-1.23]$ vs.\ no-body) and 396/652 ($-2.15$pp $[-4.60,+0.46]$). These truncated snippets inside multi-chunk prompts (290 and 271 multi-chunk pools in the body arms vs.\ 31 and 34 without bodies) do not test unrestricted code reading or trained body-aware scorers.

\paragraph{Certification shared by both scoring settings.}
Define \emph{Both} as a file win where fixed-neutral and name-blind scoring both select a changed function (excluding the own scorer). Each 14B system yields just 35 Both-positive records versus 310 and 319 own-scorer hits (roughly a 9$\times$ ratio), illustrating how strongly certification depends on the chosen scorer. Across all 1{,}574 function-evaluation records per system, 813 (Codeaware) and 834 (no-augmentation) are file wins, of which 164 and 182 lack qualified function labels. Using all file wins as denominator gives Both rates of 4.31\% ($[2.95,5.78]$) and 4.20\% ($[2.88,5.64]$) versus own-scorer rates of 38.13\% and 38.25\%; restricting to the 649 and 652 label-available wins gives 5.39\% and 5.37\% versus 47.77\% and 48.93\%. Nominal Krippendorff agreement between the two shared scorers under the all-win convention is 0.178 and 0.194 (0.185 pooled; 0.205 across all three scorers; 100{,}000 bootstrap replicates over repository--issue keys).

\paragraph{Paired comparisons between file rerankers.}
For each shared-scorer comparison, we restrict to issues on which both file rerankers rank a gold file first and both function labels are defined. A single shared scorer then evaluates the functions from each reranker's top-10 files. Paired percentile bootstraps resample repository--issue keys ($B=10{,}000$, seed 42), preserving both outcomes in each pair, with exact two-sided McNemar tests on discordant outcomes.

On the matched bf16/top-100 path-only snapshot reported in Table~\ref{tab:shared_scorer_mde}C, the fixed-neutral scorer selects a changed function on 144/472 Codeaware records versus 145/472 path-only records (13 gains, 14 losses; Codeaware minus path-only $-0.21$ points, 95\% CI $[-2.33,1.91]$, exact McNemar $p=1.000$) and on 141/481 no-augmentation records versus 142/481 path-only records (13 gains, 14 losses; $-0.21$ points, $[-2.29,1.87]$, $p=1.000$). Under candidate-name masking on the same intersections, Codeaware versus path-only gives 40/472 versus 44/472 (7 gains, 11 losses; $-0.85$ points, $[-2.54,0.85]$, $p=0.481$), and no augmentation versus path-only gives 43/481 versus 43/481 (8 gains, 8 losses; $0.00$ points, $[-1.66,1.66]$, $p=1.000$). All four cross-reranker paired intervals include zero (inconclusive paired evidence rather than an equivalence claim), whereas every within-scorer name-masking interval excludes zero.

As a separate legacy mixed-protocol sensitivity using the 4-bit/top-200 path-only snapshot (603 label-available wins, yielding 468 and 476 common wins), the fixed-neutral scorer gives 145/468 Codeaware hits versus 156/468 path-only hits (14 gains, 25 losses; $-2.35$ points, $[-4.91,0.21]$, $p=0.108$) and 143/476 no-augmentation hits versus 155/476 path-only hits (14 gains, 26 losses; $-2.52$ points, $[-5.04,0.00]$, $p=0.081$). Both the file-ranking precision/candidate pool and the common-win intersections differ from the matched bf16/top-100 comparison above.

\subsection{Transfer on Path-Fragile and Stable Wins}

\paragraph{Path-fragile vs.\ path-stable function transfer.}
We compare stable and fragile file hits to ask whether file top-1 hits that flip
under PathSwap also show lower function-level transfer. \emph{Eligible}
instances are normal-path file top-1 hits with a defined function-transfer label.
A hit is \emph{fragile} if it flips to a miss under PathSwap and \emph{stable}
otherwise. The metric is file-qualified function-level Hit@1.

We evaluate five rerankers from two model families on the GREPO function subset: four Qwen rerankers (14B/7B $\times$ Codeaware/no-augmentation) and a Llama-3.1-8B Codeaware replication. We report the difference in function-level Hit@1 between stable and fragile file hits. Because the two groups are disjoint subsets of the eligible pool, their CIs use a two-group instead of a paired bootstrap ($B{=}10{,}000$). Per-system CIs resample eligible hits directly. The \emph{pooled} row combines \emph{system and instance pairs} across the four Qwen rerankers, so the same issue can occur once per system. Its \emph{issue-clustered} bootstrap therefore resamples issue clusters and re-partitions the stable/fragile groups each iteration. Table~\ref{tab:fragility} reports the results.

Pooling the four Qwen rerankers gives $826$ fragile and $1{,}530$ stable system and instance pairs over $792$ issue clusters. The Llama row reports that system alone and is not pooled. The pooled gap is $\Delta{=}10.93$pp (issue-clustered 95\% CI $[6.09,15.81]$, naive unclustered CI $[6.72,15.13]$). The confidence interval excludes zero for three of the four Qwen rerankers and includes zero for 7B Codeaware. Path-fragile file hits therefore show lower function-level transfer across these observed PathSwap groups.

\begin{table}[!htbp]
\centering
\caption{Path-fragile file hits show weaker function-level transfer. On the GREPO function subset, eligible instances are normal-path file top-1 hits with defined function-transfer labels. Entries report function Hit@1 for path-\emph{stable} hits (which remain file hits under PathSwap) versus path-\emph{fragile} hits (which become file misses). $\Delta$ is stable minus fragile with a 95\% two-group bootstrap CI (resample $+$ re-partition, not paired, $B{=}10{,}000$). Per-system rows count instances. The pooled row$^{\dagger}$ counts \emph{system and instance pairs} across the four Qwen rerankers (excluding Llama) and uses an \emph{issue-clustered} bootstrap ($792$ issue clusters). $^{\star}$CI excludes 0.}
\label{tab:fragility}
\small
\setlength{\tabcolsep}{6pt}
\resizebox{\linewidth}{!}{%
\begin{tabular}{@{}lrrrrr@{}}
\toprule
{\colhead{\textbf{Reranker}}} & {\colhead{\textbf{stable $n$}}} & {\colhead{\textbf{stable Hit@1}}} & {\colhead{\textbf{fragile $n$}}} & {\colhead{\textbf{fragile Hit@1}}} & {\colhead{\textbf{$\Delta$ (95\% CI)}}} \\
\midrule
14B Codeaware      & 526 & 50.57\% & 123 & 35.77\% & $+14.80$ $[+5.30,+24.32]^{\star}$ \\
14B no-augmentation & 413 & 54.48\% & 239 & 39.33\% & $+15.15$ $[+7.16,+23.11]^{\star}$ \\
7B Codeaware       & 342 & 40.06\% & 178 & 36.52\% & $+3.54$ $[-5.34,+12.25]$ \\
7B no-augmentation  & 249 & 47.79\% & 286 & 38.46\% & $+9.33$ $[+0.98,+17.76]^{\star}$ \\
Llama-3.1-8B Codeaware & 477 & 46.54\% & 133 & 34.59\% & $+11.95$ $[+2.85,+20.93]^{\star}$ \\
\midrule
\textbf{Pooled}$^{\dagger}$ & 1{,}530 & 48.82\% & 826 & 37.89\% & $\mathbf{+10.93}$ $\mathbf{[+6.09,+15.81]^{\star}}$ \\
\bottomrule
\end{tabular}
}%
\end{table}

\section{Executed-Repair Outcomes}
\label{app:repair_outcomes}

\subsection{Inputs, units and fixed source exclusions}
On SWE-bench Lite, each 14B localizer supplies a binary label indicating whether its own function scorer selected a gold function. In whole-file repair the fixer chooses the edit site; in function-conditioned repair it rewrites the supplied function. All in-house supplied-location repair cohorts in this appendix (the whole-file, function-conditioned, pre-execution proxy, and matched-view cohorts) use localization rankings produced prior to the left-padding index correction of Appendix~\ref{app:metric_validity_confusion}, with repair-input function labels audited directly against the actual supplied inputs; published external systems (below) use their own localization pipelines, and the free-location agent (Appendix~\ref{app:free_location}) receives no localizer-supplied file or function. Our supplied-location panels therefore measure associations under fixed historical inputs and do not estimate the end-to-end causal effect of the selector correction on repair. We define resolution as passing the official tests, counting empty outputs and untested patches as unresolved in all-attempt rates.

The primary whole-file cohort has 184 source-checked attempts across 105 unique issues (79 issues appearing twice under distinct 14B localizers, 26 once; no duplicate localizer--issue key), including 126 non-empty patches. Resolution was 18/67 for function-correct inputs and 15/117 otherwise (difference $+14.0$pp, issue-clustered 95\% CI $[-1.2,29.7]$, 10{,}000 draws pooling all attempts per sampled issue). The 44 and 77 within-group issue counts overlap. Because attempts are clustered within issues, no attempt-level McNemar test is applied to the pooled 184 records. The sensitivity retains all 210 attempts on 115 issues: its 26 additional records comprise 24 uncertified input mismatches, where the fixer received the file containing the highest-ranked function instead of the declared top-ranked file, and two untested attempts with no mismatch classification. All-record resolution is 18/68 versus 16/142 ($+15.2$pp, $[0.4,30.6]$); its 141 non-empty patches give 18/52 versus 16/89 ($+16.6$pp, $[-2.3,36.1]$).

\paragraph{Repair-input function labels.} Checking the selected functions in the actual repair inputs against the verified gold-function mapping corrects two unresolved records without changing their selected functions or test outcomes (one empty patch moves from function-correct to function-incorrect and one non-empty patch moves in the opposite direction, neither involving an input mismatch). This gives all-attempt groups of 67 and 117 (18 and 15 resolutions) and non-empty groups of 51 and 75 (18/51 vs.\ 15/75, $+15.3$pp $[-4.9,35.4]$, versus $+16.3$pp $[-4.1,36.7]$ before correction). This check on repair-input labels is separate from the GREPO logit-index correction; Table~\ref{tab:corrected_free_full} uses the checked repair-input labels.

\begin{table}[htbp]\centering\small
\caption{Whole-file repair after source exclusions (184 attempts across 105 issues). Cells show resolved over eligible attempts by checked own-scorer certification; differences are certified minus uncertified (pp). Primary uncertainty follows the issue-clustered 95\% bootstrap CI (10{,}000 draws pooling all attempts per sampled issue); the separate permutation $p$-value tests an exchangeability null by shuffling issue-level certification labels within localizer and therefore need not match the bootstrap interval. Table~\ref{tab:realres}A also reports the 210-attempt sensitivity.}\label{tab:corrected_free_full}
\begin{tabular}{lrrrr}\toprule
{\colhead{Set}} & {\colhead{Cert. resolved}} & {\colhead{Uncert. resolved}} & {\colhead{$\Delta$ [95\% CI]}} & {\colhead{Perm. $p$}} \\ \midrule
All attempts & 18/67 & 15/117 & $14.0$ $[-1.2,29.7]$ & .027 \\
Non-empty patches (diagnostic) & 18/51 & 15/75 & $15.3$ $[-4.9,35.4]$ & .069 \\
\bottomrule\end{tabular}\end{table}

\paragraph{Retrospective selector-label regrouping and equal-issue-weight sensitivities.}
Two descriptive sensitivities re-examine the exact same 184 archived whole-file repair attempts and unchanged test outcomes without generating new patches or running new tests. First, joining \S\ref{sec:metric_validity}'s corrected own-scorer function labels onto the same 184 localizer--issue pairs reclassifies 46 attempts from incorrect to correct and 9 from correct to incorrect (leaving 71 incorrect and 58 correct), moving the all-attempt contrast from 18/67 vs.\ 15/117 ($+14.0$pp, $[-1.2,29.7]$) to 20/104 vs.\ 13/80 ($+2.98$pp, issue-clustered 95\% CI $[-11.66,17.24]$; non-empty 20/79 vs.\ 13/47). Holding supplied files and outcomes fixed, this exploratory regrouping leaves a wide interval compatible with substantial effects in either direction and cannot inform function-conditioned or 262-record repair, where changing the selector would change the supplied function itself. Second, an explicitly post-hoc equal-issue-weight sensitivity weights each retained attempt from issue $i$ by $1/k_i$ ($k_i=2$ for 79 issues, $k_i=1$ for 26) under the original checked labels: weighted resolution is $27.78\%$ vs.\ $12.32\%$ ($+15.46$pp, whole-issue bootstrap 95\% CI $[-0.15,31.06]$), matching the primary interval in spanning zero.

\subsection{Original-attempt routing and residual diagnostics}
\paragraph{Standardization weights.}
Let $C$ be own-scorer certification, $Z$ indicate that the chosen edit function is gold, and $Y$ indicate executed resolution. All probabilities below are empirical proportions on the original 210 attempts, before source exclusion.
\begin{align}
\mathrm{COMP}&=\sum_{z\in\{0,1\}}[\widehat P(Z=z\mid C=1)-\widehat P(Z=z\mid C=0)]\widehat P(Y=1\mid Z=z),\\
\mathrm{RESID}&=\sum_{z\in\{0,1\}}\widehat P(Z=z)[\widehat P(Y=1\mid C=1,Z=z)-\widehat P(Y=1\mid C=0,Z=z)].
\end{align}
COMP compares how often certified and uncertified attempts select a gold or a non-gold edit function. For each edit-function outcome, it weights the difference in these proportions by the pooled resolution rate of attempts with that outcome. RESID compares certified and uncertified resolution rates within each edit-function outcome, weighting each difference by that outcome's pooled share of attempts. Multiplying either quantity by 100 gives percentage points. The indicator $Z$ records whether the function chosen after localization is gold.
\paragraph{Edit-site routing and within-stratum contrasts.}
On all 210 free-fixer attempts, the certified minus uncertified gap is $+15.2$pp (issue-clustered 95\% CI $[0.4,30.6]$). The standardization stratifies on the fixer's observed edit-function correctness and reports two contrasts: the difference in edit-function selection, COMP $=+7.7$pp (CI $[3.2,13.6]$), and the within-stratum residual, RESID $=+6.4$pp (CI $[-5.3,18.4]$). Because the two contrasts use different pooled weights, they do not sum to the $+15.2$pp total gap. Of these routing and residual contrasts, only COMP excludes zero.

The descriptive ratio COMP/G, where G is the total gap before exclusion, has a point estimate of $0.51$ and a CI of $[0.17,2.42]$. Because the reported contrasts do not partition the total gap, the interval for this ratio extends above one. Its denominator changes with the provenance exclusion: after the 26-attempt exclusion, the total gap is $+14.0$pp and its interval $[-1.2,29.7]$ includes zero.

Within the correct-function stratum, the residual between certified and uncertified attempts is $+11.6$pp (CI $[-7.9,31.4]$), and its interval includes zero. To compare the two repair settings, let $\Delta_{\mathrm{COND}}$ be the resolution-rate gap between certified and uncertified records among \mbox{Qwen3-14B} function-conditioned non-empty patches, and let $\Delta_{\mathrm{FF}}$ be the all-attempt gap in the free-fixer arm. The de-conditioning ratio $(\Delta_{\mathrm{COND}}-\Delta_{\mathrm{FF}})/\Delta_{\mathrm{COND}}$ expresses the difference between these two gaps as a fraction of the function-conditioned gap. It is $0.470$ when $\Delta_{\mathrm{FF}}$ is computed before the 26-attempt exclusion and $0.511$ when it is computed after that exclusion. On conditioned rows, certification agrees at a rate of $0.996$ with whether the conditioning function is gold, leaving almost no uncertified records in the correct-function stratum.

\subsection{Function-conditioned outcomes and execution coverage}
\begin{table}[!htbp]
\centering
\caption{Resolution among non-empty patches. The first three rows supply a localized function; the final row supplies a whole file after source exclusions. Both Qwen3 fixers cover twelve Lite repositories and Llama covers seven (all-attempt estimates appear in Table~\ref{tab:realres}). Differences are certified minus uncertified, with issue-clustered bootstrap 95\% CIs (100{,}000 draws for function-conditioned rows) and within-localizer issue-level permutation tests.}
\label{tab:realres_full}
\small
\setlength{\tabcolsep}{5pt}
\begin{tabularx}{\linewidth}{@{}Xrrrr@{}}
\toprule
{\colhead{\textbf{Fixer}}} & {\colhead{\textbf{Cert.}}} & {\colhead{\textbf{Uncert.}}} & {\colhead{\textbf{$\Delta$ (95\% CI)}}} & {\colhead{\textbf{Perm. $p$}}} \\
\midrule
Qwen3-14B (Qwen3) & 23/77 (29.9\%) & 1/85 (1.2\%) & $+28.7$ $[+15.7,+42.5]$ & $<10^{-4}$ \\
Qwen3-8B (Qwen3) & 20/66 (30.3\%) & 0/57 (0.0\%) & $+30.3$ $[+16.0,+45.3]$ & $<10^{-4}$ \\
Llama-3.1-8B (Llama) & 17/68 (25.0\%) & 1/125 (0.8\%) & $+24.2$ $[+11.3,+38.1]$ & $<10^{-4}$ \\
\midrule
Whole-file source-checked (Qwen3-14B) & 18/51 (35.3\%) & 15/75 (20.0\%) & $+15.3$ $[-4.9,+35.4]$ & $0.069$ \\
\bottomrule
\end{tabularx}
\end{table}
\paragraph{Repair execution and resolution.}
Table~\ref{tab:realres_full} reports executed repair under the supplied-location settings of \S\ref{sec:consequence}. In the function-conditioned arms, each 14B reranker's top-1 (file, function) pair is given to the fixer to rewrite, and the resulting patch is submitted to the official SWE-bench harness. A record is \emph{certified} if the reranker's top-1 function is a gold changed function and \emph{uncertified} otherwise. The table also reports the primary source-checked whole-file cohort, in which the fixer chooses where to edit.

We run three \emph{conditioned} fixers, each of which rewrites the localized function. They span two model families: Qwen3-14B and Qwen3-8B differ in capacity within one family, and Llama-3.1-8B provides a cross-family check. Each fixer run uses the same greedy decoding protocol.

The Qwen3-14B arm covers all twelve Lite repositories and includes $n{=}266$ eligible file-wins, including 2 records on one \texttt{requests} instance whose test runs exceeded the time limit, and both count as unresolved. The Qwen3-8B fixer covers all twelve repositories, while Llama-3.1-8B covers seven of the twelve repositories.

\paragraph{Execution coverage.}
Among the 184 whole-file attempts, the certified group had 16 empty patches, 2 untested patches, 31 applied patches that failed the tests and 18 resolutions. The uncertified counts in the same order were 42, 6, 54 and 15. In the function-conditioned populations, the all-attempt counts were 23/90 versus 1/176 for Qwen3-14B, 20/85 versus 0/172 for Qwen3-8B, and 17/68 versus 1/136 for Llama. The corresponding certified/uncertified counts for empty patches were 13/91, 19/115 and 0/11. Every non-empty patch applied. The numbers of applied patches that failed tests were 53/83, 45/56 and 48/114 for the three fixers.

Across these four populations, 25 non-empty patches had no completed test result: 8 in the 184-attempt whole-file population, 2 for each Qwen3 function-conditioned fixer and 13 for Llama. Nineteen test-environment builds failed on \texttt{matplotlib} issues. Four test runs timed out and two returned a container error on \texttt{requests} issues. Excluding the untested patches gives the following certified-minus-uncertified differences (points, issue-clustered 95\% CIs), with the all-attempt estimate listed before the non-empty-patch estimate. For the 184 whole-file attempts, they are $+14.2$ $[-1.5,30.0]$ and $+15.0$ $[-6.2,35.9]$. For supplied functions, they are $+25.3$ $[13.7,37.7]$ and $+29.1$ $[15.9,43.1]$ for Qwen3-14B, $+23.8$ $[12.1,36.4]$ and $+30.8$ $[16.3,46.0]$ for Qwen3-8B, and $+25.4$ $[11.9,39.7]$ and $+25.3$ $[11.8,39.7]$ for Llama. These exclusions apply to the 184-attempt source-checked cohort and three function-conditioned populations.

Among non-empty patches, certified file-wins are resolved at rates of $29.9\%$/$30.3\%$/$25.0\%$ for the three fixers, compared with $1.2\%$/$0\%$/$0.8\%$ for uncertified records. All issue-clustered CIs exclude zero, and all issue-level permutation tests give $p{<}10^{-4}$.

For Qwen3-14B, the uncertified group contains almost no successful repairs, which causes quasi-separation in an unpenalized logistic fit. We therefore use Firth penalization in the difficulty-adjusted model, which gives an average marginal effect of $+22.6$pp and a penalized odds ratio of $34$ (95\% CI $[6.8,169]$, $p{=}1.7{\times}10^{-5}$).

Under the gold-file restriction, the Qwen3-14B comparison contains $n{=}149$ non-empty patches. Repairs succeed for 23/76 certified inputs and 1/73 uncertified inputs ($+28.9$pp, CI $[15.7,42.9]$). The same restriction gives $+30.3$pp (CI $[16.1,45.3]$) for Qwen3-8B and $+24.4$pp (CI $[11.4,38.5]$) for Llama-3.1-8B. Three fixers are restricted to rewriting the localized function, and certification nearly coincides with whether that function is gold. Rewriting a non-gold function leaves the fixer little room to edit the fault site, which fits the near-zero resolution among uncertified records. 

The primary source-checked whole-file analyses are given in Table~\ref{tab:realres}A and Table~\ref{tab:corrected_free_full}. Restricting the fixer to a supplied function measures a different association. Agreement between certification and whether the supplied function is gold was 0.996, making it difficult to separate localization and edit-site choice in that arm.

\paragraph{Additional supplementary checks.} A separate within-issue descriptive analysis retained only issues on which the two 14B localizers received opposite certification labels. After the fixed source exclusions, 16 such issues remain in the whole-file all-attempt cohort, with 2 resolved under both systems, 14 resolved under neither, and 0 discordant outcomes (paired difference $0.0$pp, exact McNemar $p{=}1.0$). This observational subset targets a distinct within-issue estimand with zero discordance and provides no evidence of equivalence.

Separately, we extended the conditioned-arm Firth model with an indicator for whether the issue text names any gold function exactly. The certification coefficient remains positive in this model (OR $45.5$, 95\% CI $[8.6,241.3]$, $p{=}7.2{\times}10^{-6}$, AME $+25.6$pp). A conservative specification that omits the number of predicted functions, a covariate that may be determined after localization, gives similar results (OR $45.5$, CI $[8.4,246.8]$, $p{=}9.7{\times}10^{-6}$). Both models adjust for direct lexical mention of a gold function in the issue text.

\subsection{Published-system outcome checks}
\paragraph{Published-system checks (no localization conditioning).}
Three analyses of released systems use logistic regression, resolved $\sim$ file-correct $+$ function-correct, to relate each system's own localization correctness to resolution under official test execution. These analyses use neither our scorer nor repair conditioning, and in them the function-correctness coefficient is the counterpart of certification. We report all three.

These analyses differ in benchmark and system version. On Verified, Agentless-1.5 has few discordant cases in which the file is correct but the function is incorrect. The v0.1 check also uses a different repair stage and derives function labels from submitted patches rather than from a released localization stage.

For Agentless-1.5 on Lite, the analysis uses 274 instances with function-level gold labels. File correctness is determined by the first returned file, and function correctness holds when any localized bare function name matches a gold changed-function name.
\begin{table}[!htbp]
\centering
\caption{Published-system checks. Odds ratio of function correctness for real resolution, controlling for file correctness, estimated from released materials. }
\label{tab:published_checks}
\small
\begin{tabular}{@{}lrrrl@{}}
\toprule
{\colhead{\textbf{System / benchmark}}} & {\colhead{$n$}} & {\colhead{\textbf{Func OR}}} & {\colhead{$p$}} & {\colhead{\textbf{Reading}}} \\
\midrule
Agentless-1.5 / Lite       & 274 & 1.96 & 0.032 & significant \\
Agentless-1.5 / Verified   & 496 & 1.40 & 0.096 & not significant \\
Agentless v0.1 (GPT-4o) / Lite & 265 & 0.86 & 0.70 & not significant (patch-derived labels) \\
\bottomrule
\end{tabular}
\end{table}

One of the three published-system checks shows a positive association on Lite, while the other two are not statistically significant.

The within-system analysis compares own-scorer certification with recorded repair outcomes. The published-system checks instead use each released system's own localization procedure, candidate set, and harness outcomes.

\subsection{Pre-execution edit-localization proxy}

For the 14B Codeaware and 7B no-augmentation localizers, we supply the top-ranked file and its function list, together with the highest-ranked retained function from that file, when one is present, and that function's available body. A repair model run separately (Qwen3-14B, single greedy generation) names the function it would edit. Certification refers to the global function outcome of the localizer's own scorer over the top 10 files.

Among 261 file-hit records, the named target matches a gold changed-function name in 53/63 (84.1\%) of \emph{certified} records and 90/198 (45.5\%) of \emph{uncertified} records ($\Delta=38.7$pp, 95\% CI $[27.1,49.6]$, significant for both systems). The model overrides the supplied function suggestion on 86.9\% of uncertified records, while a copy of that suggestion matches a gold name on 4.0\%.

The difference remains positive across the three file-function-count strata: at most 17, 18--43, and more than 43 functions in the supplied file ($n=90$, $84$, and $87$), with differences of $+29.2$, $+34.8$, and $+49.5$pp, respectively. This check scores the proposed edit location prior to patch generation and test execution.

\subsection{Function selection and executed repair on matched records}
We compared views of the same localized file on 264 SWE-bench Lite file-hit records from two 14B localizers (at most two distinct localizer records per issue, using pre-index-correction upstream rankings). Two records exceeded the input limit in at least one view and were excluded from every view for both models. The primary function-selection set has 219 unambiguous records across 126 observed issues (93 issues with two localizer records, 33 with one). Adding 43 records where several functions share the top-labelled name gives 262 records across 142 observed issues (120 double-record, 22 single-record). The original 264-record collection defines the 144-issue resampling frame.

Given the issue report, Qwen3-14B and Qwen3-8B each named a function to edit from the complete file, a narrow view showing only functions matching the localizer's top-labelled name, or the complete file with the three top-ranked functions listed as hints; a fourth view supplies the reference-changed functions as a manipulation check (greedy decoding, 256-token output cap). On the 219 primary records, Qwen3-14B selected a changed function on 97/219 (44.29\%) with the complete file, 49/219 (22.37\%) with the narrow view, and 111/219 (50.68\%) with hints; Qwen3-8B reached 70/219 (31.96\%), 35/219 (15.98\%), and 86/219 (39.27\%). Hints raised selection over the complete file by 6.39 points ($[0.47,12.50]$) and 7.31 points ($[1.41,13.68]$), respectively (2{,}000 resamples of the 144-issue frame at a fixed seed; frame issues absent from a subset contribute no record when drawn). On all 262 records, the gains were 6.11 points ($[0.00,12.21]$) and 7.25 points ($[1.90,13.23]$). The underlying upstream top label was correct on 73/219 primary records.

Qwen3-14B also generated a unified diff under each of the three non-oracle views on the same 262 records (142 issues), using greedy decoding and a 4{,}096-token limit. It chose the edit site with the complete file or complete file plus hints, and was restricted to supplied functions in the narrow view. Resolution counts were 5/262 with the complete file, 6/262 with the narrow view, and 4/262 with the complete file plus hints; 85 (32.4\%), 114 (43.5\%), and 92 (35.1\%) patches failed to apply, while 172, 142, and 166 applied patches failed tests. With only four to six resolutions per arm and aggregate arm counts without per-attempt paired keys, this severely floor-limited comparison has low resolution headroom and supports neither a repair benefit from hints nor equivalence across views.

\subsection{Free-location repair}\label{app:free_location}
On the 105 issues of the primary whole-file cohort, mini-swe-agent~\citep{minisweagent2025} (v2.4.1) paired with Qwen3-14B navigates the pre-fix repository and selects its own edit location without any localizer-supplied file or function. Across all 105 issues, 21 runs produce non-empty patches and 3 resolve their issue (two in Django and one in Pytest), with the 84 empty outputs counted as unresolved. For context on the exact same 105 issues without localizer input, a separate stronger free-location configuration built on \texttt{gpt-6.1-sol} (using its own agent adapter, context trimming, and generation settings) resolves 80/105 (3 both, 77 strong-only, 0 weak-only, 25 neither; paired difference $+73.33$pp, 95\% CI $[64.76,81.90]$, exact McNemar $p=1.3\times10^{-23}$). Because this contrast compares two complete fixer configurations rather than varying localization, it underscores that the 3/105 rate reflects the weak Qwen3-14B configuration rather than a ceiling on free-location repair.

\section{Training-Recipe and Code-Preview Controls}
\label{app:cross_cut}
This appendix reports the matched 14B Normal comparison, an independent 7B factorial experiment, and reference-recipe results. Each uses the checkpoint family and data specified in its own section.

\subsection{Matched 14B Code-Preview Comparison}
\label{app:codeprompt_14b_matched}
We ask whether a code-preview recipe improves file localization when context length and the evaluation pool are matched. This compares two trained LoRAs, rather than removing the preview from one fixed checkpoint.

The comparison pairs the primary unaugmented Qwen2.5-14B code=YES checkpoint with a newly trained code=NO LoRA (seed 42, aug$=0$). Both use the same 5,916 training groups, eight BM25-hard negatives, fresh base-model initialization, and a 768-token training limit; their matched configurations differ only in code visibility. Code=YES uses up to 50 source lines and code=NO receives no source preview. Evaluation is BF16 on the same top-100 fixed pool with 768-token prompts and, for code=YES, a 50-line/1500-character preview cap. Table~\ref{tab:hyperparams} gives the shared optimizer settings. Lite has no repository--issue overlap with the complete training set ($0/300$) and has one gold per issue, so fractional Recall@1 equals binary Hit@1 here.

Each arm selects one checkpoint using the mean of the most recent
50 training-loss values, without using test or evaluation metrics. Only that checkpoint is scored; the final and other unselected checkpoints are excluded.

\begin{table}[!htbp]
\centering
\caption{Matched 14B code-preview contrasts on the fixed Lite pool. The Normal contrast is primary; the PathSwap contrast is secondary.}
\label{tab:matched_14b_codepreview}
\begin{tabular}{lrrr}
\toprule
{\colhead{Fixed Lite pool}} & {\colhead{code=YES}} & {\colhead{code=NO}} & {\colhead{YES$-$NO}} \\
\midrule
Normal (primary) & 171/300 (57.00\%) & 167/300 (55.67\%) & $+1.33$pp \\
PathSwap (secondary) & 119/300 (39.67\%) & 1/300 (0.33\%) & $+39.33$pp \\
\bottomrule
\end{tabular}
\end{table}

The primary Normal contrast is $+1.33$pp (paired 95\% CI $[-3.00,+5.67]$pp, exact McNemar $p=0.644$, $b=23$, $c=19$), showing no detectable difference on original paths. Under PathSwap, however, the gap jumps to $+39.33$pp, yielding a recipe-level perturbation-loss interaction of $+38.00$pp ($[+31.00,+44.67]$). Because the same unaugmented 14B checkpoint appears under four Lite evaluation runs in this study---171/300 Normal and 119/300 PathSwap here, 172/300 and 122/300 in the single-snapshot batch-16 PathAlias reference (\S\ref{sec:shortcut}), 169/300 and 118/300 in the 4-bit seed-42 recipe table (Appendix~\ref{app:codeaware_figure}), and 187/300 and 140/300 under per-issue base-commit rebuilds (\S\ref{sec:shortcut})---every contrast stays paired within its own evaluation run. The aug$=0.5$, code$=$NO cell is not included in this single-seed 14B pair.

\subsection{7B Three-Seed Factorial Comparison}
\label{sec:lever}
\label{sec:codeaware}

We ask how code previews and path augmentation separately change Normal-input localization and PathSwap sensitivity in a fully crossed Qwen2.5-7B comparison on single-snapshot Lite ($n=300$). All four SWE-only cells use fresh Qwen2.5-7B LoRAs under matched hyperparameters with seeds 42/43/44. Code-present cells use 30-line training previews; Lite scoring uses up to 50 lines, 1{,}500 characters, and a 768-token prompt on the same fixed BM25 top-100 pool. Each training group uses eight negatives with configured BM25/graph/random sampling proportions of 50/25/25; fallback sampling fills unavailable pools. At augmentation 0.5, half of the training batches use shuffled filenames with code fixed. These checkpoints are distinct from the single-seed operating points in Tables~\ref{tab:codeaware_main}--\ref{tab:codeaware_invariance}. Table~\ref{tab:cross_cut} reports three-seed means, seed standard deviations, and effect rows; seed-42 paired-instance tests appear separately in Table~\ref{tab:mcnemar}. Paired bootstrap intervals concern fractional Recall@1 differences, while McNemar tests concern binary any-hit or explicitly named flip outcomes.

\begin{table}[!htbp]
\centering
\caption{Fully crossed 2$\times$2 code-preview and path-augmentation factorial (Qwen2.5-7B, SWE-only training, single-snapshot Lite $n{=}300$). Cells report fractional Recall@1 mean$\pm$std over three seeds (42/43/44). Effect rows use three-seed means; paired McNemar code-preview contrasts in Table~\ref{tab:mcnemar} use seed 42 alone ($54.00$/$1.00\%$ for (0,no) and $55.67$/$33.00\%$ for (0,yes), giving $+1.67$pp Normal and $+32.00$pp PathSwap vs.\ the three-seed mean differences of $+1.33$pp and $+28.78$pp).}
\label{tab:cross_cut}
\small
\setlength{\tabcolsep}{6pt}
\renewcommand{\arraystretch}{1.2}
\begin{tabular}{l l r r S[table-format=+2.2] S[table-format=+2.1]}
\toprule
{\colhead{\textbf{Cell}}} & {\colhead{\textbf{Recipe}}} & {\colhead{{\textbf{Normal}}}} & {\colhead{{\textbf{PathSwap}}}} & {\colhead{{\textbf{$\Delta_{\text{abs}}$}}}} & {\colhead{{\textbf{$\Delta_{\text{rel}}$ (\%)}}}} \\
\midrule
(0, no)    & aug$=$0,\ \ \ code$=$No   & {\(53.00\pm0.88\)} &  {\(1.44\pm0.51\)} & -51.56 & -97.3 \\
(0, yes)   & aug$=$0,\ \ \ code$=$Yes  & {\(54.33\pm1.20\)}& {\(30.22\pm3.24\)}& -24.11 & -44.4 \\
(0.5, no)  & aug$=$0.5, code$=$No      & {\(50.00\pm1.00\)}&  {\(3.11\pm1.58\)}& -46.89 & -93.8 \\
(0.5, yes) & aug$=$0.5, code$=$Yes     & {\(48.56\pm2.59\)}& {\(36.78\pm4.86\)}& -11.78 & -24.3 \\
\midrule
\multicolumn{6}{@{}l}{\itshape\footnotesize Effects} \\
\addlinespace[1pt]
\multicolumn{2}{@{}l}{code-effect at aug$=$0:\quad (0,yes)$-$(0,no)}      & {$+1.33$} & {$+28.78$} & \multicolumn{2}{c}{} \\
\multicolumn{2}{@{}l}{code-effect at aug$=$0.5:\quad (0.5,yes)$-$(0.5,no)}& {$-1.44$} & {$+33.67$} & \multicolumn{2}{c}{} \\
\multicolumn{2}{@{}l}{aug-effect at code$=$No:\quad (0.5,no)$-$(0,no)}    & {$-3.00$} & {$+1.67$}  & \multicolumn{2}{c}{} \\
\multicolumn{2}{@{}l}{aug-effect at code$=$Yes:\quad (0.5,yes)$-$(0,yes)} & {$-5.77$} & {$+6.56$}  & \multicolumn{2}{c}{} \\
\multicolumn{2}{@{}l}{\itshape code$\times$aug interaction (PathSwap, descriptive)} & {} & {$+4.89$} & \multicolumn{2}{c}{} \\
\bottomrule
\end{tabular}
\end{table}

The three-seed means in Table~\ref{tab:cross_cut} identify the largest observed PathSwap contrast with the code-preview factor: $+28.78$pp at augmentation 0 and $+33.67$pp at augmentation 0.5. Their Normal-input counterparts are $+1.33$pp and $-1.44$pp, respectively. Augmentation without code adds $1.67$pp under PathSwap and costs $3.00$pp under Normal paths; with code, it adds $6.56$pp under PathSwap at a $5.77$pp Normal cost. Thus the grid describes a sensitivity--accuracy trade-off, with a descriptive four-cell PathSwap interaction of $+4.89$pp. Table~\ref{tab:mcnemar} separately reports paired-instance tests for seed 42.

The seed-42 paired tests in Table~\ref{tab:mcnemar} answer a narrower, instance-level question. At augmentation 0, the PathSwap code-preview difference on that seed is $+32.00$pp, with 96 discordant hits favoring code and none favoring no code ($p\approx2.5\times10^{-29}$). The Normal contrast on that seed is $+1.67$pp and is not significant. The corresponding arm-total flip counts are 81 with code and 159 without code ($p\approx1.2\times10^{-17}$ for the paired flip comparison). These results are not uncertainty estimates for the three-seed mean contrasts above.

\subsection{Seed-42 Paired Comparisons}
\label{sec:codeaware:mcnemar}

 A flip is a Normal top-1 hit becoming a PathSwap miss; the 14B flip-rate difference is no-augmentation minus Codeaware. The 14B paired contrasts in this subsection evaluate the 4-bit seed-42 single-snapshot checkpoints of Tables~\ref{tab:codeaware_main}--\ref{tab:codeaware_invariance}.

\paragraph{14B recipe comparison on Lite.}
Paired bootstrap 95\% CIs use $B{=}10{,}000$ resamples of the same 300 Lite instances; changes are in percentage points. The Codeaware Normal-to-PathSwap drop is $6.67$ $[3.0,10.7]$;
its Normal contrast against no-augmentation is $4.00$ $[-0.01,8.3]$ (includes zero);
the corresponding PathSwap contrast is $14.33$ $[9.3,19.3]$;
the no-augmentation Normal-to-PathSwap drop is $17.00$ $[11.3,22.7]$;
and the flip-rate difference is $13.67$ $[9.0,18.3]$.

\paragraph{Verified paired bootstrap 95\% CIs (14B Codeaware vs.\ no-aug, $B{=}10{,}000$, paired over $n{=}500$ Verified instances):}
Recall contrasts and their intervals use fractional Recall@1. The accompanying McNemar $p$-values use binary any-hit, except for the explicitly named binary flip-rate test. Codeaware's Normal-to-PathSwap drop is $+2.27$ $[+0.22, +4.47]$, with McNemar $p{=}0.099$ (borderline). The Normal contrast of Codeaware versus no-augmentation has paired difference $+0.31$ (bootstrap mean $+0.32$, CI $[\!-\!1.87, +2.47]$), with $p{=}0.871$. That interval includes zero, so no gain is detected. The corresponding PathSwap contrast is $+12.48$ $[+9.25, +15.82]$, with $p{\approx}1.2{\times}10^{-12}$. The no-augmentation Normal-to-PathSwap drop is $+14.43$ $[+10.73, +18.18]$, with $p{\approx}8.6{\times}10^{-15}$. The \textbf{flip-rate difference} is $+12.2$ $[+8.6, +15.8]$, with McNemar $p{\approx}7{\times}10^{-16}$ (Codeaware flips $19/500$, no-augmentation flips $80/500$). The flip-rate difference is detected on both splits (Lite $p{\approx}3{\times}10^{-8}$).

\paragraph{7B code-preview comparison on Lite.}
On Lite, the paired (0,yes) minus (0,no) contrast in Table~\ref{tab:cross_cut} is $+1.67$pp $[-2.0,+5.3]$ on Normal (includes zero) and $+32.00$pp $[+26.7,+37.3]$ under PathSwap.

Here $b,c$ count the two discordant directions of paired binary outcomes, reported as $(b+c,|b-c|)$; the 14B flip row labels discordant counts, and the 7B row labels arm-total flips.

\begin{table}[!htbp]
\centering
\caption{McNemar paired tests of binary top-1 any-hit or the explicitly named flip indicators on single-snapshot SWE-bench Lite ($n{=}300$). The first five rows compare the 4-bit seed-42 14B Codeaware and 14B no-augmentation checkpoints (Tables~\ref{tab:codeaware_main}--\ref{tab:codeaware_invariance}); the flip-rate difference has $p{\approx}3{\times}10^{-8}$. The final three rows isolate the 7B code-preview contrast at fixed aug$=$0 (seed-42 checkpoints from the three-seed factorial in \S\ref{sec:lever}): all discordant PathSwap instances favor code-in-prompt ($p{\approx}2.5{\times}10^{-29}$), while the Normal contrast is not significant. Bold rows highlight the headline PathSwap and flip-rate tests.}
\label{tab:mcnemar}
\footnotesize
\setlength{\tabcolsep}{4pt}
\renewcommand{\arraystretch}{1.15}
\begin{tabularx}{\linewidth}{@{}X r r >{\raggedright\arraybackslash}p{0.27\linewidth}@{}}
\toprule
{\colhead{\textbf{Comparison}}} & {\colhead{$\boldsymbol{(b{+}c, |b{-}c|)}$}} & {\colhead{\textbf{McNemar $p$}}} & {\colhead{\textbf{Interpretation}}} \\
\midrule
\multicolumn{4}{@{}l}{\itshape\small 14B Codeaware vs.\ 14B no-augmentation} \\
\addlinespace[1pt]
Codeaware: Normal vs.\ PathSwap             & (36, 20)          & $1.2{\times}10^{-3}$            & PathSwap drop: significant \\
Codeaware vs.\ no-augmentation (Normal)     & (44, 12)          & $9.6{\times}10^{-2}$            & $+4.0$pp Lite gain: n.s. \\
Codeaware vs.\ no-augmentation (PathSwap)   & (65, 43)          & $6.0{\times}10^{-8}$            & Codeaware higher: *** \\
No-augmentation: Normal vs.\ PathSwap        & (87, 51)          & $3.3{\times}10^{-8}$            & PathSwap drop: *** \\
\textbf{Flip-rate difference}               & \textbf{(57, 41)} & $\boldsymbol{2.7{\times}10^{-8}}$ & \makecell[l]{\textbf{Discordant flips:}\\\textbf{49 vs.\ 8; ***}} \\
\midrule
\multicolumn{4}{@{}l}{\itshape\small 7B code-preview contrast at aug$=$0 ((0,no) seed=42 vs.\ (0,yes) seed=42)} \\
\addlinespace[1pt]
Code-in-prompt (Normal)                     & (31, 5)           & $0.473$                          & Normal contrast: n.s. \\
\textbf{Code-in-prompt (PathSwap)}          & \textbf{(96, 96)} & $\boldsymbol{2.5{\times}10^{-29}}$ & \makecell[l]{\textbf{Discordant hits:}\\\textbf{96 vs.\ 0; ***}} \\
\textbf{Flip-count contrast}                & \textbf{(94, 78)} & $\boldsymbol{1.2{\times}10^{-17}}$ & \makecell[l]{\textbf{Total flips:}\\\textbf{81 vs.\ 159; ***}} \\
\bottomrule
\end{tabularx}
\end{table}

\paragraph{PathSwap flip rates and recipe contrast.}
In the 4-bit seed-42 14B comparison, unaugmented 14B flips from a Normal hit to a PathSwap miss on 23.0\% of Lite issues versus 9.3\% for Codeaware (paired difference $+13.67$pp, $[9.00,18.33]$, $p\approx 3{\times}10^{-8}$), with the same direction on Verified ($p\approx 7{\times}10^{-16}$). On original paths, by contrast, Codeaware's difference is modest on Lite ($+4.0$pp, $p=0.096$) and near zero on Verified ($+0.31$pp, $[-1.87,+2.47]$, $p=0.871$; discordant counts in Table~\ref{tab:mcnemar}). Overall, the unaugmented absolute Recall@1 loss under PathSwap is $2.6\times$ Codeaware's loss on Lite and $6.4\times$ on Verified, leaving Codeaware's relative drops at $-11\%$ on Lite and $-4.6\%$ on Verified.

\subsection{Body-Deletion Content Probe}
To probe body content, we replace 80\% of body lines with \texttt{\# REDACTED} placeholders, retaining imports, class/def headers and the first non-blank line. Only the candidate-code preview changes; the Codeaware LoRA and data stay fixed.

\begin{table}[!htbp]
\centering
\caption{Body-deletion content probe (7B Codeaware LoRA, code-in-prompt, BM25 top-100 pool). This is the same-base 7B Codeaware checkpoint, distinct from the 14B recipe-table model; only the candidate-code preview is redacted.}
\label{tab:body_deletion}
\small
\setlength{\tabcolsep}{5pt}
\renewcommand{\arraystretch}{1.12}
\begin{tabular}{lrrrr}
\toprule
{\colhead{\textbf{Split / Condition}}} & {\colhead{\makecell{Baseline\\R@1}}} & {\colhead{\makecell{Redacted\\R@1}}} & {\colhead{\textbf{Delta}}} & {\colhead{\makecell{McNemar\\$p$}}} \\
\midrule
Lite Normal & 49.33 & 48.33 & $-1.00$ & 0.77 \\
Verified Normal & 43.32 & 41.80 & $-1.52$ & 0.26 \\
Lite PathSwap & 40.67 & 26.67 & $-14.00$ & $6.5{\times}10^{-6}$ \\
Verified PathSwap & 32.23 & 24.53 & $-7.70$ & $2.2{\times}10^{-5}$ \\
\bottomrule
\end{tabular}
\end{table}

Body deletion changes little under Normal paths and substantially lowers Recall@1 under PathSwap. This is consistent with a larger role for body content under opaque identifiers.

\subsection{Legacy recipe results and precision}
\label{app:codeaware_figure}

This appendix groups the fixed-snapshot recipe results and their visualization; it does not merge the separate training protocols.

\subsubsection{Recipe Operating Points}

\paragraph{Scope of the legacy 7B comparison.}
The combined-data row and SWE-only Codeaware row in Tables~\ref{tab:codeaware_main}--\ref{tab:codeaware_invariance} share a base model and evaluation pool, while differing in negative sampling, context length, checkpoint-save interval, and training data. This qualification applies to those two legacy 7B rows; the 14B contrasts and the separate three-seed $2\times2$ factorial (\S\ref{sec:lever}) have their own stated scope.

\FloatBarrier

\begin{table}[!t]
\centering
\caption{Codeaware and same-base reference recipes under the single-snapshot 4-bit BM25 top-100 evaluation pool on SWE-bench Lite ($n=300$) and Verified ($n=500$, adaptation-overlap sensitivity). Values are fractional Recall@1 (\%). \textit{aug}: path-augmentation fraction at training time; \textit{code}: short source preview; the 7B SWE-only Codeaware and combined-data rows use 50-line evaluation previews. The 30-line training configurations apply only to their specified runs, not to every row. PathSwap results for these rows appear in Table~\ref{tab:codeaware_invariance}; paired tests appear in \S\ref{sec:codeaware:mcnemar}.}
\label{tab:codeaware_main}
\small
\setlength{\tabcolsep}{4pt}
\renewcommand{\arraystretch}{1.2}
\begin{tabularx}{\linewidth}{@{}X>{\raggedright\arraybackslash}p{.20\linewidth}rlrr@{}}
\toprule
{\colhead{\textbf{Method}}} & {\colhead{\textbf{Train}}} & {\colhead{\textbf{aug}}} & {\colhead{\textbf{code}}} & {\colhead{{\textbf{Lite}}}} & {\colhead{{\textbf{Verified}}}} \\
\midrule
\multicolumn{6}{@{}l}{\itshape\small External baselines (different backbones)} \\
\addlinespace[1pt]
SweRankEmbed-Large & {--}            & {--} & {--}                  & 37.00 & {--}  \\
SweRankLLM-Small   & {--}            & {--} & {--}                  & 47.93 & {--}  \\
\midrule
\multicolumn{6}{@{}l}{\itshape\small Same-base 7B reference recipes (Qwen2.5-7B)} \\
\addlinespace[1pt]
SWE-only unaugmented recipe & SWE 1.9k & 0.0 & Yes                   & 53.67 & 46.12 \\
Codeaware          & SWE 1.9k        & 0.5 & Yes                   & 48.33 & 44.98 \\
Combined-data path-augmented recipe & SWE+GREPO 5.9k & 0.5 & Yes     & 56.67 & 48.13 \\
\midrule
\multicolumn{6}{@{}l}{\itshape\small Qwen2.5-14B scale-up (same recipe, larger backbone)} \\
\addlinespace[1pt]
14B no-aug                  & SWE+GREPO 5.9k  & 0.0 & Yes & 56.33 & 48.62 \\
\rowcolor{ours}
14B Codeaware (path-aug recipe) & SWE+GREPO 5.9k  & 0.5 & Yes & \bfseries 60.33 & 48.93 \\
\bottomrule
\end{tabularx}
\end{table}

\paragraph{Normal-path scores.} The Normal columns describe operating-point scores under the stated recipes. 

\begin{table}[!t]
\centering
\caption{PathSwap with opaque component identifiers (single-snapshot, 4-bit seed-42 operating points): 14B Codeaware has the smallest relative drop among the unaugmented and path-augmented recipe rows in the seed-42 table. The no-augmentation absolute PathSwap drop is $2.6\times$ the Codeaware drop on Lite (17.00 vs.\ 6.66pp) and $6.4\times$ the Codeaware drop on the Verified adaptation-overlap sensitivity (14.44 vs.\ 2.26pp). Entries use the BM25 top-100 pool from Table~\ref{tab:codeaware_main}. PathSwap assigns an opaque identifier to every path component while leaving file contents unchanged. $\Delta_{\text{abs}}=$\,PathSwap${-}$Normal (pp); $\Delta_{\text{rel}}=\Delta_{\text{abs}}/$Normal. \textsuperscript{$\dagger$}\,\emph{full repository-identifier cleaning}: 16 repository-specific identifier tokens in the 50-line snippet are consistently replaced with opaque identifiers; cleaned $\Delta_{\text{rel}}$ is $-5.9\%$ versus $-11.0\%$ uncleaned, consistent with residual code-identifier association.}
\label{tab:codeaware_invariance}
\small
\setlength{\tabcolsep}{4pt}
\renewcommand{\arraystretch}{1.2}
\begin{tabularx}{\linewidth}{@{}Xrrrrr@{}}
\toprule
{\colhead{\textbf{Method}}} & {\colhead{\textbf{aug}}}   & {\colhead{{\textbf{Normal}}}} & {\colhead{{\textbf{PathSwap}}}}   & {\colhead{{\textbf{$\Delta_{\text{abs}}$ (pp)}}}} & {\colhead{{\textbf{$\Delta_{\text{rel}}$ (\%)}}}} \\
\midrule
\multicolumn{6}{@{}l}{\itshape\small SWE-bench Lite ($n{=}300$)} \\
\addlinespace[1pt]
SWE-only unaugmented recipe & 0.0 & 53.67          & 33.33          & -20.34          & -37.9 \\
Codeaware SWE               & 0.5 & 48.33          & 41.00          &  -7.33          & -15.2 \\
Combined-data path-augmented recipe & 0.5 & 56.67     & 47.33          &  -9.34          & -16.5 \\
14B no-aug                  & 0.0 & 56.33          & 39.33          & -17.00          & -30.2 \\
\rowcolor{ours}
14B Codeaware (path-aug recipe) & 0.5 & \bfseries 60.33 & \bfseries 53.67 & \bfseries -6.66 & \bfseries -11.0 \\
\addlinespace[3pt]
\multicolumn{6}{@{}l}{\itshape\small Full repository-identifier cleaning control (14B, Lite):} \\
\addlinespace[1pt]
14B Codeaware clean$^\dagger$    & 0.5 & 56.67          & 53.33          &  -3.34          &  -5.9 \\
\midrule
\multicolumn{6}{@{}l}{\itshape\small SWE-bench Verified ($n{=}500$)} \\
\addlinespace[1pt]
SWE-only unaugmented recipe & 0.0 & 46.12          & 26.43          & -19.69          & -42.7 \\
Codeaware SWE               & 0.5 & 44.98          & 36.02          &  -8.96          & -19.9 \\
Combined-data path-augmented recipe & 0.5 & 48.13     & 45.02          &  -3.11          &  -6.5 \\
14B no-aug                  & 0.0 & 48.62          & 34.18          & -14.44          & -29.7 \\
\rowcolor{ours}
14B Codeaware (path-aug recipe) & 0.5 & \bfseries 48.93 & \bfseries 46.67 & \bfseries -2.26 & \bfseries -4.6 \\
\bottomrule
\end{tabularx}
\end{table}

\paragraph{Path sensitivity.} The PathSwap losses describe path sensitivity under the stated 4-bit seed-42 recipes; paired contrasts and uncertainty are reported in \S\ref{sec:codeaware:mcnemar}.

\begin{figure}[!htbp]
\centering
\includegraphics[width=\linewidth]{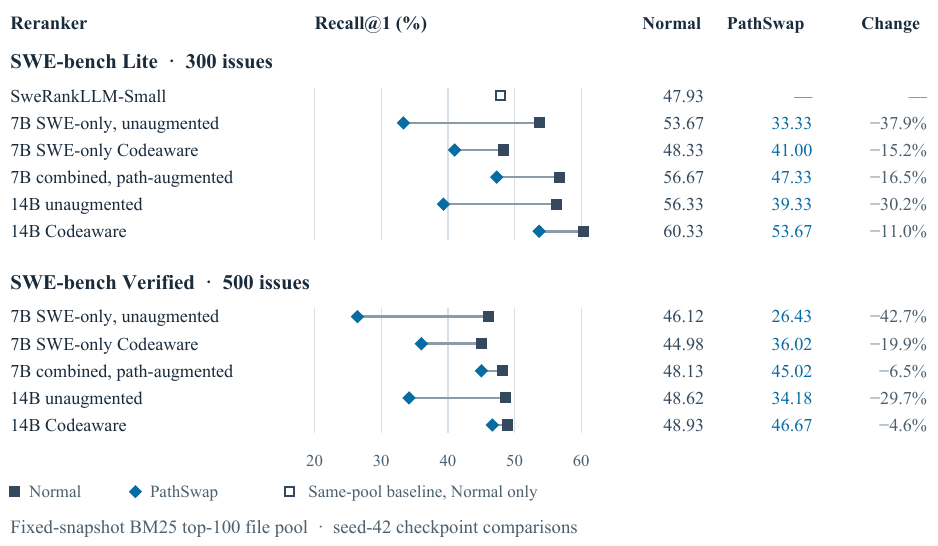}
\caption{PathSwap sensitivity by reranker. File ranking with original paths and opaque identifiers. Squares and diamonds show fractional Recall@1 on fixed BM25 top-100 candidate pools for 300 Lite issues and 500 Verified issues. For the trained pointwise 7B and 14B LoRA rerankers, the change is $100 \times (\mathrm{PathSwap} - \mathrm{Normal}) / \mathrm{Normal}$ at the seed-42, 4-bit operating point. The open square shows the listwise SweRankLLM-Small reranker on the same pool with original paths, 47.93\% on Lite. On Verified the matched 14B Normal difference, Codeaware minus unaugmented, is $+0.31$ percentage points (paired bootstrap 95\% CI $-1.87$ to $+2.47$). The markers show point estimates for selected rows from the fixed-snapshot study. Tables~\ref{tab:codeaware_main} and~\ref{tab:codeaware_invariance} give the corresponding results.}
\label{fig:codeaware_main}
\end{figure}

Figure~\ref{fig:codeaware_main} visualizes the same fixed-snapshot diagnostic reported in Tables~\ref{tab:codeaware_main}--\ref{tab:codeaware_invariance}.

\subsubsection{Precision Sanity Check}
\label{app:precision_sanity}

Comparing bf16 against 4-bit inference under identical prompts, candidates, and seed across three 7B recipes (SWE-only unaugmented, Codeaware, and combined-data path-augmented) on Normal and PathSwap Lite yields small absolute differences ($|\Delta|{\leq}2.0$pp, mean $1.0$pp) and preserves the qualitative ranking across all six cells: relative PathSwap drops are $-37.9\%$ (4-bit) vs.\ $-36.1\%$ (bf16) for unaugmented, $-15.2\%$ vs.\ $-19.5\%$ for Codeaware, and $-16.5\%$ vs.\ $-16.0\%$ for combined-data path-augmented, with the unaugmented drop roughly $2$--$2.5\times$ larger in either precision. This six-cell check covers 7B models only and does not isolate precision for 14B checkpoints or for GREPO's 603-versus-598 path-only snapshots (where precision and candidate pool changed together, Appendix~\ref{app:metric_validity_confusion}). Throughout the paper, 4-bit numbers appear only in Tables~\ref{tab:codeaware_main}--\ref{tab:codeaware_invariance} and their companion rows, while PathAlias, organic drift, and GREPO function transfer use bf16 as noted in each section. Small Lite Normal differences between the bf16 PathAlias harness (Table~\ref{tab:pathalias_cells}) and 4-bit rows (at most 1pp for 14B; 56.0\% vs.\ 53.67\% for 7B SWE-only) reflect separate evaluation runs rather than an isolated precision effect.

\FloatBarrier
\section{Metadata and Code-Use Diagnostics}
\label{app:path_diagnostics}
\begin{table}[!htbp]
\centering\small
\caption{Displayed-path controls for path-only GREPO cross-encoders ($n=1{,}704$, fractional Recall@1 in \%). Parentheses give the relative change from each model's single-checkpoint Normal baseline. Delexicalize replaces path tokens also present in the issue with opaque identifiers. Historical-rename calibration on 862 renamed GREPO issues with code-in-prompt 14B recipes appears in Table~\ref{tab:organic_drift}.}
\label{tab:path_controls_main}\label{tab:path_perturb}
\newcommand{\pathcell}[2]{\makebox[2.7em][r]{#1}\hspace{.2em}\makebox[3.4em][l]{#2}}
\begin{tabularx}{\linewidth}{@{}Xrrr@{}}
\toprule
{\colhead{Intervention}} & {\colhead{Qwen2.5-7B}} & {\colhead{Llama-3.1-8B}} & {\colhead{Qwen3-8B}}\\\midrule
None (Normal) & \pathcell{27.01}{} & \pathcell{28.35}{} & \pathcell{29.38}{} \\
Remove module names & \pathcell{27.70}{($+3\%$)}  & \pathcell{25.48}{($-10\%$)}  & \pathcell{26.43}{($-10\%$)} \\
Flatten to single level & \pathcell{20.93}{($-23\%$)}  & \pathcell{0.63}{($-98\%$)}  & \pathcell{0.67}{($-98\%$)} \\
Swap leaf directories & \pathcell{17.19}{($-36\%$)}  & \pathcell{1.93}{($-93\%$)}  & \pathcell{2.03}{($-93\%$)} \\
Shuffle directories & \pathcell{14.72}{($-46\%$)}  & \pathcell{2.17}{($-92\%$)}  & \pathcell{2.17}{($-93\%$)} \\
Delexicalize & \pathcell{8.85}{($-67\%$)}  & \pathcell{3.72}{($-87\%$)}  & \pathcell{3.90}{($-87\%$)} \\
Shuffle filenames & \pathcell{7.24}{($-73\%$)}  & \pathcell{7.11}{($-75\%$)}  & \pathcell{7.60}{($-74\%$)} \\
PathSwap (opaque IDs) & \pathcell{1.06}{($-96\%$)}  & \pathcell{1.30}{($-95\%$)}  & \pathcell{1.38}{($-95\%$)} \\
\bottomrule
\end{tabularx}
\end{table}
\subsection{Static Metadata Associations}

\paragraph{Paths predict graph structure.}
Logistic regression on path features alone predicts co-change edges with AUC$=0.667$ and import edges with AUC$=0.728$ (Appendix~\ref{app:path_predicts}).
The most predictive features are Jaccard path-component overlap ($+$5.01 log-odds for import) and leaf-filename similarity ($+$1.43 for co-change).
These path features predict held-out edge labels above chance.

\paragraph{Path overlap tracks reranker gains in the tested strata.}
Stratified by the token overlap between the bug report and file paths, high-overlap examples gain $+$19.0pp from reranking, medium $+$12.8pp, and low only $+$6.2pp, a pattern consistent with the reranker \emph{amplifying} path-associated signal. Input-gradient attribution ($|\nabla_{\mathbf{e}} \cdot \mathbf{e}|$) on 300 GREPO examples provides a concordant, model-specific diagnostic: path tokens receive $4.8\times$ higher per-token attribution density than issue tokens (22.9\% of total attribution from $\sim$4\% of tokens; Table~\ref{tab:gradient_attribution}). Large path-perturbation sensitivity recurs across the tested model families, benchmarks, and languages; the gradient-attribution statistic itself is from one Qwen2.5-7B model. The SweRankEmbed row is a separate path-document encoder diagnostic (\S\ref{sec:generalization}; Tables~\ref{tab:cross_architecture} and~\ref{tab:swerank_perturb}).

\subsection{Filename and Directory Perturbations}

Table~\ref{tab:path_perturb} applies seven separate path perturbations against each path-only model's GREPO baseline, while Table~\ref{tab:pathswap_diagnostic} adds filename-shuffling controls across additional scoring architectures. Although all three cross-encoders collapse under filename shuffling ($-73$ to $-75\%$) and PathSwap ($-95$ to $-96\%$), they diverge sharply on directory structure: under directory flattening, leaf-directory swaps, and directory shuffling, Qwen2.5-7B retains 14.7--20.9\% Recall@1 whereas Llama-3.1-8B and Qwen3-8B drop to 0.6--2.2\%. Across model scales from 0.5B to 7B on the graph candidate pool, PathSwap pushes Recall@1 down to 1--2\% everywhere, with relative loss deepening from $-91.8\%$ at 0.5B to $-96.1\%$ at 7B (Table~\ref{tab:pathswap}).

\begin{table}[!htbp]
\centering
\caption{Filename shuffling diagnostic on GREPO-Test (1704 examples). Gap = Standard $-$ Shuffled. Ratio = Shuffled / Standard. Lower Gap and higher Ratio indicate less path dependence. Cross-encoder and code-residual rows use fractional Recall@1; generative and bi-encoder use Hit@1 (binary top-1 hit).}
\label{tab:pathswap_diagnostic}
\small
\begin{tabular}{l S[table-format=2.2] S[table-format=2.2] S[table-format=2.2] S[table-format=1.2]}
\toprule
{\colhead{\textbf{Model}}} & {\colhead{{\textbf{Standard}}}} & {\colhead{{\textbf{Shuffled}}}} & {\colhead{{\textbf{Gap $\downarrow$}}}} & {\colhead{{\textbf{Ratio $\uparrow$}}}} \\
\midrule
Generative (Qwen 7B)              & 41.55 & 19.54 & 22.01 & 0.47 \\
Path-only (Qwen3-8B)              & 29.38 &  7.60 & 21.78 & 0.26 \\
Path-only (Qwen2.5-7B)            & 27.01 &  7.24 & 19.77 & 0.27 \\
Path-only (Llama-3.1-8B)          & 28.35 &  7.11 & 21.24 & 0.25 \\
Code-centric reranker             & 28.74 &  6.62 & 22.12 & 0.23 \\
Combined candidates (no reranker) & 22.51 & {not evaluated} & {not applicable} & {not applicable} \\
Code-residual reranker            & 22.68 &  6.04 & 16.64 & 0.27 \\
Bi-encoder (E5-large)             &  9.50 &  2.92 &  6.58 & 0.31 \\
\bottomrule
\end{tabular}
\end{table}

The additional rows change scoring procedure as well as model, so their magnitudes are read within the stated metric family.

\subsection{Code-input and selected-subset diagnostics}
\label{app:decomposed_reranker}

We separate prompt-content comparisons, trained code-based scorers, code-availability subsets and additional protocol controls. Aggregate results and selected subsets answer different questions; each comparison retains its stated checkpoint and pool.

\subsubsection{Scorer Prompt and Training Comparisons}

Table~\ref{tab:content_ablation} records prompt-content and training-input comparisons under their stated configurations.

\begin{table}[!htbp]
\centering
\caption{Scorer prompt and training configurations. The rows compare complete training and prompting configurations. The path-only row has the highest listed Recall@1; a separate 3-seed path-only mean is 28.92$\pm$0.56\% (Table~\ref{tab:seeds}). }
\label{tab:content_ablation}
\small
\begin{tabular}{l S[table-format=2.2] S[table-format=2.2]}
\toprule
{\colhead{\textbf{Condition}}} & {\colhead{{\textbf{Recall@1}}}} & {\colhead{{\textbf{R@5}}}} \\
\midrule
Path only (default)                          & \bfseries 27.01 & \bfseries 49.17 \\
Path + content summary                       & 23.01 & 39.89 \\
Path + content (init from path-only)         & 22.94 & 39.47 \\
Content training $\to$ path-only eval        & 22.28 & 41.28 \\
\bottomrule
\end{tabular}
\end{table}

\paragraph{Path-associated signal is large in the tested scorers.}
Adding file content does not displace the path signal: a content-aware reranker
that augments paths with file-content summaries achieves Recall@1$=23.01\%$, lower
than the path-only baseline despite having strictly more information, and all
content-integration variants underperform the same baseline
(Table~\ref{tab:content_ablation}; we examine feature-defined subsets
in \S\ref{sec:where_code}).

\subsubsection{Decomposed and Content-Based Scorers}

\paragraph{Decomposed reranker setup.}
Pointwise scoring is a black box: it can route on any combination of path tokens and code tokens. To disentangle the channels, we train a code-residual reranker that receives paths whose components use opaque identifiers plus full file code content, encouraging reliance on code signals rather than path identity. This model achieves Recall@1 = 22.68\%, substantially below the path-only model (28.92$\pm$0.56\%).

\paragraph{Alpha-sweep fusion.}
We linearly fuse the code-residual and path-only scores: $s_{\text{fuse}} = \alpha s_{\text{code-residual}} + (1-\alpha) s_{\text{path}}$, sweeping $\alpha \in \{0, 0.1, \ldots, 1\}$ on the held-out validation split. The optimal $\alpha$ on the validation set is \textbf{$\alpha = 0$} (\emph{pure path-only}), so no additive validation benefit from code content is detected in this configuration (Table~\ref{tab:decomposed_reranker}).

\begin{table}[!htbp]
\centering
\caption{Decomposed reranker: path-only vs.\ code-residual scores and linear fusion on GREPO ($n=1{,}704$). The validation $\alpha$-sweep selects $\alpha{=}0$; no additive validation gain is detected for this residual. Code-residual and fusion rows are single-seed; the 6+pp gap exceeds the path-only three-seed standard deviation ($\pm0.56$).}
\label{tab:decomposed_reranker}
\small
\begin{tabular}{l r}
\toprule
{\colhead{\textbf{Model}}} & {\colhead{\textbf{Recall@1 (\%)}}} \\
\midrule
Path-only (Qwen2.5-7B, 3-seed)             & $\mathbf{28.92 \pm 0.56}$ \\
Code-residual reranker (opaque-identifier paths + code) & $22.68$ \\
Best fusion ($\alpha$ sweep on val)        & $28.92$ \ ($\alpha{=}0$ selected) \\
\bottomrule
\end{tabular}
\end{table}

\paragraph{Counterfactual and code-centric controls.}
A matched-compute code-centric scorer (real paths + 50 lines of code + AST signatures) achieves Recall@1 = 28.74\%, no better than path-only ($28.92\pm0.56\%$, $p > 0.20$), and collapses equally under filename shuffling (Table~\ref{tab:code_centric}). Path-only wins on all subsets except issue-path mismatch ($2.4\%$ of the 1{,}704-example test ($n{=}41$), $+7.2$pp for code-centric); training with 50\% shuffled paths achieves 6.54\% under shuffling.

\paragraph{Code inputs during training and file scoring.}
Selective-retrieval training supplies up to three functions per candidate file, ranked by TF--IDF cosine similarity of issue and function text using a smoothed two-document inverse frequency (up to 2{,}000 characters, with a file-head fallback). Long-context training supplies up to 200 lines (6{,}000 characters). Both training sequences have a 2{,}048-token limit. For the file-ranking runs summarized here, both types of trained checkpoint receive only the issue and candidate path in a 512-token prompt. Full-test fractional Recall@1 is $28.44\pm0.62\%$ for selective retrieval (two seeds), $28.56\pm0.23\%$ for long context (three seeds), and $28.92\pm0.56\%$ for path-only (three seeds). Table~\ref{tab:selective-code-main} gives the separate binary outcomes for the 116-issue subset.

\begin{table}[!htbp]
\centering
\caption{Training recipes and file-ranking results on GREPO ($n=1{,}704$). For selective retrieval and long context, code-token counts describe training inputs; both score the issue and candidate path at inference. Multi-seed entries report mean $\pm$ SD. The path-only filename shuffle was measured on a single checkpoint ($27.01$ to $7.24$, $-73\%$, Tables~\ref{tab:path_perturb} and~\ref{tab:cross_architecture}); the three-seed path-only row reports its Normal mean.}
\label{tab:strong_code_baselines}
\small
\begin{tabular}{l l r r r}
\toprule
{\colhead{\textbf{Model}}} & {\colhead{\makecell{\textbf{Training}\\\textbf{code tokens}}}} & {\colhead{{\textbf{Recall@1 (\%)}}}} & {\colhead{{\textbf{Shuffle Recall@1}}}} & {\colhead{\textbf{Drop}}} \\
\midrule
Path-only (3-seed)                & 0          & $28.92\pm0.56$ & ---        & --- \\
Code-centric (50L, 768 ctx)       & $\sim$375  & $28.74$        & $6.62$     & $-77\%$ \\
Selective retrieval (top-3, 2-seed)& $\sim$800 & $28.44\pm0.62$ & $7.47$     & $-74\%$ \\
Long-context code (200L, 3-seed)  & $\sim$1500 & $28.56\pm0.23$ & $7.37\pm0.06$ & $-74\%$ \\
\makecell[l]{Code-residual long-ctx\\(opaque-identifier paths)}& $\sim$1500& $25.19$ & $7.38$ & $-71\%$ \\
\bottomrule
\end{tabular}
\end{table}

\subsubsection{A feature-defined subset with weak path cues}

\label{sec:where_code}

\paragraph{Code-Crucial.}
These 116 GREPO issues (6.8\% of the 1,704-issue test set) have bottom-quartile overlap between the issue and gold-file path tokens. At least one gold file shares its filename stem with another file, or no gold file ranks in the top ten under path-token Jaccard similarity. The gold files must be parseable, at least half the candidate files must be parseable, and code-only BM25 must place a gold file in the top twenty. This subset defines a setting in which source text is available.

On the 116 fixed issues, the path-only reference has 41 top-one file hits, selective training with seed 42 has 47, and seed 123 has 37. These binary rates (35.34\%, 40.52\% and 31.90\%) are distinct from fractional Recall@1 when several files are gold. The corresponding fractional Recall@1 values from these same rankings are 21.59\%, 25.99\% and 19.12\%, respectively. At evaluation both selective checkpoints saw only the report and path, and the direction of change varies across the two training seeds.

\paragraph{Procedures that score other inputs.}
On the same 116 issues, the separate residual procedure gives 24/116 binary hits (20.69\%; fractional Recall@1 is 12.97\%). A separate zero-shot procedure actually provides file code during scoring (up to 200 lines, capped at 2{,}000 characters); it gives 15/116 hits (12.93\%). Their metrics and scoring inputs are identified separately to avoid comparing a fractional score with a binary hit rate. Across seven broader slice definitions the residual procedure's fractional Recall@1 is 1.5--5.1 points below its path reference (Table~\ref{tab:cc_sensitivity}). The feature-defined subset has Jaccard overlap 0.18 with an earlier model-defined set. Of 25 inspected failures, 15 gold files were judged code-identifiable. The repository path-descriptiveness check below reports the separate median-overlap comparison.

\paragraph{Per-example decomposition.}
Path-only is uniquely correct on 17.1\% of examples (2.4$\times$ code's 7.2\%); code helps on a small mismatch slice ($2.4\%$ of the 1{,}704-example test ($n{=}41$), $+7.2$pp), but that slice is too rare to shift aggregates. Further decomposition of code signals (identifier obfuscation, attention analysis, issue-path mismatch slicing) appears in Table~\ref{tab:code_centric} and Appendix~\ref{app:counterfactual}.

\paragraph{Repository Path Descriptiveness}
\label{app:opaque_descriptive}

We classify each repository as ``descriptive'' or ``opaque'' by its median path--issue token overlap, splitting at the across-repository median. Path-only reaches 35.8\% Recall@1 on repositories above that median and 21.4\% on repositories below the median path--issue token overlap ($-$14.4pp). This split partly explains the association.

\subsubsection{Additional Protocol Controls}

\paragraph{Code-only and pool-scope controls.}
Removing paths entirely and scoring by code content alone (``content-replace'') collapses to 6.85\% Recall@1. Post-scoring graph propagation is also negative, and mixed-retriever-pool training drops Recall@1 from 27.0\% to 20.0\%. By contrast, scoring \textbf{all Python source files} per repository (avg.\ 541 files) yields path-only Recall@1\,$=$\,29.71\%, within one point of the BM25-based procedure ($28.92\pm0.56\%$), showing that the high path-only score is not confined to the BM25 candidate pool.

\paragraph{Hierarchical localization.}
Does code matter more when the search scope is narrowed? We compare function-name shuffling in gold-directory and gold-file-conditioned candidate pools (Table~\ref{tab:hierarchical}).

\begin{table}[!htbp]
\centering
\caption{Conditioned-scope function-name shuffling controls on GREPO. The gold-directory and gold-file rows restrict candidates to functions within the gold directory or gold file, respectively, and are interpreted as separate conditioned scopes.}
\label{tab:hierarchical}
\small
\begin{tabular}{l S[table-format=2.2] S[table-format=2.2] r r}
\toprule
{\colhead{\textbf{Scope}}} & {\colhead{{\textbf{Baseline Recall@1 (\%)}}}} & {\colhead{{\textbf{Shuffle Recall@1}}}} & {\colhead{\textbf{Drop}}} & {\colhead{\textbf{Avg cands}}} \\
\midrule
Gold directory                 & 13.76 & 1.62 & $-88\%$ & $\sim$2033 \\
Gold file                      & 17.91 & 4.47 & $-75\%$ & 204         \\
\bottomrule
\end{tabular}
\end{table}

Shuffling function names drops Recall@1 by 88\% within the gold directory and by 75\% within the gold file, showing lower but still large sensitivity in the narrower conditioned pool.

\begin{table}[!htbp]
\centering
\small
\caption{PathSwap across tested model scales on GREPO-Test ($n=1{,}704$), using stable structure-preserving opaque identifiers on the graph candidate pool. No random baseline is included.}
\label{tab:pathswap}
\begin{tabular}{llrrrrrr}
\toprule
{\colhead{\textbf{Model}}} & {\colhead{\textbf{Setting}}} & {\colhead{\textbf{Recall@1}}} & {\colhead{$\boldsymbol{\Delta}$}} & {\colhead{\textbf{R@5}}} & {\colhead{$\boldsymbol{\Delta}$}} & {\colhead{\textbf{R@10}}} & {\colhead{$\boldsymbol{\Delta}$}} \\
\midrule
0.5B & original & 20.02 & reference & 38.69 & reference & 47.46 & reference \\
0.5B & PathSwap & 1.65 & $-91.8\%$ & 4.97 & $-87.2\%$ & 8.50 & $-82.1\%$ \\
1.5B & original & 21.54 & reference & 41.16 & reference & 48.74 & reference \\
1.5B & PathSwap & 1.63 & $-92.4\%$ & 4.52 & $-89.0\%$ & 8.16 & $-83.3\%$ \\
3B & original & 23.37 & reference & 43.58 & reference & 51.24 & reference \\
3B & PathSwap & 1.55 & $-93.4\%$ & 3.70 & $-91.5\%$ & 6.86 & $-86.6\%$ \\
7B & original & 27.01 & reference & 49.17 & reference & 56.45 & reference \\
7B & PathSwap & 1.06 & $-96.1\%$ & 4.09 & $-91.7\%$ & 6.86 & $-87.8\%$ \\
\bottomrule
\end{tabular}
\end{table}

\subsection{Path Controls and Attribution Diagnostics}

Readable-path examples, opaque controls, attribution and slice diagnostics answer separate questions.

\begin{table}[!htbp]
\centering\small
\setlength{\tabcolsep}{4pt}
\caption{Readable paths and the same hierarchy with opaque component identifiers.}
\label{tab:pathswap_examples}
\begin{tabularx}{\linewidth}{@{}Xl@{}}
\toprule
{\colhead{Readable path}} & {\colhead{Same hierarchy with opaque component identifiers}} \\
\midrule
\nolinkurl{cirq/value/linear\_dict.py} & \nolinkurl{d\_f4e8/d\_a406/m\_008e.py} \\
\nolinkurl{cirq/value/linear\_dict\_test.py} & \nolinkurl{d\_f4e8/d\_a406/m\_d8be.py} \\
\nolinkurl{cirq/\_\_init\_\_.py} & \nolinkurl{d\_f4e8/m\_3414.py} \\
\nolinkurl{cirq/ops/\_\_init\_\_.py} & \nolinkurl{d\_f4e8/d\_0cf0/m\_3414.py} \\
\nolinkurl{cirq/circuits/circuit\_test.py} & \nolinkurl{d\_f4e8/d\_1dc6/m\_921d.py} \\
\nolinkurl{cirq/google/sim/xmon\_sim\_test.py} & \nolinkurl{d\_f4e8/d\_d493/d\_0736/m\_7fbe.py} \\
\nolinkurl{cirq/sim/\_\_init\_\_.py} & \nolinkurl{d\_f4e8/d\_0736/m\_3414.py} \\
\bottomrule
\end{tabularx}
\end{table}

\begin{table}[!htbp]
\centering\small
\setlength{\tabcolsep}{4pt}
\caption{Gradient attribution on 300 GREPO examples (mean $\pm$ SD); density is $\times10^{3}$.}
\label{tab:gradient_attribution}
\begin{tabular}{lrrrr}
\toprule
{\colhead{Category}} & {\colhead{Mean tokens}} & {\colhead{Token share}} & {\colhead{Attribution}} & {\colhead{Density}} \\
\midrule
Path & 6.2 & 3.7\% & $22.9\pm11.8\%$ & 40.0 \\
Issue & 140.0 & 82.5\% & $51.2\pm18.5\%$ & 8.3 \\
Template & 23.6 & 13.9\% & $25.9\pm11.0\%$ & 11.0 \\
\bottomrule
\end{tabular}
\end{table}

\begin{table}[!htbp]
\centering\small
\setlength{\tabcolsep}{4pt}
\centering
\caption{Path predictability (1{,}321 issues, 66K triples, GroupKFold).}
\label{tab:path_predictability}
\begin{tabular}{lr}
\toprule
{\colhead{Feature set}} & {\colhead{AUC}} \\
\midrule
Path lexical overlap with issue & 0.636 \\
Path TF--IDF & 0.640 \\
Issue + Path TF--IDF & 0.589 \\
Path structure only & 0.587 \\
\bottomrule
\end{tabular}
\end{table}

\begin{table}[!htbp]
\centering\small
\setlength{\tabcolsep}{4pt}
\centering
\caption{Code-Crucial sensitivity; the 116-example definition in Section~\ref{sec:where_code} is stricter.}
\label{tab:cc_sensitivity}
\begin{tabular}{lrrrr}
\toprule
{\colhead{Definition}} & {\colhead{$N$}} & {\colhead{Path}} & {\colhead{Code-residual}} & {\colhead{$\Delta$}} \\
\midrule
Original (frac$<0.1$, witness) & 265 & 19.1 & 17.5 & $-1.6$ \\
Strict (frac$<0.05$, witness) & 215 & 22.2 & 19.8 & $-2.4$ \\
Loose (frac$<0.2$, witness) & 489 & 20.8 & 17.6 & $-3.2$ \\
No witness (frac$<0.1$) & 454 & 11.7 & 10.2 & $-1.5$ \\
No witness (frac$<0.2$) & 727 & 14.4 & 11.8 & $-2.6$ \\
Jaccard$<0.05$, witness & 962 & 31.7 & 26.5 & $-5.1$ \\
Jaccard$<0.1$, witness & 1271 & 31.8 & 26.9 & $-4.9$ \\
\bottomrule
\end{tabular}
\end{table}

\paragraph{Path and attribution conclusions.}
\phantomsection\label{app:pathswap_examples}
PathSwap replaces each directory and filename stem with deterministic opaque identifiers, preserving depth, extension, content, graph topology, and labels while changing token familiarity and distribution. PathAlias instead tests readable names from unrelated repositories. Identical \texttt{\_\_init\_\_} stems and shared directory prefixes retain package and co-location regularities. Applying opaque identifiers to all components changes filename and directory components jointly; across 0.5B--7B graph and hybrid pools (including 0.5B hybrid $24.12\to1.36$, 3B hybrid $28.82\to0.92$, and 7B sparse hybrid $20.00\to0.39$), the collapse is 92--98\%, compared with 53--77\% for filename shuffling.

\phantomsection\label{app:gradient_attribution}
On 300 GREPO examples, path tokens are about 4\% of tokens but receive 22.9\% of attribution, a
$4.8\times$ per-token density advantage over issue tokens; the density columns are ordinal rather than
arithmetically recoverable from aggregate means.
\phantomsection\label{app:path_predictability}
Grouped path-only TF--IDF reaches AUC 0.640, and adding issue text lowers it to 0.589.
\phantomsection\label{app:cc_sensitivity}
Across seven broader Code-Crucial definitions, code-residual is $-1.5$ to $-5.1$pp below path-only; the 116-example definition in Section~\ref{sec:where_code} is stricter.

\begin{table}[htbp]
\centering
\small
\caption{Code-aware versus path-only on GREPO; CC is real paths plus code and CR is PathSwap plus code. The Path-only full-test entry reports the three-seed mean $\pm$ SD. Cells marked not evaluated were not run for that model.}
\label{tab:code_centric}
\adjustbox{max width=\linewidth}{%
\begin{tabular}{lrrrr}
\toprule
{\colhead{Test condition}} & {\colhead{Path-only}} & {\colhead{CC (50L)}} & {\colhead{CC (200L)}} & {\colhead{CR (100L)}} \\
\midrule
Full test ($n=1{,}704$) & $28.92\pm0.56$ & 28.74 & 28.94 & 24.19 \\
Shuffle filenames & 7.24 & 6.62 & not evaluated & 6.33 \\
Same-stem ($n=904$) & 24.04 & 23.08 & not evaluated & 17.67 \\
Hard ($n=904$) & 14.98 & 14.18 & not evaluated & 11.01 \\
Issue-path mismatch ($n=41$) & 27.59 & 34.81 & not evaluated & not evaluated \\
\bottomrule
\end{tabular}}
\end{table}
\paragraph{Code-signal and counterfactual conclusions.}
\phantomsection\label{app:code_decomposition}
Path-only is uniquely correct on 17.1\%, code on 7.2\%, both on 32.3\%, and neither on 43.5\%;
the tested fusion selects $\alpha=0$. On 41
extreme-mismatch examples, code-centric is 34.81\% versus path-only 27.59\% ($+7.2$pp): the only
diagnostic point estimate on which code-centric is higher, and on a $2.4\%$ slice of the 1{,}704-example test. Identifier obfuscation with
PathSwap changes 22.68\% to 17.39\% ($-5.3$pp), versus 1.1pp with real paths. Last-layer attention density is $12.2\times$ higher for path
than code tokens despite code being 53\% of input.

\phantomsection\label{app:counterfactual}
On the 1{,}546 records with valid swap pairs, the original condition and all three code-swap conditions supply the path-only scorer with the same issue text and candidate paths. Only code fields change, and this scorer does not receive them. Its 32.84\% result is therefore unchanged across these conditions by construction, providing a consistency check (not evidence that changing code never matters). Examples without valid swap pairs are excluded from this subset. Path-only wins or
ties within seed variance on the displayed code-aware subsets except issue-path mismatch.

In the GREPO-only, no-code-in-prompt cell, 50\% shuffle augmentation scores 26.01\% in the normal setting, versus $28.92\pm0.56\%$, and 6.54\% in the shuffled setting, versus 7.24\%.

The 100-line and 30-line code configurations score 24.60\% and 24.09\%, respectively. Across the full de-leaked GREPO test split ($n=2{,}175$, before restricting to the 1{,}704 Python-changing issues) and four path-selected subsets of the 1{,}704-issue test set---same-stem ($n=904$), hard ($n=904$), same-directory confused ($n=334$), and path-misled ($n=229$)---path-only scores 18.05\%, 24.04\%, 14.98\%, 9.04\%, and 8.45\%, respectively, versus 28.92\% on the 1{,}704-example test. Post-hoc selection prevents a causal reading.

\subsection{Path Features Predict Graph Structure}
\label{app:path_predicts}

To quantify how well path-derived features predict repository graph edges, we train logistic regression classifiers to predict co-change and import edges from path-derived features alone.
For each repository, we construct positive pairs (file pairs connected by the target edge type) and sample 5$\times$ negative pairs (unconnected files).
Four features are computed per pair:

\begin{enumerate}
    \item \textbf{Jaccard path components}: Jaccard similarity over the set of directory/file components in each path.
    \item \textbf{Shared directory depth}: number of shared leading path components.
    \item \textbf{Leaf filename similarity}: character-level Jaccard similarity of the filename (without extension).
    \item \textbf{Path edit distance}: Levenshtein distance between full paths, normalized by max length.
\end{enumerate}

\begin{table}[!htbp]
\centering
\caption{Path features predicting graph edges. AUC averaged across repositories with $\geq$10 positive pairs.}
\label{tab:path_pred}
\small
\begin{tabular}{l S[table-format=1.3] l}
\toprule
{\colhead{\textbf{Edge type}}} & {\colhead{{\textbf{AUC}}}} & {\colhead{\textbf{Top feature (log-odds)}}} \\
\midrule
Co-change & 0.667 & Leaf filename similarity ($+1.43$) \\
Import    & 0.728 & Jaccard path components ($+5.01$) \\
\bottomrule
\end{tabular}
\end{table}

The path-feature classifier predicts held-out edge labels above chance.

\section{Scope and Candidate-Pool Sensitivity}
\label{app:scope_studies}

\subsection{Breadth across Tested Settings}
\label{sec:generalization}

We ask whether filename-shuffle sensitivity is confined to one benchmark, scoring interface, or programming language. Each table retains its own population and metric family.

\phantomsection\label{par:api_metric_boundary}
Table~\ref{tab:cross_architecture} consolidates filename-shuffle drops across scoring architectures (cross-encoder, bi-encoder, generative listwise, and frontier API models prompted with up to 100 lines of code) and languages (Python and Python-to-Java transfer on BeetleBox~\citep{chakraborty2025blaze}), while SWE-bench Lite transfer and path-document SweRank probes follow in Tables~\ref{tab:zero_shot_transfer} and~\ref{tab:swerank_perturb}. Additional single-seed scope cells cover 1.5B--14B scales (Table~\ref{tab:model_scale_sensitivity}), Qwen/Llama/Mistral families (Table~\ref{tab:cross_family_sensitivity}), and in-domain Java data from BeetleBox and Bench4BL (Tables~\ref{tab:java_beetlebox_sensitivity} and~\ref{tab:java_bench4bl_sensitivity}); Verified sensitivity appears separately in Tables~\ref{tab:codeaware_main}--\ref{tab:codeaware_invariance}.

On the 100-line top-20 BM25 GREPO protocol ($N=152$, binary Hit@1), filename shuffling lowers \texttt{gpt-6.1-sol} from 88/152 (57.89\%) to 63/152 (41.45\%, $-28\%$; paired 95\% CI $[-23.0,-9.9]$ points, exact McNemar $p=1.6\times10^{-6}$) and GPT-4.1-mini from 78/152 (51.32\%) to 54/152 (35.53\%, $-31\%$; $[-23.0,-9.2]$, $p=1.9\times10^{-5}$). Note that this 152-issue GREPO fixed-preview protocol is separate from the 193-issue Lite tool-navigation study in \S\ref{sec:revision-source-tests} and supplies no readable-name tool comparison. GREPO-trained models transfer zero-shot to SWE-bench Lite at 48--51\% Hit@1, comparable to the SWE-adapted reranker (50.67\%; Table~\ref{tab:zero_shot_transfer}). When applied to file-path strings as retrieval documents, SweRankEmbed-Large drops by 56--78\% under filename, directory, and flatten perturbations (Table~\ref{tab:swerank_perturb}); SweRank's native function-body workflow is documented separately in Appendix~\ref{app:swerank_repro}. Applied zero-shot to 740 Java bugs from BeetleBox~\citep{chakraborty2025blaze} (BM25 top-200), the GREPO-trained Python cross-encoder reaches 14.43\% Recall@1 (29.89\% R@5) and drops by $-60\%$ to 5.79\% under filename shuffling.

\begin{table}[!htbp]
\centering
\caption{Filename-shuffling perturbation across scoring interfaces (GREPO $n=1{,}704$ unless noted). Cross-encoder rows and the Python-to-Java cell use fractional Recall@1; bi-encoder and generative listwise rows use binary Hit@1. Rows marked $^\ddagger$ report separate historical frontier-API evaluations on GREPO subsets ($N=77$--$152$) whose retained logs do not uniquely distinguish binary Hit@1 from fractional Recall@1 and are not aggregated across metric families. Rows marked $^\S$ apply the 100-line top-20 BM25 protocol ($N=152$) with verified binary Hit@1. No cross-metric aggregation is reported.}
\label{tab:cross_architecture}
\small
\setlength{\tabcolsep}{5pt}
\renewcommand{\arraystretch}{1.15}
\begin{tabular}{l l S[table-format=2.2] S[table-format=2.2] r}
\toprule
{\colhead{\textbf{Architecture}}} & {\colhead{\textbf{Training}}} & {\colhead{{\textbf{Baseline (\%)}}}} & {\colhead{{\textbf{Shuffle (\%)}}}} & {\colhead{\textbf{Drop}}} \\
\midrule
Cross-encoder (Qwen2.5-7B)               & Fine-tuned & 27.01 &  7.24 & $-73\%$ \\
Cross-encoder (Llama-3.1-8B)             & Fine-tuned & 28.35 &  7.11 & $-75\%$ \\
Cross-encoder (Qwen3-8B)                 & Fine-tuned & 29.38 &  7.60 & $-74\%$ \\
\midrule
Cross-encoder (Qwen2.5-14B)              & Zero-shot  & 16.21 &  4.70 & $-71\%$ \\
Bi-encoder (E5-large-v2)                 & Zero-shot  &  9.50 &  2.92 & $-69\%$ \\
Generative listwise (Qwen2.5-7B)         & Zero-shot  & 41.55 & 19.54 & $-53\%$ \\
\midrule
Frontier API (DeepSeek-V3.2)$^\ddagger$  & Zero-shot  &  5.06 &  2.72 & $-46\%$ \\
Frontier API (Qwen-Max)$^\ddagger$       & Zero-shot  & 22.45 & 12.57 & $-44\%$ \\
Frontier $+$ code (Qwen3-Max, 100L)$^\ddagger$ & Zero-shot & 55.26 & 30.26 & $-45\%$ \\
Agentic $+$ code (Qwen3-Max, CoT, $N{=}77$)$^\ddagger$ & Zero-shot & 54.55 & 20.78 & $-62\%$ \\
Frontier $+$ code (\texttt{gpt-6.1-sol}, 100L)$^\S$     & Zero-shot & 57.89 & 41.45 & $-28\%$ \\
Frontier $+$ code (GPT-4.1-mini, 100L)$^\S$    & Zero-shot & 51.32 & 35.53 & $-31\%$ \\
\midrule
\multicolumn{5}{@{}l}{\itshape\footnotesize Cross-language (Python$\to$Java, BeetleBox, 740 examples)} \\
\addlinespace[1pt]
Cross-encoder (Qwen2.5-7B)               & Fine-tuned & 14.43 &  5.79 & $-60\%$ \\
\bottomrule
\end{tabular}
\end{table}

\begin{table}[!htbp]
\centering
\caption{Top: SWE-bench Lite adaptation and zero-shot transfer (the adapted row is not zero-shot). Bottom: SWE-bench Lite path perturbation; the Qwen2.5 row uses the SWE-adapted checkpoint, while Llama3.1 and Qwen3 use GREPO checkpoints. Filename shuffling causes a 77--80\% collapse across all three model families. Bold marks the overall best (top) and per-column best (bottom).}
\label{tab:zero_shot_transfer}
\footnotesize
\renewcommand{\dmidp}[1]{{$(#1\%)$}}
\renewcommand{\dsevp}[1]{{$(#1\%)$}}
\setlength{\tabcolsep}{4pt}
\renewcommand{\arraystretch}{1.2}
\begin{minipage}[t]{\linewidth}
\vspace{0pt}
\centering
\begin{tabular}{@{}l S[table-format=2.2]@{}}
\toprule
{\colhead{\textbf{Model}}} & {\colhead{{\textbf{Hit@1 (\%)}}}} \\
\midrule
Qwen2.5 \emph{adapted} & 50.67 \\
Qwen2.5 GREPO          & 48.00 \\
Llama3.1 GREPO         & 48.67 \\
Qwen3 GREPO            & \bfseries 51.33 \\
\bottomrule
\end{tabular}
\end{minipage}
\par\medskip
\begin{minipage}[t]{\linewidth}
\vspace{0pt}
\centering
\begin{tabular}{@{}l r r r@{}}
\toprule
{\colhead{\textbf{Model}}} & {\colhead{\textbf{Filename}}} & {\colhead{\textbf{Dirs}}} & {\colhead{\textbf{Flatten}}} \\
\midrule
Qwen2.5 adapted &     $11.0$  \dsevp{-78} &           $14.7$  \dsevp{-71} &           $30.7$  \dmidp{-39} \\
Llama3.1 & $\phantom{0}9.7$  \dsevp{-80} & $\mathbf{22.0}$   \dsevp{-55} &           $34.3$  \dmidp{-29} \\
Qwen3    & $\mathbf{11.7}$   \dsevp{-77} &           $20.0$  \dsevp{-61} & $\mathbf{39.0}$   \dmidp{-24} \\
\bottomrule
\end{tabular}
\end{minipage}
\end{table}

\paragraph{Directory sensitivity across benchmarks.}
On SWE-bench Lite (Table~\ref{tab:zero_shot_transfer}, bottom), the directory-level pattern inverts relative to GREPO (Table~\ref{tab:path_perturb}). Qwen2.5 is now the \emph{most} sensitive model to directory shuffling ($-$71\%) and flattening ($-$39\%), while Llama-3.1 and Qwen3 are comparatively more robust. Directory sensitivity is benchmark-dependent in these cells. Filename-level vulnerability, by contrast, is consistently large in the tested cells ($-$77\% to $-$80\% on Lite; GREPO rows are in Table~\ref{tab:path_perturb}).

\paragraph{Repository-level breadth.}
Filename shuffling causes a $>$20\% Recall@1 drop in 94\% of repositories (44/47 repos with $n \geq 5$ in this BM25 top-100 subset), with a mean drop of 69\% (median 75\%). Even among repositories in the bottom quartile of baseline Recall@1, the mean drop is 58\%. Combined with the cross-benchmark evidence (SWE-bench, 12 repositories, $-$77--80\%) and the cross-language evidence (BeetleBox Java, $-$60\%), the observed sensitivity is not confined to one repository set.

\begin{table}[!htbp]
\centering
\caption{Path-document perturbation on SweRank encoders (SWE-bench Lite, $n{=}300$). \emph{Filename shuffle} permutes file names within each directory; \emph{shuffle directories} permutes directory names; \emph{flatten directories} collapses the directory hierarchy. SweRankEmbed uses the file-path string as the retrieval document so only the path string changes. This isolates encoder sensitivity to path-string content and does not test SweRank's native function-body retrieval-and-reranking pipeline (documented separately in Appendix~\ref{app:swerank_repro}).}
\label{tab:swerank_perturb}
\small
\setlength{\tabcolsep}{5pt}
\renewcommand{\arraystretch}{1.15}
\begin{tabular}{l S[table-format=2.2] r S[table-format=2.2] r}
\toprule
& \multicolumn{2}{c}{\colhead{\textbf{Small (137M)}}} & \multicolumn{2}{c}{\colhead{\textbf{Large (7B)}}} \\
\cmidrule(lr){2-3}\cmidrule(lr){4-5}
{\colhead{\textbf{Perturbation}}} & {\colhead{{\textbf{Recall@1}}}} & {\colhead{\textbf{Drop}}} & {\colhead{{\textbf{Recall@1}}}} & {\colhead{\textbf{Drop}}} \\
\midrule
None (baseline)        & 26.33 & {reference}    & 37.00 & {reference}    \\
Shuffle filenames      &  6.00 & $-77\%$  &  8.33 & $-77.5\%$\\
Shuffle directories    & {not evaluated} & {not applicable} & 15.00 & $-59.5\%$\\
Flatten directories    & {not evaluated} & {not applicable} & 16.33 & $-55.9\%$\\
\bottomrule
\end{tabular}
\end{table}

The encoder responds differently to filename, directory, and flattening interventions when given path strings. These path-document cells remain separate from the native function-body workflow in Appendix~\ref{app:swerank_repro}.

\subsection{Retrieval Rebuild under PathSwap}
To address fixed-pool conditioning, we rebuild BM25 retrieval over the corpus with PathSwap identifiers (opaque filenames with intact content), regenerate top-500 candidates, and evaluate historical Codeaware and no-augmentation checkpoints (distinct from the rerank-only checkpoints). Rebuilt top-100 oracle recall is lower on Lite (79.00\% vs.\ original 86.67\%) and higher on Verified (82.80\% vs.\ original 66.60\%, 333/500 binary any-hit).

\begin{table}[!htbp]
\centering
\caption{End-to-end PathSwap retrieval rebuild on Lite ($n=300$) and Verified ($n=500$, adaptation-overlap sensitivity). Rebuilt-pool and rerank-only columns use different checkpoint families; on Verified, the rerank-only gap is binary Hit@1 rather than fractional Recall@1.}
\label{tab:retrieval_rebuild}
\footnotesize
\setlength{\tabcolsep}{3pt}
\renewcommand{\arraystretch}{1.12}
\begin{tabular}{lrrrrr}
\toprule
{\colhead{Split}} & {\colhead{Codeaware Recall@1}} & {\colhead{no-aug Recall@1}} & {\colhead{Delta}} & {\colhead{McNemar $p$}} & {\colhead{Rerank-only gap}} \\
\midrule
Lite e2e & 36.67 & 27.33 & $+9.33$ & $3{\times}10^{-4}$ & $+14.33$pp ($p=6{\times}10^{-8}$) \\
Verified e2e & 37.72 & 27.20 & $+10.52$ & $6.9{\times}10^{-8}$ & $+13.40$pp ($p=1{\times}10^{-12}$) \\
\bottomrule
\end{tabular}
\end{table}

\subsection{Held-out repositories}
\begin{table}[htbp]
\centering
\small
\caption{Inductive evaluation on 560 GREPO examples from 12 held-out repositories.}
\label{tab:inductive}
\begin{tabular}{lrrrr}
\toprule
\addlinespace[2pt]
{\colhead{\textbf{Method}}} & {\colhead{\textbf{Recall@1}}} & {\colhead{\textbf{R@5}}} & {\colhead{\textbf{R@10}}} & {\colhead{$\boldsymbol{p}$\textbf{-value}}} \\
\midrule
Combined pool (BM25 ordering) & 20.52 & 38.47 & 46.69 & reference \\
Inductive reranker & 20.77 & 46.70 & 55.37 & R@5/R@10: $<0.0001$ *** \\
\bottomrule
\end{tabular}
\end{table}
\phantomsection\label{app:inductive}

\subsection{Model-Scale Sensitivity Study}
\vspace{-4pt}
We extend the Codeaware recipe down to smaller scales (1.5B and 3B QLoRA) to test whether reduced PathSwap sensitivity also appears at smaller model scales. Each model is trained with the same recipe (augmentation 0.5 with 50\% path-shuffled examples, code-in-prompt with 30 source lines, LoRA rank 32, two epochs) on the SWE-bench train split (n=1{,}870 issues), single seed.

\begin{table}[!htbp]
\centering
\caption{Single-seed Codeaware scale sensitivity (single-snapshot, 4-bit): fractional Recall@1 on Lite ($n=300$) and Verified ($n=500$, adaptation-overlap sensitivity) across 1.5B--14B.}
\label{tab:model_scale_sensitivity}
\begin{tabularx}{\linewidth}{@{}Xrrrrr@{}}
\toprule
\addlinespace[2pt]
{\colhead{Model}} & {\colhead{\makecell{Lite\\Normal}}} & {\colhead{\makecell{Lite\\PathSwap}}} & {\colhead{\makecell{Lite\\drop pp}}} & {\colhead{\makecell{Verified\\Normal}}} & {\colhead{\makecell{Verified\\PathSwap}}} \\
\midrule
1.5B Codeaware & 42.00 & 36.67 & $-$5.33 & 41.42 & 36.80 \\
3B Codeaware & 45.67 & 38.00 & $-$7.67 & 43.60 & 39.02 \\
7B Codeaware & 48.33 & 41.00 & $-$7.33 & 44.98 & 36.02 \\
14B Codeaware (seed 42) & 60.33 & 53.67 & $-$6.66 & 48.93 & 46.67 \\
\bottomrule
\end{tabularx}
\end{table}

The Codeaware cells have smaller PathSwap drops than the stated no-aug comparisons at each tested scale. Relative PathSwap drops on Lite are 12.7\% (1.5B), 16.8\% (3B), 15.2\% (7B), and 11.0\% (14B); on Verified they are 11.2\% (1.5B), 10.5\% (3B), 19.9\% (7B), and 4.6\% (14B). All of these are below the comparison drops (29.7--42.7\% relative across the 7B/14B comparisons), and the paired no-aug controls for 1.5B and 3B appear in Table~\ref{tab:small_scale_controls}. The relative drops are nonmonotonic across these single-seed cells and show the pattern at each tested scale.

\begin{table}[!htbp]
\centering\small
\caption{Paired no-augmentation controls at 1.5B and 3B (single-snapshot, 4-bit, single seed). Recall@1 columns report fractional Recall@1 on Lite ($n=300$) and the Verified adaptation-overlap sensitivity ($n=500$); PathSwap $\Delta$ and McNemar $p$ use paired binary top-1 any-hit outcomes on matched instances.}
\label{tab:small_scale_controls}
\setlength{\tabcolsep}{4pt}
\begin{tabular}{@{}llrrr>{\raggedright\arraybackslash}p{0.30\linewidth}@{}}
\toprule
{\colhead{Model}} & {\colhead{Split}} & {\colhead{CA Normal}} & {\colhead{no-aug Normal}} & {\colhead{CA PathSwap}} & {\colhead{no-aug PathSwap (binary-hit $\Delta$, $p$)}} \\
\midrule
1.5B & Lite     & 42.00 & 44.67 & 36.67 & 21.33 ($+15.33$, $p{=}2.3{\times}10^{-7}$) \\
1.5B & Verified & 41.42 & 44.32 & 36.80 & 12.64 ($+26.20$, $p{=}5.3{\times}10^{-32}$) \\
3B   & Lite     & 45.67 & 48.33 & 38.00 & 16.00 ($+22.00$, $p{=}2.4{\times}10^{-13}$) \\
3B   & Verified & 43.60 & 46.23 & 39.02 & 12.84 ($+28.40$, $p{=}4.6{\times}10^{-35}$) \\
\bottomrule
\end{tabular}
\end{table}

On the original paths, the fractional Recall@1 cost of Codeaware relative to the matched no-augmentation model is small ($-$2.6 to $-$2.9pp, n.s.\ to marginal). The single-seed results describe a trade-off between original-path accuracy and PathSwap sensitivity.

\subsection{Cross-Family Sensitivity Check}
\vspace{-4pt}
To test whether the PathSwap-sensitivity pattern is confined to Qwen, we trained Llama-3.1-8B-Instruct and Mistral-7B Codeaware LoRAs with the same configuration (path augmentation 0.5, 30 code lines, LoRA rank 32, two epochs) on a fixed 5{,}916-example combined training set selected before evaluation from three source slices (including 1{,}870 SWE-bench rows), using seed 42.

\begin{table}[!htbp]
\centering\small\setlength{\tabcolsep}{4pt}
\caption{Single-seed Qwen/Llama/Mistral Codeaware sensitivity (single-snapshot, 4-bit): fractional Recall@1 on Lite ($n=300$) and Verified ($n=500$, adaptation-overlap sensitivity); Lite drops are relative.}
\label{tab:cross_family_sensitivity}
\begin{tabularx}{\linewidth}{@{}Xrrrrr@{}}
\toprule
\addlinespace[2pt]
{\colhead{Model}} & {\colhead{\makecell{Lite\\Normal}}} & {\colhead{\makecell{Lite\\PathSwap}}} & {\colhead{\makecell{Lite\\drop}}} & {\colhead{\makecell{Verified\\Normal}}} & {\colhead{\makecell{Verified\\PathSwap}}} \\
\midrule
Qwen2.5-7B Codeaware & 48.33 & 41.00 & $-$15.2\% & 44.98 & 36.02 \\
Llama-3.1-8B Codeaware & 50.00 & 44.67 & $-$10.7\% & 46.15 & 43.35 \\
Mistral-7B Codeaware & 46.67 & 40.00 & $-$14.3\% & 44.92 & 39.70 \\
\bottomrule
\end{tabularx}
\end{table}

Llama-3.1-8B and Mistral-7B both show the same descriptive PathSwap-sensitivity direction (Lite relative PathSwap drops 10.7\% and 14.3\%; Verified 6.1\% and 11.6\%); both point estimates are below the Qwen-7B row. These single-seed cells show that the observed pattern is not confined to the tested Qwen checkpoint family.

\subsection{Multi-file (6+ files) Scope}
\vspace{-4pt}
A Codeaware--no-aug PathSwap contrast is detected for the Verified single-file subset ($n=430$, drop
ratio 0.50, $p=0.0001$), while the tested high-file-count subsets are floor-limited and show no such
detected contrast. On SWE-bench Verified 6+-file fixes (n=2), both Codeaware and no-aug produce 0\% Recall@1; on SWE-bench full 6+-file fixes (n=51), again both 0\% Recall@1; on SWE-bench-Live 6+ (n=116, fallback subset due to repo availability), Recall@1 is 4--5\% Normal and 2--3\% PathSwap with no detectable contrast (McNemar p=1.0). We report an observed top-1 outcome floor for these high-file-count subsets and candidate pools; we recommend evaluating cross-module repair at function-level or with R@K (K$\geq$5) in future work.

\begin{table}[!htbp]
\centering
\caption{Scope of the high-file-count checks. Every 6+-file subset sits at or near the observed top-1 outcome floor, so these rows delimit the multi-file interpretation without estimating a contrast.}
\label{tab:high_file_scope}
\small
\begin{tabular}{lrrl}
\toprule
{\colhead{\textbf{Subset}}} & {\colhead{\textbf{n}}} & {\colhead{\textbf{Observed result}}} & {\colhead{\textbf{Reading}}} \\
\midrule
Verified single-file & 430 & drop ratio 0.50, p=0.0001 & detected contrast \\
Verified 6+ files & 2 & 0\% Codeaware / 0\% no-aug & too small, no top-1 hits \\
SWE-bench full 6+ files & 51 & 0\% Codeaware / 0\% no-aug & top-1 floor in this pool \\
SWE-bench-Live 6+ files & 116 & 4-5\% Normal; 2-3\% PathSwap & no detected contrast, p=1.0 \\
\bottomrule
\end{tabular}
\end{table}

\textbf{Full SWE-bench-Live n=231:} We also evaluate the full set of 231 instances. Top-1 file localization remains near the floor on the full 231-instance set: Codeaware Normal 4.96\%, PathSwap 3.62\%; no-aug Normal 5.32\%, PathSwap 3.55\% (no detectable contrast, paired McNemar p$\approx$1.0).

\subsection{In-domain Java PathSwap Sensitivity Check}
\vspace{-4pt}
To complement the negative zero-shot Java result, we trained an in-domain Java Codeaware LoRA using BeetleBox Java data (n=1{,}174 paired examples after corpus filtering, drawn from Elasticsearch and Apache DolphinScheduler), single seed. Both Codeaware (aug=0.5) and a no-aug control (aug=0) were trained with the same recipe as our 7B Python Codeaware. On the held-out 100-instance Java PathSwap test set, both in-domain trained models have a 0\% PathSwap score, matching the
zero-shot floor. At $n_{\mathrm{train}}{=}1{,}174$, this comparison detects no
Codeaware--no-aug PathSwap contrast.

\textbf{Bench4BL at n=7{,}836:} we use an in-domain Java corpus (10 Java projects, 10{,}017 raw reports; 7{,}836 train / 872 test after filtering). Its 6.7$\times$ larger training set is confounded with corpus, project, and test-set changes.

\begin{table}[!htbp]
\centering
\caption{BeetleBox Java sensitivity: fractional Recall@1 on $n=100$.}
\label{tab:java_beetlebox_sensitivity}
\begin{tabular}{lrr}
\toprule
\addlinespace[2pt]
{\colhead{\textbf{Cell}}} & {\colhead{\textbf{Normal Recall@1}}} & {\colhead{\textbf{PathSwap Recall@1}}} \\
\midrule
Java Codeaware (aug=0.5, in-domain trained) & 6.37\% & 0.00\% \\
Java no-aug (aug=0.0, in-domain trained) & 9.00\% & 0.00\% \\
Java zero-shot Python-Codeaware & 9.64\% & 0.00\% \\
Java zero-shot Python-no-aug & 11.73\% & 0.00\% \\
\bottomrule
\end{tabular}
\medskip

\caption{Bench4BL Java sensitivity: fractional Recall@1 on $n=872$.}
\label{tab:java_bench4bl_sensitivity}
\begin{tabular}{lrr}
\toprule
\addlinespace[2pt]
{\colhead{\textbf{Cell (Bench4BL, n=872)}}} & {\colhead{\textbf{Normal Recall@1}}} & {\colhead{\textbf{PathSwap Recall@1}}} \\
\midrule
Codeaware (aug=0.5) & 20.70\% & 3.94\% \\
no-aug (aug=0.0) & 19.84\% & 3.00\% \\
\bottomrule
\end{tabular}
\end{table}

On Bench4BL, the fractional Recall@1 difference under PathSwap is $+0.94$ percentage points. The paired binary any-hit difference is $+1.61$ points ($28$ versus $14$ discordant outcomes among $872$ instances; exact McNemar $p=0.044$). The binary test applies to the binary outcome, not to the fractional contrast. BeetleBox
($n_{\text{train}}{=}1{,}174$) and Bench4BL ($n_{\text{train}}{=}7{,}836$) differ in corpus,
projects, and splits as well as in training-set size, so the two benchmarks differ across multiple factors.

\section{Training and Inference Interventions}
\label{app:exploratory}
These studies vary training, instructions, or inference-time scoring to test changes in localization. Each uses its own comparison and stopping rule; we report passing screens, null contrasts, and failed screens separately. None of these studies establishes a link from shared-scorer certification to executed repair.

\subsection{Test-Time Rescoring and Rank Fusion}
\label{app:testtime_rescoring}
A separate reranking check asks which file and function to return after both sets of scores have been computed. On 502 repository-held-out test issues, a development-tuned combination of file and function scores for 14B Codeaware yields 148 joint hits versus 126 under a file-first rule ($+4.38$ percentage points; 95\% paired interval $[2.19,6.77]$). Selecting the highest-scoring function across all ten files, then its file, already yields 149 joint hits on those issues. Rank-based reranking gives 148 versus this 149-hit reference ($-0.20$, $[-1.99,1.59]$). All four rank-based intervals include zero against their stronger function-first references. The denominator here is the 502 held-out issues; the 2{,}356 file wins in Table~\ref{tab:file_func_confusion} answer a different question. Separately, in a Reciprocal Rank Fusion comparison combining Codeaware with SweRankEmbed-Small on SWE-bench Lite, validation selects $\alpha=1.0$ (pure Codeaware, 52.33\%, vs.\ 34.00\% for the dense retriever alone on that candidate pool), showing no fusion gain over pure Codeaware.

\subsection{Training-time PathMask}
\label{app:pathmask}

The path-augmentation in \emph{Codeaware} corrupts paths by within-group \emph{shuffling} (each candidate's path is replaced by another candidate's path, preserving the path-token distribution). A complementary training-time intervention replaces readable path content with stable, structure-preserving opaque PathSwap identifiers. We call this \textbf{PathMask}; unlike PathSwap, which is a held-out test-time diagnostic, PathMask is applied during training only. The training-time PathMask of Table~\ref{tab:pathmask_invariance} is distinct from the test-time MaskPath of Table~\ref{tab:maskpath_v8}.

\paragraph{Setup.} We train two 7B SWE-only LoRAs (matched in data and hyperparameters to the Codeaware row of Table~\ref{tab:codeaware_main}, single seed=42) varying only the masking fraction: \textbf{PathMask-50} masks 50\% of training groups; \textbf{PathMask-100} masks 100\%. Both use code-in-prompt, no path-shuffle augmentation, and the same number of training steps as Codeaware.

\begin{table}[!htbp]
\centering
\caption{PathMask vs.\ no-augmentation and 50-line Codeaware (7B SWE-only, single-snapshot, 4-bit). Same BM25 top-100 pool and reporting convention as Table~\ref{tab:codeaware_invariance}; Verified ($n=500$) is an adaptation-overlap sensitivity (384/500 overlap). Against the matched 30-line Codeaware run of Table~\ref{tab:body_deletion}, PathMask-100 has no detected difference on Verified (paired McNemar $p{=}0.30$ Normal, $p{=}0.72$ PathSwap; 80\% MDE 4.71pp for the fractional PathSwap comparison) and is significantly higher than the SWE-only unaugmented recipe on Verified PathSwap ($p{\approx}1.0{\times}10^{-5}$).}
\label{tab:pathmask_invariance}
\small
\setlength{\tabcolsep}{6pt}
\renewcommand{\arraystretch}{1.15}
\begin{tabular}{@{}l l
                S[table-format=2.2]
                S[table-format=2.2]
                S[table-format=+2.2]
                S[table-format=+2.1]@{}}
\toprule
{\colhead{\textbf{Method}}} & {\colhead{\textbf{aug / mask}}}   & {\colhead{{\textbf{Normal}}}} & {\colhead{{\textbf{PathSwap}}}}   & {\colhead{{\textbf{$\Delta_{\text{abs}}$}}}} & {\colhead{{\textbf{$\Delta_{\text{rel}}$ (\%)}}}} \\
\midrule
\multicolumn{6}{@{}l}{\itshape\footnotesize SWE-bench Lite ($n{=}300$)} \\
\addlinespace[1pt]
SWE-only unaugmented recipe & 0.0    & 53.67 & 33.33 & -20.34 & -37.9 \\
Codeaware (50-line)   & 0.5 shuffle  & 48.33 & 41.00 &  -7.33 & -15.2 \\
PathMask-50           & 0.5 mask     & 53.33 & 34.00 & -19.33 & -36.2 \\
\textbf{PathMask-100} & 1.0 mask     & 45.00 & 39.33 &  -5.67 & \bfseries -12.6 \\
\midrule
\multicolumn{6}{@{}l}{\itshape\footnotesize SWE-bench Verified ($n{=}500$)} \\
\addlinespace[1pt]
SWE-only unaugmented recipe & 0.0    & 46.12 & 26.43 & -19.69 & -42.7 \\
Codeaware (50-line)   & 0.5 shuffle  & 44.98 & 36.02 &  -8.96 & -19.9 \\
PathMask-50           & 0.5 mask     & 44.60 & 30.32 & -14.28 & -32.0 \\
\textbf{PathMask-100} & 1.0 mask     & 41.82 & 32.74 &  -9.08 & -21.7 \\
\bottomrule
\end{tabular}
\end{table}

\paragraph{Results.} Table~\ref{tab:pathmask_invariance} reports Recall@1 on the same BM25 top-100 pool used for Codeaware. PathMask-100 narrows the no-aug PathSwap relative drop (Lite $-$12.6\% drop vs.\ no-aug $-$37.9\% and Codeaware $-$15.2\%; Verified $-$21.7\% drop vs.\ no-aug $-$42.7\% and Codeaware $-$19.9\%), at a $-$3 to $-$4pp Normal-path cost relative to Codeaware. Paired McNemar (recall$>$0 binary): \textbf{Against the matched 30-line Codeaware run of Table~\ref{tab:body_deletion}, PathMask-100 has no detected difference on Verified} on either Normal ($n_{10}{=}19, n_{01}{=}27$, $p{=}0.30$) or PathSwap ($n_{10}{=}36, n_{01}{=}32$, $p{=}0.72$); the fractional-Recall@1 PathSwap delta is $+0.51$pp against the matched 30-line Codeaware run of Table~\ref{tab:body_deletion}, but the 80\%-power minimum detectable effect (MDE) is 4.71pp, so no difference is detected. PathMask-100 versus the SWE-only unaugmented recipe on Verified PathSwap: paired-binary $+$8.80pp ($p{\approx}1.0{\times}10^{-5}$); aggregate Recall@1 $+$6.31pp. After Holm-Bonferroni correction within the PathMask family, 6/14 tests survive at 0.05.

\paragraph{Interpretation.} The matched Verified comparison uses $n{=}500$. The $-$3 to $-$4pp Normal-path cost relative to Codeaware is consistent with masking being a more aggressive intervention than shuffling (the model sees no real paths, versus 50\% real / 50\% shuffled). The PathMask-50 underperformance vs.\ Codeaware on PathSwap ($-$6.7pp, Verified, paired binary any-hit, $p{=}0.017$) is compatible with residual reliance on the unmasked half.

\subsection{In-Domain Path-Trust Replication}
\label{app:prompttrust_indomain}

Before testing reliability instructions in another domain, we test them directly on our code-localization path channel. The unaugmented Qwen2.5-14B code-in-prompt checkpoint reranked the fixed single-snapshot top-100 pools for the 500 SWE-bench Verified sensitivity issues (384/500 adaptation overlap; no Lite replication was run) under two prompts keeping the base relevance question verbatim while adding a 13-token instruction stating that displayed paths are either reliable (T) or unreliable (D). Issue text, candidate pools, 50-line code previews, weights, and scoring stayed fixed (Table~\ref{tab:prompttrust_indomain}).

\begin{table}[!htbp]
\centering
\caption{In-domain path-trust check on the 500-issue Verified adaptation-overlap sensitivity panel (fractional Recall@1 in \%; $\Delta$ is Distrust minus Trust with paired bootstrap 95\% CIs).}
\label{tab:prompttrust_indomain}
\small
\begin{tabular}{@{}lrrr@{}}
\toprule
{\colhead{\textbf{Surface}}} & {\colhead{\textbf{Trust (T)}}} & {\colhead{\textbf{Distrust (D)}}} & {\colhead{\textbf{$\Delta$ (pp)}}} \\
\midrule
Normal & 49.22 & 49.22 & $0.00\ [-0.80,0.80]$ \\
PathSwap & 35.32 & 35.37 & $+0.05\ [-1.10,1.20]$ \\
PathAlias (seed 42) & 32.38 & 31.68 & $-0.70\ [-2.20,0.70]$ \\
\bottomrule
\end{tabular}
\end{table}

Instructional distrust leaves the Normal--PathSwap gap virtually untouched ($(N_T-P_T)-(N_D-P_D)=+0.05$pp, $[-1.30,1.45]$) with zero Normal-path drift ($D-T=0.00$pp), and changes the Normal--PathAlias gap by $-0.70$pp ($[-2.40,0.90]$). Thus, a brief reliability prompt does not meaningfully alter path dependence in this fine-tuned reranker.

\subsection{Cross-Domain Title-Trust Check}
\label{app:channeltrust}

PathAlias establishes that path dependence does not require opaque identifier tokens. This pilot asks the
narrower construct question of whether the damage from readable false metadata depends on how much
the scorer is instructed to trust the metadata channel. We used a fixed BEIR-SciFact top-20 pool for 75 queries
and a zero-shot Qwen3-8B pointwise reranker. The displayed title took one of three forms: unchanged (Normal);
replaced by a form-matched readable title from an unrelated corpus (five mappings); or
replaced by punctuation-only dot/dash noise preserving whitespace and word lengths. The passage
tokens, query, document identities, pool, and user message were fixed. Prompt T stated that the title
was reliable metadata and should be used with the passage; prompt D stated that titles might be
unreliable and relevance should be judged from passage content. A disjoint title-use probe
passed before scoring.

After 64-token truncation, 1{,}499 of 1{,}500 query--document pairs per template preserved exact form matching, with one retained exception shared identically across both prompts. All 14 cells completed (1{,}500 query--document pairs per cell; 21{,}000 scores). Defining $A_p$ as the readable-title drop minus the obvious-noise drop under prompt $p$, the prompt interaction $A_T-A_D$ is positive on both metrics ($+4.97$pp $[1.77,8.80]$ on nDCG@10 and $+6.31$pp $[0.36,13.24]$ on Recall@1; query-level paired bootstrap, Table~\ref{tab:channeltrust}), showing that readable unrelated titles disrupt scoring far more when the prompt tells the model to trust titles than when it warns against them.

\begin{table}[!htbp]
\centering\small
\caption{Cross-domain title-trust results (95\% paired-bootstrap CIs; nDCG on a percent scale). $A_p$ is the readable-title drop minus the punctuation-noise drop under prompt $p$; the interaction is $A_T-A_D$.}
\label{tab:channeltrust}
\begin{tabular}{@{}lrr@{}}
\toprule
{\colhead{Contrast}} & {\colhead{nDCG@10}} & {\colhead{Recall@1}} \\
\midrule
T readable drop & $+6.194\ [2.461,10.240]$ & $+9.178\ [2.667,16.089]$ \\
$A_T$ & $+6.966\ [3.507,10.659]$ & $+10.178\ [4.400,16.222]$ \\
$A_D$ & $+1.995\ [-0.011,4.186]$ & $+3.867\ [-0.533,8.533]$ \\
$A_T-A_D$ & $+4.971\ [1.765,8.799]$ & $+6.311\ [0.356,13.244]$ \\
\bottomrule
\end{tabular}
\end{table}

Thus, in this SciFact pilot, explicit reliability instructions strongly moderate susceptibility to misleading titles, whereas the same instruction format had no detectable effect on the fine-tuned 14B code localizer above (Appendix~\ref{app:prompttrust_indomain}).

\subsection{Paired-view continuation pilots}
\label{app:paired_consistency}

Finally, we test whether short continued training with consistency regularization reduces path dependence without changing the candidate pool or adding inference calls. Starting from the same Qwen2.5-7B checkpoint on 1{,}682 training groups (106 updates, identical optimizer schedule), we compare a standard cross-entropy continuation ($\mathrm{CE}(\text{real})$) against two paired-view continuations minimizing $\mathrm{CE}(\text{real})+0.5\,\mathrm{CE}(\text{perturbed})+0.5\,\mathrm{KL}(\operatorname{stopgrad}(p_{\text{real}})\Vert p_{\text{perturbed}})$, using either deterministic opaque paths (\emph{opaque-view}) or readable unrelated aliases (\emph{readable-alias}) as the second view while holding source code and labels fixed. All arms are scored in bf16 on fixed BM25 top-100 pools with 30-line previews (Table~\ref{tab:paired_consistency}).

\begin{table}[!htbp]
\centering
\caption{Paired-view consistency pilots (single-snapshot, bf16, fractional Recall@1 in \%). Each $\Delta$ is the paired continuation minus the control continuation in percentage points. PathAlias uses the preselected seed-42 alias map on Lite ($n=300$) and the Verified adaptation-overlap sensitivity ($n=500$).}
\label{tab:paired_consistency}
\small
\setlength{\tabcolsep}{4pt}
\begin{tabular}{@{}llrrrrr@{}}
\toprule
{\colhead{\textbf{Split}}} & {\colhead{\textbf{View}}} & {\colhead{\makecell{\textbf{Control}\\\textbf{continuation}}}} & {\colhead{\makecell{\textbf{Opaque-view}\\\textbf{continuation}}}} & {\colhead{\textbf{$\Delta_{H}$}}} & {\colhead{\makecell{\textbf{Readable-alias}\\\textbf{continuation}}}} & {\colhead{\textbf{$\Delta_{R}$}}} \\
\midrule
Lite & Normal & 53.67 & 52.00 & $-1.67$ & 53.00 & $-0.67$ \\
     & PathSwap & 40.33 & 43.33 & $+3.00$ & 40.33 & $0.00$ \\
     & PathAlias (seed 42) & 38.00 & 41.00 & $+3.00$ & 43.00 & $+5.00$ \\
\midrule
Verified & Normal & 48.80 & 47.53 & $-1.27$ & 48.28 & $-0.52$ \\
         & PathSwap & 37.60 & 39.65 & $+2.05$ & 35.00 & $-2.60$ \\
         & PathAlias (seed 42) & 38.90 & 42.03 & $+3.13$ & 40.68 & $+1.78$ \\
\bottomrule
\end{tabular}
\end{table}

Relative to the control continuation, the opaque-view continuation improves both opaque PathSwap ($+3.00$pp on Lite, $+2.05$pp on Verified) and readable PathAlias ($+3.00$pp $[0.00,6.33]$ on Lite; $+3.13$pp $[1.17,5.13]$ on Verified) while staying within 1.7 points on original paths. The readable-alias continuation raises PathAlias further on Lite ($+5.00$pp $[1.67,8.67]$; $+1.78$pp $[-0.70,4.23]$ on Verified) with smaller Normal costs ($-0.67$ and $-0.52$pp), yet regresses on Verified PathSwap ($-2.60$pp $[-4.90,-0.40]$) and therefore fails its screening rule against opaque paths. Both pilots use a single training seed and one alias map and remain exploratory.

\section{Native SweRank Reproduction and Boundary}
\label{app:swerank_repro}
\label{sec:codeaware:swerank_repro}

This appendix records the unperturbed reproduction of SweRank's native function-body retrieval-and-reranking pipeline and states its boundary relative to our path-perturbation checks. The separate path-document encoder probe (where SweRankEmbed encodes file-path strings on Lite) appears in Appendix~\ref{sec:generalization} (Table~\ref{tab:swerank_perturb}).

\paragraph{Native retrieval-and-reranking reproduction.}
In SweRank's native procedure~\citep{reddy2025swerank}, SweRankEmbed-Large (7B) retrieves a top-100 function pool from the full-repository function index, and SweRankLLM-Small (7B) listwise-reranks that pool. On the 274-issue SWE-bench Lite function benchmark, SweRankEmbed-Large retrieval reaches 199/274 (72.63\%) strict file Acc@1, and SweRankLLM-Small reranking reaches \textbf{216/274 (78.83\%)} strict file Acc@1, reproducing the published 72.63\% retrieval and 78.10\% reranking results under the same benchmark and metric.

\paragraph{Comparison boundary.}
We did not run PathSwap or PathAlias on this reproduced 78.83\% native SweRank pipeline, so this paper makes no claim about PathSwap sensitivity or robustness for SweRank's native function-body workflow. (An earlier misconfigured run at 20.8\% Acc@1 does not represent SweRank's native operating point and is excluded.) On our separate fixed BM25 top-100 file-pool benchmark (single-snapshot Lite, $n=300$, fractional Recall@1), 14B Codeaware reaches 60.33\% versus 47.93\% for SweRankLLM-Small ($+12.4$pp; Table~\ref{tab:codeaware_main}).

\section{CodeGRIP: Implementation and Reference Results}
\label{app:impl}
\begin{figure}[!ht]
\centering
\includegraphics[width=\linewidth]{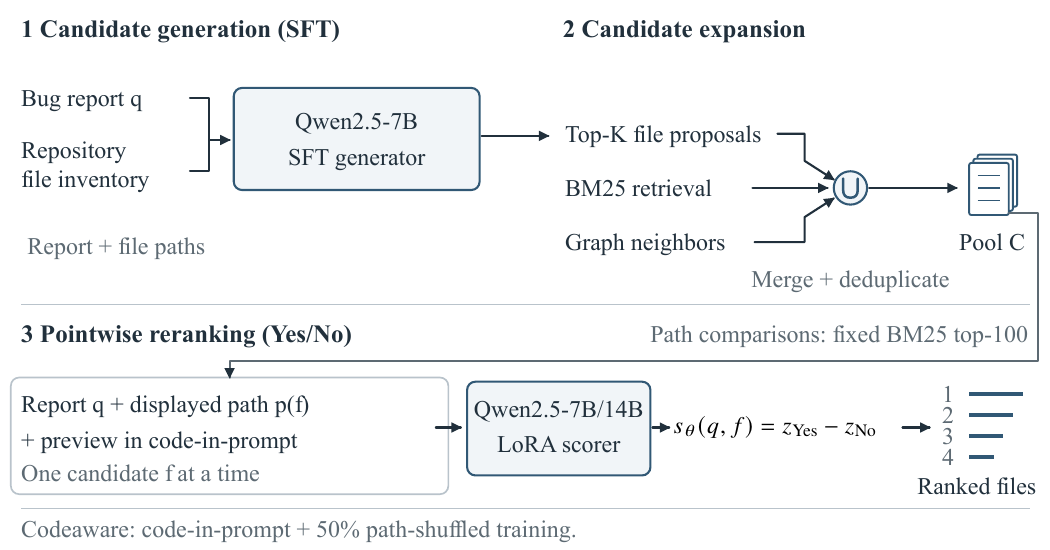}
\caption{The CodeGRIP localization workflow. An SFT generator proposes file paths; retrieval and graph neighbors supply additional candidates. After merging and deduplication, a LoRA scorer ranks files by their Yes/No logit difference, optionally using a code preview. Codeaware is the code-in-prompt recipe with 50\% path-shuffled training examples. The primary path comparisons instead fix a BM25 top-100 pool and vary only displayed paths.}
\label{fig:method}
\end{figure}
\label{app:group_training_objective}
\paragraph{Group training objective.}
Pointwise inference is distinct from learning: the reranker is trained on groups of positive and negative files with a listwise ranking loss. The primary and matched 14B recipes use eight BM25-hard negatives per positive on the combined 5,916-group training set. The 7B factorial uses eight negatives with configured BM25/graph/random proportions of 50/25/25 and fallback sampling when a pool is short. The early 7B path-only adaptation uses 32 negatives, configured as 75\% BM25-hard and 25\% random, and starts from a GREPO-trained checkpoint. The historical GREPO configuration uses 16 negatives with a 50/25/25 BM25/graph/random mixture. Table~\ref{tab:hyperparams} separates these recipes.

For a training group with positive file $f_0$ and negatives $f_1,\ldots,f_m$, the implementation uses the group-softmax loss
\begin{equation}
\ell(q;f_0,\ldots,f_m)=-\log\frac{\exp s_\theta(q,f_0)}{\sum_{j=0}^{m}\exp s_\theta(q,f_j)}.
\label{eq:group_softmax}
\end{equation}
Each report--candidate pair is processed separately; group normalization couples scores during training, whereas inference again scores files individually. The loss uses no temperature. Code-present variants retain the original candidate's source preview when the training path is shuffled; augmentation alters the displayed cue, while the loss and the original target remain unchanged. These recipe-specific negative samplers and initialization choices remain distinct.

\begin{table}[!htbp]
\centering\small
\caption{CodeGRIP reference operating points on GREPO ($n=1{,}704$). Panel A reports Recall@1 and average candidate pool size across stages. Panel B reports the candidate budget needed to reach approximately $90.7\%$ oracle recall, not runtime.}
\label{tab:testbed_summary}
\begin{tabularx}{\linewidth}{@{}Xrr@{}}
\toprule
\multicolumn{3}{l}{\colhead{\textbf{A. GREPO operating points} ($n=1{,}704$)}}\\
{\colhead{Configuration}} & {\colhead{Recall@1 (\%)}} & {\colhead{Pool size}}\\\midrule
BM25, no reranker & 10.00 & 500\\
Reranker on BM25 & 28.45 & 500\\
Candidate generation only & 21.33 & $\sim32$\\
Generation + graph/BM25 merge & 22.51 & $\sim163$\\
Complete generation--expansion--reranking & 29.38 & $\sim163$\\\midrule
\multicolumn{3}{l}{\textbf{B. Candidate budget at $\sim90.7\%$ oracle recall}}\\
Graph-expanded & \multicolumn{2}{r}{$\sim208$ candidates}\\
BM25-only & \multicolumn{2}{r}{$\sim393$ candidates}\\\bottomrule
\end{tabularx}
\end{table}

\paragraph{Input representation and preview lengths.}
The SFT candidate generator receives the flat repository file-path inventory rather than source lines. Preview lengths for rerankers are recipe-specific: primary 14B and matched code=YES training use up to 50 lines, whereas 7B factorial code-present cells train on 30 lines; both families score with a 50-line / 1{,}500-character cap within a 768-token prompt. Matched code=NO and early 7B path-only adaptation receive no code (768- and 512-token context limits, respectively). Other controls retain the lengths stated in their own sections.

\label{app:additional}
Figure~\ref{fig:method} introduces the CodeGRIP workflow. This appendix specifies candidate construction, scoring, training hyperparameters, data separation, and seed/checkpoint sensitivity. Each comparison states its configuration, candidate pool, and evaluation setting.

\subsection{Candidate construction and scoring}
\subsubsection{Prompt Templates}
\label{app:prompts}

\paragraph{Cross-encoder file reranking.}
The reranker uses a binary relevance prompt, scoring each candidate independently:

\begin{quote}
\small
\texttt{Given the bug report, is this file likely to need modification?}\\[2pt]
\texttt{Bug Report: \{issue\_text\}}\\[2pt]
\texttt{File: \{candidate\_path\}}\\[2pt]
\texttt{Answer:}
\end{quote}

The score is $s = z_{\text{Yes}} - z_{\text{No}}$, where $z_{\text{Yes}}$ and $z_{\text{No}}$ are the logits for the ``Yes'' and ``No'' tokens at the generation position after ``Answer:''.
This yields a pointwise scorer without changing the LLM architecture.

\paragraph{Candidate generation (SFT).}
The SFT model receives the bug report and a flat repository file-path inventory (no co-change or import graph in the prompt), and generates file paths:

\begin{quote}
\small
\texttt{You are analyzing a bug report for the Python repository ``\{repo\}''.}\\[2pt]
\texttt{The repository contains the following Python files:}\\
\texttt{\{file\_list\}}\\[2pt]
\texttt{Bug Report: \{issue\_text\}}\\[2pt]
\texttt{From the file list above, identify which Python files need to be modified to fix this bug. Output file paths only, one per line, most relevant first.}
\end{quote}

\subsubsection{BM25 Retrieval Details}
\label{app:bm25}

\paragraph{GREPO evaluation.}
BM25 retrieval indexes all Python files per repository, using file paths and content. The top-500 candidates form the initial retrieval pool. Candidates are merged with graph-expanded files by score-weighted interleaving, yielding $\sim$163 candidates on average.

\paragraph{SWE-bench Lite retrieval.}
We apply several benchmark-specific choices to standard BM25 retrieval:
\begin{enumerate}
    \item \textbf{Test exclusion}: Files identified as tests are removed from the candidate pool; H@1 increases by $\sim$9pp absolute in this benchmark-specific recipe.
    \item \textbf{Multi-granularity indexing}: We create parallel function-level and chunk-level indexes (500-token chunks) and fuse their rankings using reciprocal rank fusion (RRF) with $k = 60$.
    \item \textbf{Stemming}: We apply the Snowball stemmer to both queries and documents.
    \item \textbf{Post-processing}: We map the results back to files and deduplicate them by keeping the first occurrence.
\end{enumerate}

An oracle drawing on all retrieval pools (path, content, function, chunk, and their combinations) reaches H@1 = 54\%; that figure is a multi-pool envelope.

\subsection{Training configurations and data separation}
\subsubsection{Training Hyperparameters}
\label{app:hyperparams}

Table~\ref{tab:hyperparams} lists the candidate generation and file reranking hyperparameters.

\begin{table}[!htbp]
\centering
\caption{Training settings separated by experiment family. The matched 14B code=YES arm reuses the primary unaugmented checkpoint; code=NO is newly trained. Code limits apply only when code is provided. Sampling mixtures are configured proportions, with fallback for unavailable pools.}
\label{tab:hyperparams}
\footnotesize
\setlength{\tabcolsep}{4pt}
\begin{tabularx}{\linewidth}{@{}l>{\raggedright\arraybackslash}X>{\raggedright\arraybackslash}X@{}}
\toprule
{\colhead{Setting}} & {\colhead{Candidate generation SFT}} & {\colhead{Rerankers below}} \\
\midrule
Base model & Qwen2.5-7B & Qwen2.5-Instruct, size by family \\
LoRA rank / $\alpha$ / dropout & 32 / 64 / 0.05 & 32 / 64 / 0.05 \\
LoRA targets & $\mathbf{W}_{q,k,v,o}$, $\mathbf{W}_{\mathrm{gate,up,down}}$ & $\mathbf{W}_{q,k,v,o}$, $\mathbf{W}_{\mathrm{gate,up,down}}$ \\
Learning rate & $2\times10^{-5}$ & Family-specific below \\
Scheduler & Cosine, 5\% warmup & Cosine, 5\% warmup \\
Epochs & 3 & 2 \\
Train batch / accumulation & Not applicable & 1 / 16 \\
Context length & 2048 & Family-specific below \\
Optimizer & AdamW ($\beta_1=0.9$, $\beta_2=0.999$) & AdamW ($\beta_1=0.9$, $\beta_2=0.999$) \\
Gradient clipping & 1.0 & 1.0 \\
\bottomrule
\end{tabularx}
\medskip
\begin{tabularx}{\linewidth}{@{}>{\raggedright\arraybackslash}p{.18\linewidth}>{\raggedright\arraybackslash}p{.25\linewidth}>{\raggedright\arraybackslash}p{.18\linewidth}>{\raggedright\arraybackslash}X@{}}
\toprule
{\colhead{Reranker family}} & {\colhead{Training data / initialization}} & {\colhead{Negatives}} & {\colhead{Preview and context; learning rate}} \\
\midrule
Primary and matched 14B & 5,916 groups: 1,870 SWE + 4,046 GREPO; fresh base-model LoRA & 8 BM25-hard & Code=YES: 50 lines at training and scoring; code=NO: none. 768 tokens; $5\times10^{-5}$ \\
7B factorial & 1,870 SWE groups; fresh base-model LoRA; seeds 42/43/44 & 8; 50\% BM25, 25\% graph, 25\% random & Code=YES: 30 lines at training, 50 at scoring; code=NO: none. 768 tokens; $5\times10^{-5}$ \\
Early 7B adaptation & 1,874 SWE training records; GREPO-trained initialization & 32; 75\% BM25, 25\% random & No code; 512 tokens; $1\times10^{-5}$ \\
\bottomrule
\end{tabularx}
\par\smallskip\raggedright
The primary and matched 14B code-visible previews and the 7B factorial scoring previews are capped at 1500 characters. Primary 14B training uses seed 42 and path-augmentation fractions 0 or 0.5; the matched pair uses fraction 0, and the factorial crosses 0/0.5 with code visibility. The historical GREPO reference recipe uses 16 negatives with a 50/25/25 BM25/graph/random mixture. Repository--issue overlap with Lite is 0/300 for each listed input set. Verified overlap is 384/500 for the primary/matched 14B and 7B factorial inputs, and 385/500 for the early 7B adaptation input.
\end{table}

\subsubsection{Data Separation and Graph Provenance}
\label{app:leakage}

Co-change edges are computed exclusively from the GREPO training split; no test-split data is used. Import edges are derived from static AST analysis of repository snapshots, independent of any PR timestamp. The BM25 test pools use the repository file tree at the pre-fix repository snapshot, which keeps post-fix file-tree information out of those pools. The train/test split in GREPO is by issue ID, not by time; we verify that no issue IDs overlap between the train split (7,883) and the test (2,175) splits.
The primary 14B training uses the cleaned 4,046-group GREPO subset combined with 1,870 SWE-bench groups; 7,883 describes the original GREPO reference split. All three index builders explicitly restrict input to the training split. In 62.2\% of test examples, a ground-truth file appears as a node in the training co-change graph. The model can therefore use relations from training issues involving that file even though the train and test issue IDs are disjoint.

\subsubsection{SWE-bench Domain Adaptation}
\label{app:swebench_details}
\label{app:adaptation}

\paragraph{Training data.}
The cleaned SWE-bench component used by the primary 14B and 7B factorial recipes contains 1{,}870 groups from 12 repositories, excluding Lite issues and filtering for BM25-candidate availability and Python changes. The separate early 7B adaptation reference uses 1{,}874 distinct repository--issue pairs (0/300 Lite and 385/500 Verified overlaps).

\paragraph{Training setup.}
That early Qwen2.5-7B path-only reference starts from a GREPO-trained reranker and fine-tunes for 2 epochs at learning rate $1\times10^{-5}$ with 32 negatives (75\% BM25-hard, 25\% random) and a 512-token context, distinct from the fresh primary 14B LoRAs trained on 5{,}916 combined groups with eight BM25-hard negatives and a 768-token context.

\paragraph{Training-data scale.}
The early 7B adaptation uses 1{,}874 target-domain examples ($\sim$9.8K including 7{,}883 GREPO pre-training examples, compared with 67K in SweRankEmbed-Large); because the systems differ in pipeline and pool, these counts compare training-set size rather than sample efficiency. Without SWE-bench adaptation, the GREPO-trained model reaches 51.5\% H@1 on 274 matched Lite issues via hybrid fusion.

\subsection{Seed and checkpoint sensitivity}
\label{app:caveats}
\label{app:external_validity}

Because CodeGRIP serves as our controlled testbed, ablations sharing a base model and candidate pool (Appendices~\ref{app:cross_cut} and~\ref{app:decomposed_reranker}) provide the cleanest comparisons. Across five random seeds on GREPO ($n=1{,}704$, Table~\ref{tab:seeds}), graph-hard negatives average $+0.74$pp over BM25-only negatives ($25.01 \pm 1.17$ vs.\ $24.27 \pm 0.39$, paired $t$-test $p = 0.33$), while the three-seed path-only reference averages $28.92 \pm 0.56\%$.

\begin{table}[htbp]
\centering
\small
\caption{Seed robustness on GREPO ($n=1{,}704$, mean $\pm$ SD).}
\label{tab:seeds}
\begin{tabular}{lrr}
\toprule
{\colhead{Method}} & {\colhead{Recall@1 (\%)}} & {\colhead{R@5 (\%)}} \\
\midrule
Path-only (3 seeds) & $28.92 \pm 0.56$ & not reported \\
Graph-hard (5 seeds) & $25.01 \pm 1.17$ & $45.56 \pm 2.02$ \\
BM25-only (5 seeds) & $24.27 \pm 0.39$ & $44.52 \pm 0.58$ \\
\bottomrule
\end{tabular}
\end{table}

\begin{table}[htbp]
\centering
\small
\caption{Checkpoint-schedule sensitivity across candidate pools on GREPO ($n=1{,}704$). Across the 23-checkpoint dense grid, the graph-minus-hybrid Recall@1 gap has mean $0.49$pp and maximum $1.12$pp.}
\label{tab:oracle_fallacy}\label{tab:recovery}
\begin{tabular}{lrrrr}
\toprule
{\colhead{Checkpoint schedule}} & {\colhead{Graph}} & {\colhead{Hybrid}} & {\colhead{BM25}} & {\colhead{Graph$-$hybrid gap}} \\
\midrule
Sparse (every 200 training steps) & 27.01 & 19.29 & 20.08 & $+7.72$ \\
Dense best (every 10 training steps) & 29.38 & 29.38 & 28.45 & $0.00$ \\
\bottomrule
\end{tabular}
\end{table}

\paragraph{Checkpoint and pool-mismatch boundary.}
\phantomsection\label{app:coupling}
\phantomsection\label{app:delex}
With checkpoints saved every 200 steps, a single sparse checkpoint appeared to lose 7.72pp on the higher-oracle hybrid pool (BM25 + E5-large, 92.1\% oracle recall vs.\ 90.7\% for the graph-expanded pool). Evaluating all 23 checkpoints from the denser 10-step schedule eliminates that discrepancy: dense best scores 29.38\% on both pools, with a mean graph-minus-hybrid gap of 0.49pp and a maximum of 1.12pp (Table~\ref{tab:oracle_fallacy}). Delexicalization (50\% overlap-token replacement during training) and counterfactual-negative training remain within 1pp of the dense baseline ($29.46/29.61\%$ and $28.60/29.52\%$ across graph/hybrid pools). On SWE-bench, adding GREPO-style graph expansion without in-domain graph data shows no effect (Recall@1 $18.0\%$ with or without expansion).

\FloatBarrier
\section{Related Work}
\label{app:relatedwork}

\subsection{Localization granularity and outcome criteria}
\label{sec:related}

\paragraph{Related work.}
Bug localization has progressed from IR methods such as BM25~\citep{robertson2009probabilistic} and learned text--code similarity~\citep{huo2016learning,
lam2017bug} to LLM-based localization systems such as Agentless, AutoCodeRover, LocAgent, and
SweRank~\citep{xia2024agentless,zhang2024autocoderover,chen2025locagent,reddy2025swerank}, alongside the CodeRankEmbed dense retriever~\citep{suresh2025cornstack}. Training-side work fine-tunes on software-evolution trajectories~\citep{wei2025swerl,pan2025swegym}.
Our concern is complementary: contamination~\citep{liang2025swebench} and surface path overlap raise
the possibility that Recall@1 can overstate code understanding. NLI input ablations and counterfactual QA/code perturbations~\citep{poliak2018hypothesis,jia2017adversarial,yefet2020adversarial,rabin2021understanding}
motivate PathSwap and PathAlias.

Recent work overlaps with individual checks: SACL masks textual code
features and evaluates code retrieval, reranking, and SWE-bench-Lite localization~\citep{gupta2025sacl};
a representation study compares paths, source, and generated summaries under retrieval and
reranking~\citep{caumartin2026representations}; BLAgent ablates path augmentation and evaluates an Agentless
repair integration~\citep{mamun2026blagent}; and BugCerberus evaluates hierarchical file/function/statement localization with Agentless repair~\citep{chang2025bugcerberus}. Surface path/name sensitivity
is already documented in this literature~\citep{logicloc2026}. Our contribution is the joint measurement design: a fixed-pool displayed-path intervention with
readable PathAlias and an opaque reference, active tool-using repository navigation across multiple models and pipelines, a same-win scorer-referenced file-to-function transfer
check, and separate executed-repair outcome checks.

Related studies measure different quantities. RepoMirage~\citep{repomirage2026} perturbs context in end-to-end agents, LogicLoc~\citep{logicloc2026} authors
keyword-agnostic queries, and the SWE-bench illusion analysis~\citep{liang2025swebench} studies
memorization and contamination. PathAlias instead breaks path--content correspondence using readable, form- and tokenizer-matched components from unrelated repositories. This differs from those agent, query-construction and contamination analyses and from opaque corruption. Matched opaque PathSwap is the reference; strict Alias--PathSwap ordering at rank one appears only under the single-snapshot Lite checkout and remains unresolved under per-issue base-commit checkouts.

RepoMirage's runtime-target masking is closest, operating inside a multi-hop agent rather than isolating one scoring stage. Loc2Repair~\citep{loc2repair2026} and context ablations~\citep{flcontext2026} quantify repair headroom under gold localization, complementing our examination of scorer-dependent function transfer and supplied-interface repair outcomes.

\subsection{Repository Structure, Retrieval and Reranking}
Co-change history~\citep{zimmermann2004mining,ying2004predicting,zimmermann2008predicting} and dependency graphs supply signals beyond lexical retrieval. GREPO~\citep{wang2026grepo}, CodexGraph~\citep{liu2025codexgraph}, CoSIL~\citep{jiang2025cosil}, RepoGraph~\citep{ouyang2025repograph} and GRIP~\citep{feng2025grip} use structural traversal or LoRA-internalized structure. Our path interventions preserve the graph-free scoring interface and fixed candidate pool; training recipes supply supporting diagnostics.

Hard-negative mining~\citep{xiong2021approximate,qu2021rocketqa} and pool mismatch~\citep{gao2021lce,jacob2024drowning} motivate the checkpoint-dependent comparisons in Table~\ref{tab:oracle_fallacy}. Reciprocal rank fusion~\citep{cormack2009rrf} does not reproduce the code-preview PathSwap contrast. SweRank reranking yields 216/274 = 78.8\% strict file Acc@1 on unperturbed Lite issues (Appendix~\ref{sec:codeaware:swerank_repro}), and the path-document probe is reported separately.

\section{Full-File Repository Navigation and Code-Inspecting Pipelines}\label{app:revision-source-studies}
This appendix documents the 193-issue SWE-bench Lite studies of read-only repository navigation (GPT-4.1-mini and \texttt{gpt-6.1-sol}) and established code-inspecting pipelines (Agentless and Moatless). These runs perform localization only---no patch generation or test execution---and are kept separate from the matched 14B fixed-preview control and the four-view Qwen3 complete-file selection comparison.

\subsection{Active repository navigation}\label{app:revision-navigation}
\paragraph{Population and handle mapping.}
We evaluate 193 SWE-bench Lite issues across eleven repositories for which the \texttt{PathSwap-agent-v1} handle constructor produces paired original and opaque repository trees (six additional \texttt{pylint} issues were excluded because single-character opaque stems draw from 52 ASCII letters and were exhausted). Distinct from the fixed-pool \texttt{d\_f4e8/m\_008e.py} map of \S\ref{sec:setup}, this interactive constructor preserves directory hierarchy, file extensions, and shared structural names while matching opaque stem length to a companion readable-alias stem length (starting with an ASCII letter followed by alphanumerics and avoiding collisions with original or previously assigned stems). Requests referencing hidden original paths are rejected under the opaque arm. Pre-fix source bytes, embedded code identifiers, and internal strings remain identical across handle maps.

The read-only environment exposes four tools---\texttt{list\_files}, \texttt{search\_code}, \texttt{open\_file}, and \texttt{finalize}---with no shell, editing, or test execution. The model may inspect complete source files and revise its ranked file list before finalizing, with no access to reference patches or test files.

\paragraph{Budgets and incomplete runs.}
Each arm runs independently under a budget of 24 model requests and 48 tool calls per trajectory. For GPT-4.1-mini (\texttt{gpt-4.1-mini-2025-04-14}), trajectories without a returned file fall into two categories: (i) 2 original and 9 opaque runs that hit the 24-request ceiling without finalizing, and (ii) 18 original and 25 opaque runs that remained without a recorded outcome even after one systematic retry due to unresolved infrastructure errors (17 unknown request outcomes plus 1 infrastructure-estimate limit under original handles; 25 unknown request outcomes under opaque handles). Both categories stay in the fixed 193-issue denominator.

\paragraph{Outcomes, identification bounds, and finalized rates.}
Primary success is binary first-file Hit@1 in the fixed 193-issue denominator, counting non-finalized runs as misses. GPT-4.1-mini drops from 142/193 under original handles to 92/193 under opaque handles (89 shared successes, 53 original-only, 3 opaque-only, 48 shared misses; $-25.91$pp, paired 10{,}000-resample issue bootstrap 95\% CI $[-32.1,-19.2]$pp, exact two-sided McNemar $p<.001$). Leaving the 18 original and 25 opaque unresolved infrastructure outcomes completely unrestricted while keeping the 2/9 request-cap failures as misses bounds accuracy to $[142,160]/193$ original and $[92,117]/193$ opaque, giving missing-outcome identification bounds of $[-68/193,-25/193]=[-35.23,-12.95]$pp (distinct from sampling uncertainty). Among finalized trajectories alone, hit rates are 142/173 ($82.08\%$) original and 92/159 ($57.86\%$) opaque on unequal arm denominators (descriptive sensitivity, not a paired subgroup).

\paragraph{Relation to four-view complete-file selection.}
In the separate four-view prompt study (Appendix~\ref{app:repair_outcomes}; 219 unambiguous file-hit records across 126 Lite issues using pre-index-correction upstream rankings), complete-file inspection reached 97/219 (Qwen3-14B) and 70/219 (Qwen3-8B), versus 49/219 and 35/219 with the narrow top-function view and 111/219 and 86/219 with complete file plus three hints---recovering 65 and 40 of 146 upstream misses. Those fixed-file context views remain distinct from the 193-issue navigation cohort.

\subsection{\texttt{gpt-6.1-sol} in the tool-using interface}\label{app:revision-gpt61}
Evaluated under the exact same 193 issues, handle maps, read-only tools, and 24-request / 48-tool-call budget, the high-reasoning configuration logged under the deployment identifier \texttt{gpt-6.1-sol} (without a separate dated snapshot ID) finalized all 386 trajectories with zero retries needed. First-file Hit@1 is 188/193 under both original and opaque handles (187 shared successes, 1 original-only, 1 opaque-only, 4 shared misses; $0.0$pp, paired 95\% CI $[-1.55,1.55]$, $B{=}10{,}000$, exact McNemar $p=1.0$). On the 51 issues where GPT-4.1-mini missed under original handles, \texttt{gpt-6.1-sol} succeeded on 49/51 under original handles and 48/51 under opaque handles.

\subsection{Agentless and Moatless under PathSwap}\label{app:revision-code-inspecting}
\paragraph{Systems and arms.}
Agentless~\citep{xia2024agentless} (v1.5.0, commit \texttt{b150f28}) and Moatless~\citep{orwall2024moatless} (\texttt{aorwall/moatless-tools} v0.0.2, commit \texttt{a1017b78}) ran with their native prompts and localization workflows using GPT-4o (\texttt{gpt-4o-2024-08-06}) on the same 193 Lite issues and handle maps. Each pipeline was evaluated under three path views with identical pre-fix source contents: original paths, opaque handles (\texttt{PathSwap-agent-v1}), and readable shuffled filenames. Returned paths were mapped back to canonical paths before scoring.

\paragraph{Metric and diagnostics.}
First-file Hit@1 evaluates the first mapped returned file---the top file of Agentless's combined top-3 output and the first file returned by Moatless's \texttt{IdentifyCode}---counting failed, empty, or invalid first selections as misses in the 193-issue denominator. Intervals use paired 10{,}000-resample issue bootstraps alongside exact two-sided McNemar tests (Table~\ref{tab:agentless_moatless_full}).

\begin{table}[H]
\centering\small
\caption{First-file Hit@1 of Agentless and Moatless (GPT-4o) on the 193 SWE-bench Lite issues of Appendix~\ref{app:revision-navigation}. Changes are arm minus original in percentage points with paired 95\% issue-bootstrap CIs ($B{=}10{,}000$).}\label{tab:agentless_moatless_full}
\begin{tabularx}{\linewidth}{@{}Xlrlll@{}}\toprule
{\colhead{System}} & {\colhead{Arm}} & {\colhead{Hit@1}} & {\colhead{Change}} & {\colhead{95\% CI}} & {\colhead{McNemar $p$}}\\\midrule
Agentless & Original & 135/193 & --- & --- & ---\\
 & PathSwap & 0/193 & $-69.9$ & $[-76.2,-63.2]$ & $4.6\times10^{-41}$\\
 & Filename shuffle & 23/193 & $-58.0$ & $[-65.3,-50.8]$ & $1.6\times10^{-31}$\\\midrule
Moatless & Original & 143/193 & --- & --- & ---\\
 & PathSwap & 131/193 & $-6.2$ & $[-11.4,-1.0]$ & $0.029$\\
 & Filename shuffle & 129/193 & $-7.3$ & $[-13.0,-1.6]$ & $0.020$\\\bottomrule
\end{tabularx}
\end{table}

\paragraph{Component-level breakdown.}
In Agentless, all three arms completed 193/193 native retrieval runs and the unchanged combination step. Although opaque handles yielded 0/193 first-file hits, 9/193 opaque combined top-three lists still contain a gold file at ranks 2--3. Component retrieval outputs were empty on 7 original, 63 opaque, and 6 shuffled issues, while the standalone tree-localization step scored 140/193 original, 0/193 opaque, and 18/193 shuffled (returning no valid file on 1 original and 12 opaque issues). Because retained logs do not separate empty or invalid final first selections from valid incorrect ones, we report system-level outcomes and these component summaries without isolating a single internal failure mode.

In Moatless, complete first-selection statuses are recorded across all 193 issues with no API or setup failures: original yields 143 valid correct, 49 valid incorrect, and 1 empty; opaque yields 131 valid correct, 59 valid incorrect, and 3 unusable (1 no-file loop error and 2 invalid first paths; 19 original-only vs.\ 7 opaque-only hits); and filename shuffling yields 129 valid correct, 58 valid incorrect, and 6 empty. Requiring every returned path in Moatless's list to validate lowers opaque accuracy by one issue to 130/193.

\end{document}